\documentclass[aps,prc,amsfonts,superscriptaddress,showpacs,showkeys,nofootinbib,floatfix,notitlepage]{revtex4-1}

\include{usepack}
\usepackage[margin=0.7in]{geometry}
\usepackage{color,graphicx}
\usepackage{tcolorbox}
\usepackage{dcolumn}
\usepackage{bm}
\usepackage{slashed}
\usepackage{cancel}
\usepackage{multirow}
\usepackage{mathtools}
\usepackage{soul}

\newcommand{\pold}{$d^{\uparrow}$}
\newcommand{\xT}{$x_{2}$\ }
\newcommand{\xB}{$x_1$\ }

\newcommand{\dbar}{$\bar{d(x)}$\ }

\newcommand{\ND}{$\mathrm{ND}_3$\ }

\begin{document}

\title{The Transverse Structure of the Deuteron with Drell-Yan}
\author{The SpinQuest Collaboration\footnote{Contact: D. Keller (dustin@virginia.edu)}} 


\begin{abstract}
We propose to measure neutron and deuteron transversity TMDs.  The quark transversity distributions of the nucleon are decoupled from the deuteron gluon transversity in the $Q^2$ evolution due to the chiral-odd property in the transversely-polarized target.  The gluon transversity TMD only exists for targets of spin greater or equal to 1 and does not mix with quark distributions at leading twist, thereby providing a particularly clean probe of gluonic degrees of freedom.  This experiment would be the first of its kind and would probe the gluonic structure of the deuteron, investigating exotic glue contributions in the nucleus not associated with individual nucleons.  This experiment can be performed with the SpinQuest polarized target recently assembled for experiment E1039 and the spectrometer already in place in NM4.  This new experimental setup would require very minimal modification to the target system and no modification to the detector package.  An additional RF-circuit and target coil are necessary to RF-modulate across the domain of the Larmor frequency to manipulate the solid-state target spin population densities. Dedicated beam-time with this novel target system is required to achieve our physics goals.
\end{abstract}

\maketitle

\clearpage
\tableofcontents
\clearpage
\section{Introduction}
How is the quantum spin built in composite systems?  This is the quintessential question of Spin Physics.  Efforts to answer this question have resulted in the realization that hadrons and nuclei have an increasingly complex internal structure, likely involving quark orbital angular momentum (OAM) as well as gluonic and sea-quark contributions.  The depth of this structure and these dynamics is just more recently beginning to be realized, due in large part to novel experimentation.  The next generation of Spin Physics experiments is now driven by a modern understanding of spin and must leverage the techniques and technology developed in recent years to acquire new data with a broader physics reach.

The spin of nucleons and nuclei is well known, but how the internal mechanisms of motion and conservation manifest to preserve this fixed quantized spin is still not clear.  What is clear is that spin, like mass, appears to be an emergent quantity based on constituent movement and interaction with the vacuum.  Since the pivotal results provided by the EMC collaboration \cite{emc}, the particle physics community has striven to make sense of experimental results, leading to extensive theoretical development.  Decades of experimental studies on high-energy polarized-hadron reactions have been performed to clarify the origin of spin mainly through longitudinally-polarized structure functions, sparking considerable work on how to decompose the nucleon spin, see reviews \cite{rev1,rev2,rev3,rev4,rev5}.

Studying the spin structure of the nucleon and nuclei is a complex subject, as the internal motion of the partons is relativistic, and it is non-trivial to define the angular momenta. In addition, gluon spin is generally thought to be gauge dependent \cite{gi1}, but there are investigations into quark-gluon spin components and OAM contribution in a gauge invariant way \cite{gi2}.
Considering the nonpeturbative nature of these studies, calculations based solely on first principles of QCD are prohibitively challenging. 
The parton model \cite{pm} illustrates the nucleon as a collection of quasi-free quarks, antiquarks, and gluons, with longitudinal momentum distributions described by parton densities.  The formalism of collinear factorization directly connects these concepts to QCD and provides the foundational framework needed in Spin Physics, but only quantifies structure in a single spatial dimension.

To investigate partons in the plane transverse to the direction of motion of its parent nucleon requires the Generalized Parton Distributions (GPDs) and Transverse Momentum Distributions (TMDs) \cite{diehl}.  For both GPDs and TMDs, the relevant scales are in the non-perturbative domain, in contrast to the longitudinal momentum fractions on which all types of parton distributions depend.  Subject to kinematics, the TMDs and GPDs can contain much more information on non-perturbative phenomena and are critical to the interpretation of spin dependent hadron-hadron and lepton-hadron collisions, providing the advantage of a multi-dimensional exploration of the structure of nucleons and nuclei.  Through this avenue, Spin Physics studies of the strong force in its non-perturbative domain and beyond can also provide insight into color confinement as well as the origin of dynamic mass and charge density.  The culmination of Spin Physics has yet to come, but, ultimately, experiments will reveal exactly how partonic interactions manifest into hadronic and nuclear degrees of freedom.
\begin{figure}[!tbph]
  \centering
  \begin{minipage}[b]{0.35\textwidth}
    \includegraphics[width=\textwidth]{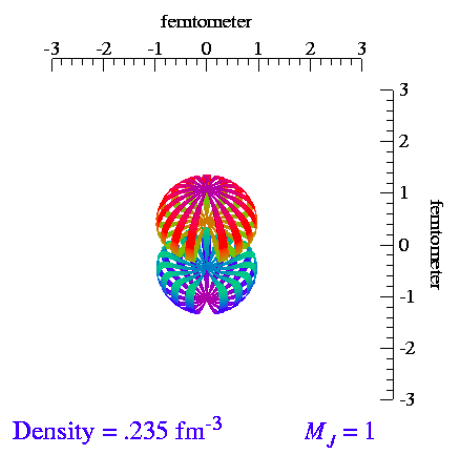}
  \end{minipage}
  \hspace{15mm}
  \begin{minipage}[b]{0.35\textwidth}
    \includegraphics[width=\textwidth]{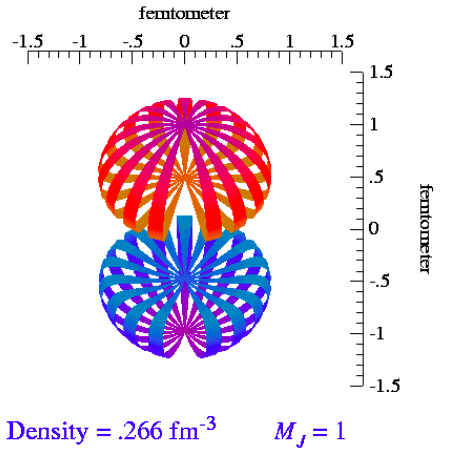}
  \end{minipage}
    \caption{Graphical representation of the shape of the deuteron for two specified equidensity surfaces.  Here, the deuteron is in the $M_J=1$ spin state.  The same is similar for $M_J=-1$.  Image from Argonne National Lab.}  
    \label{m1}
\end{figure}
The spin decomposition using lattice QCD (LQCD) \cite{lqcd1,lqcd2,lqcd3,lqcd4,lqcd5} also provides a guiding light.  Efforts have been made recently to obtain $x$-dependent parton distributions from LQCD \cite{lqcd6}.  Calculations of the nucleon spin from first principle simulations are beginning to provide results with control over all systematics \cite{lqcd7}. The best determined contributions so far are $\Sigma_q(\frac{1}{2}\Delta q)$, the quark intrinsic spin contribution with quark flavor ($q=u,~d,~s,~c$); $J_q$, the quark total angular momentum; $J_g$, the gluon total angular momentum; and $L_q$, the OAM of the quarks.  The PNMDE \cite{pnmde} collaboration have published results for $\Sigma_q(\frac{1}{2}\Delta q)$ and find $\Sigma_q=0.143(31)(36)$, consistent with the COMPASS value 0.13$<$$\frac{1}{2}\Delta\Sigma$$<$0.18 obtained at 3 GeV$^2$ \cite{comp}.  The ETMC \cite{etmc} collaboration has presented first results for $J_q$, $J_g$, and $L_q$ \cite{eng} for the OAM of quarks.  Within the next several years, improved high performance computing resources will allow much higher precision LQCD calculations, which will require much more experimental information as a basis for comparison.  In fact, the greatest opportunity to deepen our understanding will come from the intersection of consistent results from LQCD, phenomenology, and experiments over a broad range of kinematics.

The next generation of experiments must attempt to measure gluon-spin and partonic OAM contributions and further explore spin on a composite level by studying nuclei. To extract and understand this information, we need to investigate both the longitudinal spatial structure and the transverse momentum structure using novel methods. 
Though significant experimental progress has been made adding to the understanding of the spin structure of hadrons, the data frequently leaves more questions to be answered.  To understand the spin configuration of the nucleon and nuclei in terms of quarks and gluons remains one of the most challenging and critical open problems in nuclear physics \cite{jaffe1,leader}.  Vital experimental information is missing, especially around the transversely-polarized structure \cite{tran1,tran2,tran3,tran4,tran5}, with only minimal studies on quark transversity distributions \cite{trans}.  The transverse polarized target observables provide unique and crucial details on the 3D picture.  The internal workings of these observables are distinct from those of the longitudinal structure, as the quark transversity distributions are decoupled from the gluon transversity in the Q$^2$ evolution \cite{evol1,evol2,evol3} for polarized nuclei with spin $\ge$1, such as the deuteron, due to the helicity-flip (chiral-odd) property.
\begin{figure}[!tbp]
  \centering
  \begin{minipage}[b]{0.35\textwidth}
    \includegraphics[width=\textwidth]{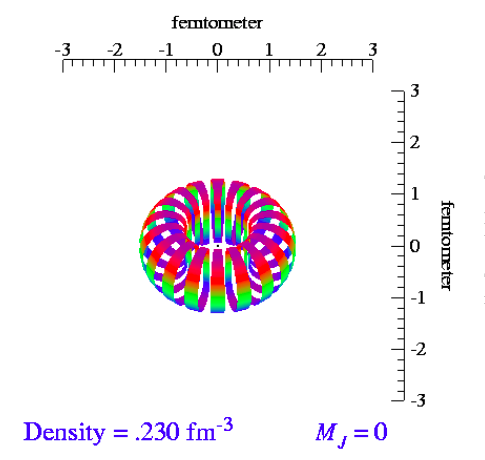}
  \end{minipage}
\hspace{15mm}
  \begin{minipage}[b]{0.35\textwidth}
    \includegraphics[width=\textwidth]{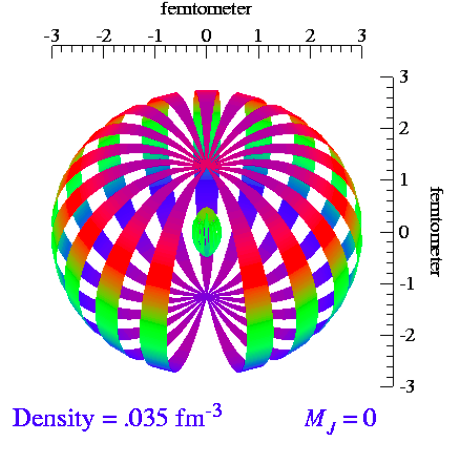}
  \end{minipage}
  \caption{Graphical representation of the shape of the deuteron for two specified equidensity surfaces.  Here the deuteron is in the $M_J=0$ spin state.  Image from Argonne National Lab.}
  \label{m0}
\end{figure}

The deuteron is the simplest spin-1 system and offers a vast array of observables to explore as we begin to build the composite spin picture of nuclei.  The deuteron initially appears as a loosely bound pair of nucleons with spins aligned (spin triplet state). However, the existence of the small quadrupole moment implies that these two nucleons are not in a pure S-state of relative orbital angular momentum and that the force between them is not central. Taking into account total spin and parity, an additional D-wave component is allowed.
There are several layers to understanding this system, starting with the tensor force.  The deuteron would simply not be bound without the tensor force, and there are geometric implications of this force on the deuteron structure which have yet to be explored on the quark and gluon level.
The spin configuration and alignment of the deuteron is a tool yet to be taken full advantage of.  If a deuteron can be aligned in such a fashion that it is in a  $M_J= \pm1$ magnetic substate (Fig. \ref{m1}), where $J$ is the spin of the deuteron, then the deuteron can have two separate equidensity surface lobes depending on the energy density.  This configuration is associated with the standard spin-up and spin-down common to the spin-1/2 nucleon, but, for spin-1, it is distinctly referred to as vector polarization.  On the other hand, if the deuteron is in the $M_J=0$ magnetic substate (Fig. \ref{m0}), then the equidensity surfaces that enclose the deuteron are toroidal in shape \cite{forest}.  The hole in the torus is due to the repulsive core of the $N$–$N$ interaction, and the overall shape is largely governed by the tensor force. It is only recently that the highly controlled manipulation of a solid-polarized target spin ensemble has allowed access to the optimally aligned high density deuteron targets, allowing increased sensitivity to the correlations between geometric properties and partonic degrees of freedom.  The use of the Transverse Momentum Distribution functions (TMDs) of polarizable nuclei offers the necessary connective bridge, allowing us to explore how these geometric properties emerge from quark and gluon dynamics.

We propose the first ever Spin-1 TMD measurements using a polarized deuteron target, including a direct measurement of gluon transversity, while also for the first time measuring the sea-quark transversity distribution of the deuteron/neutron.  The gluon transversity was first mentioned in regards to Deep Inelastic Scattering \cite{jaffe3}.  Contributions to this observable vanish identically for a nucleus made up of protons and neutrons, regardless of Fermi motion or binding corrections.  It is therefore an unambiguous probe of the gluonic components of the nuclear wavefunction, which cannot be identified with individual nucleons.
We propose to use the same SpinQuest/E1039 setup using Drell-Yan production from an unpolarized 120 GeV proton beam interacting on a transversely polarized deuteron target.  This experiment would use the exact same experimental configuration at Fermilab already setup in NM4.  With this new proposal, we suggest implementing polarized target technology in a dedicated run to optimize and separate the tensor polarized observables from the vector contributions, making these challenging measurements viable.  The unique beam cycle of the high intensity proton beam at Fermilab allows for special characteristics of the thermal properties of the solid-state polarized target system to be employed.  This provides significant improvement over any other facility to run intense proton beams on RF-manipulated target systems with the spin ensemble held outside of a thermal equilibrium state. The combination of high luminosity, large $x$-coverage, and a high-intensity beam with significant time between proton spills makes Fermilab \textbf{\textit{the best}} place for this novel approach to measuring polarized target asymmetries in Drell-Yan scattering with high precision. Using the antiquark selectivity of Drell-Yan, we will make the first ever determination of several observables, providing multiple constraints and significant advancement in the understanding of High-Energy/Nuclear Spin Physics.

In summary, this new proposal suggests taking full advantage of the new SpinQuest infrastructure by embarking on a Spin Physics program to measure multiple polarized observables in the deuteron within the range of $0.1< x_B <0.5$.  Here, we propose for the first time a way to probe exotic gluonic components in the target using transverse momentum distribution functions (TMDs).  This experiment would be highly complementary to the approved experiment E1039 \cite{E1039}, which will measure the Sivers function of the sea-quarks using both a polarized proton and deuteron target.  The physics presented here is also suggestive of other experiments to gain even further insight.
 
It is also important to note that the proposed measurement is the only currently planned experiment which will cleanly access the sea-quark and gluon transversity.  Fermilab provides a unique and complimentary kinematics with virtuality $Q^2 \sim 10$~GeV$^2$ and transverse momentum $q_T$ in the few GeV region.  This experiment is made possible by the SpinQuest polarized target and supporting infrastructure, as well as the technology that is required to optimize the deuteron target to access linearly polarized gluons.  It is necessary to measure the vector polarized asymmetry with zero tensor polarization and alternate with an enhanced tensor polarization and unpolarized target.  The technology to achieve these types of RF manipulated target systems has recently been developed at the University of Virginia and would require only small hardware modification to the SpinQuest target insert.  While this project would be a continuation of the SpinQuest effort, this proposal has its own physics goals that require dedicated beam-time on a specialized target.  All of the recent modification in the NM4 experimental hall are required.  The installation of the polarized target and closed loop liquid helium system, the modifications to the beamline to protect the target superconducting coils, and changes to the shielding around the target area and the first magnet are all still necessary in the exact same way for the proposed experiment.  There are no additional installation costs required for the proposed run.

\section{Motivation}

\subsection{The Spin-1 Target in Drell-Yan}
High energy scattering experiments are required to probe the quark and gluon structure of hadrons and hadronization processes.  This has made parton distribution functions (PDFs) and fragmentation functions (FFs) crucial tools for hadron and particle physicists for years. More recently, transverse-momentum-dependent distribution functions (TMDs) and fragmentation functions (TMD FFs) have been a primary focus in Spin Physics, both experimentally and theoretically.  At leading-twist, the internal transverse-momentum-dependent quark structure of spin-half hadrons is expressed in terms of six time-reversal even ($T$-even) and two time-reversal odd ($T$-odd) TMDs. After integrating over the transverse momenta of quarks, there remain three PDFs: the unpolarized, helicity, and transversity PDFs. However, for spin-one hadrons such as the deuteron, the spin degrees of freedom require three additional leading-twist $T$-even TMDs.  Before venturing into the specifics of the deuteron target, we look again at the nucleon observables so they can be sorted out on the experimental level.

The Drell-Yan process \cite{DY} describes the hadron-hadron collisions, where, at tree level, a quark from one particle annihilates with an antiquark from the other particle, creating a virtual photon.  The virtual photon subsequently decays into two leptons.  The Drell-Yan process is one of the cornerstone perturbative QCD processes that cleanly probes the internal structure of the colliding hadrons, has low background, and is free of fragmentation uncertainties.

 For this proposed experiment, we will use $p+d^\uparrow \to \mu^+ + \mu^- +X$ with a transverse vertically pointing deuteron. To lowest order, the cross-section for the Drell-Yan process depends on the product of the quark and antiquark distributions $q,~\bar{q}$, in the beam $x_1$ and in the target $x_2$, where $x_1$, $x_2$ are the Bjorken-$x$ of the process and express the fraction of the longitudinal momentum of the hadron carried by the quark.  The Drell-Yan cross-section can be written as,
\begin{equation}
\frac{d\sigma}{dx_1dx_2} = \frac{4\pi\alpha^{2}}{9sx_1x_2}\sum_{i}e^{2}_{i}(q_{i}^{B}(x_1,Q{^2})\bar{q}_{i}^{T}(x_2,Q{^2}) + \bar{q}_{i}^{B}(x_1,Q{^2})q_{i}^{T}(x_2,Q{^2}) \;, 
\label{DYF}
\end{equation}
where $s$ is the square of the center of mass energy and is given by $s = 2m_{T}E_{Beam} +m_{T}^{2} +m_{B}^{2}$,  $E_{Beam}$ is the beam energy, and $m_{B,T}$ are the rest masses of the beam and target particles.  Measuring the two decay leptons in the spectrometer allows one to determine the virtual photon center of mass momentum $p_{\|}^{\gamma}$ (longitudinal) and $p_{T}^{\gamma}$ (transverse) as well as the mass $M_{\gamma}$. From these quantities, one can deduce the momentum fractions of the quarks through:
\begin{equation}
x_{F}  =  \frac{p_{\|}^{\gamma}}{p_{\|}^{\gamma,max}} = x_1 - x_2 \, , \qquad
x_1x_2  =  M^{2}_{\gamma} \, .
\end{equation}
If one chooses the kinematics of the experiment such that $x_F > 0$ and \xB\ is large, the contributions from the valence quarks in the beam dominate. 

In this case, in Eq. \ref{DYF}, the second term becomes negligible, and the cross-section can be written as
\begin{equation}
\frac{d\sigma}{dx_1dx_2} \approx \frac{4\pi\alpha^{2}}{9sx_1x_2}\sum_{i}e^{2}_{i}q_{i}^{B}(x_1,Q{^2})\bar{q}_{i}^{T}(x_2,Q{^2}) \, .
\end{equation}
For a proton beam on a proton target, the process is dominated by the $u(x_1)$ distribution due to the charge factor $e^{2}_{i}$.  However, on a neutron (or deuteron) target, the process is $d(x_1)$ dominant.

To explore the sea-quark observables in the neutron, we consider the general form of the hadronic tensor from \cite{arnold} to express the full angular distribution of the Drell-Yan cross-section as,
\begin{equation}
\begin{gathered}
d \sigma d^{4} q d \Omega=\frac{\alpha_{e m}^{2}}{F q^{2}} \times\left\{\left(\left(1+\cos ^{2} \theta\right) F_{U U}^{1}+\left(1-\cos ^{2} \theta\right) F_{U U}^{2}+\sin 2 \theta \cos \phi F_{U U}^{\cos \phi}+\sin ^{2} \theta \cos 2 \phi F_{U U}^{\cos 2 \phi}\right)\right. \\
+S_{a L}\left(\sin 2 \theta \sin \phi F_{L U}^{\sin \phi}+\sin ^{2} \theta \sin 2 \phi F_{L U}^{\sin 2 \phi}\right)+S_{b L}\left(\sin 2 \theta \sin \phi F_{U L}^{\sin \phi}+\sin ^{2} \theta \sin 2 \phi F_{U L}^{\sin 2 \phi}\right) \\
+\left|\vec{S}_{a T}\right|\left[\sin \phi_{a}\left(\left(1+\cos ^{2} \theta\right) F_{T U}^{1}+\left(1-\cos ^{2} \theta\right) F_{T U}^{2}+\sin 2 \theta \cos \phi F_{T U}^{\cos \phi}+\sin ^{2} \theta \cos 2 \phi F_{T U}^{\cos 2 \phi}\right)\right. \\
\left.+\cos \phi_{a}\left(\sin 2 \theta \sin \phi F_{T U}^{\sin \phi}+\sin ^{2} \theta \sin 2 \phi F_{T U}^{\sin 2 \phi}\right)\right] \\
+\left|\vec{S}_{b T}\right|\left[\sin \phi_{b}\left(\left(1+\cos ^{2} \theta\right) F_{U T}^{1}+\left(1-\cos ^{2} \theta\right) F_{U T}^{2}+\sin 2 \theta \cos \phi F_{U T}^{\cos \phi}+\sin ^{2} \theta \cos 2 \phi F_{U T}^{\cos 2 \phi}\right)\right. \\
\left.+\cos \phi_{b}\left(\sin 2 \theta \sin \phi F_{U T}^{\sin \phi}+\sin ^{2} \theta \sin 2 \phi F_{U T}^{\sin 2 \phi}\right)\right] \\
+S_{a L} S_{b L}\left(\left(1+\cos ^{2} \theta\right) F_{L L}^{1}+\left(1-\cos ^{2} \theta\right) F_{L L}^{2}+\sin ^{2} 2 \theta \cos \phi F_{L L}^{\cos \phi}+\sin ^{2} \theta \cos 2 \phi F_{L L}^{\cos 2 \phi}\right) \\
+\left|\vec{S}_{a T}\right| S_{b L}\left[\cos \phi_{a}\left(\left(1+\cos ^{2} \theta\right) F_{T L}^{1}+\left(1-\cos ^{2} \theta\right) F_{T L}^{2}+\sin 2 \theta \cos \phi F_{T L}^{\cos \phi}+\sin ^{2} \theta \cos 2 \phi F_{T L}^{\cos 2 \phi}\right)\right. \\
\left.+\sin _{a}\left(\sin 2 \theta \sin \phi F_{T L}^{\sin \phi}+\sin ^{2} \theta \sin 2 \phi F_{T L}^{\sin 2 \phi}\right)\right] \\
+S_{a L}\left|\vec{S}_{b T}\right|\left[\cos \phi_{b}\left(\left(1+\cos ^{2} \theta\right) F_{L T}^{1}+\left(1-\cos ^{2} \theta\right) F_{L T}^{2}+\sin 2 \theta \cos \phi F_{L T}^{\cos \phi}+\sin ^{2} \theta \cos 2 \phi F_{L T}^{\cos 2 \phi}\right)\right. \\
\left.+\sin _{b}\left(\sin 2 \theta \sin \phi F_{L T}^{\sin \phi}+\sin ^{2} \theta \sin 2 \phi F_{L T}^{\sin 2 \phi}\right)\right]+
\cr \left|\vec{S}_{a T}\right|\left|\vec{S}_{b T}\right|\left[\cos \left(\phi_{a}+\phi_{b}\right)\left(\left(1+\cos ^{2} \theta\right) F_{T T}^{1}+\left(1-\cos ^{2} \theta\right) F_{T T}^{2}+\sin 2 \theta \cos \phi F_{T T}^{\cos \phi}+\sin ^{2} \theta \cos 2 \phi F_{T T}^{\cos 2 \phi}\right)\right.\\
+ \cos \left(\phi_{a}-\phi_{b}\right)\left(\left(1+\cos ^{2} \theta\right) \bar{F}_{T T}^{1}+\left(1-\cos ^{2} \theta\right) \bar{F}_{T T}^{2}+\sin 2 \theta \cos \phi \bar{F}_{T T}^{\cos \phi}+\sin ^{2} \theta \cos 2 \phi \bar{F}_{T T}^{\cos 2 \phi}\right)
\cr +\sin \left(\phi_{a}+\phi_{b}\right)\left(\sin 2 \theta \sin \phi F_{T T}^{\sin \phi}+\sin ^{2} \theta \sin 2 \phi F_{T T}^{\sin 2 \phi}\right)
\left.\left.+\sin \left(\phi_{a}-\phi_{b}\right)\left(\sin 2 \theta \sin \phi \bar{F}_{T T}^{\sin \phi}+\sin ^{2} \theta \sin 2 \phi \bar{F}_{T T}^{\sin 2 \phi}\right)\right]\right\}.\nonumber
\end{gathered}
\end{equation}

\noindent The notation $a$ and $b$ are used in \cite{arnold} to differentiate between the two interacting hadrons.  Here, there are 48 structure functions that can play some type of role in the observables.  In order to shorten the notation, the indices for the angles which characterize the lepton momenta and the transverse spin vectors of the hadrons are left out. Also, the components of the spin vectors can be understood in different frames, such as the rest frame of one of the hadrons, the cm-frame, or the dilepton rest frame.  Below, we will specify the frame as needed.  

For the additional structure functions that surface from the spin-1 target, see \cite{kum0,kum2,kum1,kumano1,kumano2,kumano3,kumano4}.  In particular, there are 108 structure functions for the Spin-1 target with 60 of them being associated with the tensor structure of the deuteron.

Summing over the polarizations of the produced leptons, the expression for the Drell-Yan cross-section using a transversely polarized nucleon target contains five transverse spin-dependent asymmetries.  This part of
the differential cross-section can be expressed as \cite{siss2},
\begin{equation}
\frac{d \sigma}{d q^{4} d \Omega} \propto \hat{\sigma}_{U}\left\{1+S_{T}\left[D_{1} A_{T}^{\sin \varphi_s} \sin \varphi_{s}\right.\right.
+D_{2}\left(A_{T}^{\sin \left(2 \varphi_{cs}-\varphi_{s}\right)} \sin \left(2 \varphi_{cs}-\varphi_{s}\right)\right.
\left.\left.\left.+A_{T}^{\sin (2 \varphi_{cs}+\varphi_{cs})} \sin \left(2 \varphi_{cs}+\varphi_{s}\right)\right)\right]\right\},
\end{equation}
where $q$ is the four-momentum of the virtual photon, $\hat{\sigma}_U=(F_U^1+F_U^2)(1+\lambda \text{cos}^2\theta_{cs})$, $F^1_U$, $F_U^2$ are the polarization and azimuth-independent structure functions, and polar asymmetry $\lambda$ is given as $\lambda=(F^1_U-F^2_U)/(F_U^1+F^2_U)$.  $D_1=(1+\text{cos}^2\theta_{cs})/(1+\lambda\text{cos}^2\theta_{cs})$ and $D_2=\text{sin}^2\theta_{cs}/(1+\lambda\text{cos}^2\theta_{cs})$.  The angles $\varphi_{cs}$, $\theta_{cs}$, and $\Omega$, the solid angle of the lepton, are defined in the Collins-Soper frame \cite{arnold}.  Naturally, $S_T$ is the transverse part of the nucleon spin, and the azimuthal angle $\varphi_S$ is the transverse
spin orientation $S_T$ of the target (determined in the target rest frame).

With the deuteron, one can measure observables from the spin-1/2 neutron-proton pair; with our kinematics, these are specific to the sea-quarks.  It is also possible to measure observables specific to the spin-1 target as a whole.  To extract the transverse spin TMDs in the deuteron, one has to measure the $p+$\pold~transverse spin asymmetry with the target either in the vector or tensor polarization state.  The polarization of the solid-state target can be manipulated with RF-techniques.  The RF spin manipulation to orient the target ensemble specific to a particular observable can be achieved in-between beam spills, optimizing the figure of merit for the beam-target interaction time.  The time between beam spills required by the FNAL main injector (55.6 sec) is an advantage, in this case, allowing time to selectively optimize the target to isolate specific sea-quark and gluon observables of interest using novel RF polarized target technology.  The time between beam spills also allows target spin flips per spill, reducing the time-dependant drifts in the target asymmetry measurements.

In Drell-Yan lepton-pair production with transversely polarized nucleons in the initial state, the TSA $A^{\text{sin}\varphi_s}_T$ is related to the Sivers TMD by a convolution, and the QCD predicted sign-change can be measured in the Drell-Yan process when compared at the same kinematics to the semi-inclusive deep inelastic scattering process (SIDIS).  The other two asymmetries, $A_T^{\text{sin}(2\varphi_{cs}-\varphi_s)}$ and $A_T^{\text{sin}(2\varphi_{cs}+\varphi_s)}$, are related to convolutions of the beam Boer-Mulders ($h^\perp_{1}$) and the target transversity ($h_{1}$) or pretzelosity ($h^\perp_{1T}$) such that, 
\begin{eqnarray}
\text{Boer-Mulders}&\otimes&  \text{Boer-Mulders}:~~~~A_{T}^{\cos 2 \varphi_{c s}} \propto h_{1}^{\perp q} \otimes h_{1}^{\perp q}\\
\text{Unpolarized}&\otimes&  \text{Sivers}:~~~~~~~~~~~~~~~A_{T}^{\sin \varphi_{s}} \quad \propto f_{1}^{q} \otimes f_{1 T}^{\perp q}\\
\text{Boer-Mulders}&\otimes&  \text{Transversity}:~~~~~~A_{T}^{\sin \left(2 \varphi_{cs}-\varphi_{s}\right)} \propto h_{1}^{\perp q} \otimes   h_{1}^{q} \\
\text{Boer-Mulders}&\otimes&  \text{Pretzelosity}:~~~~~~~A_{T}^{\sin \left(2 \varphi_{c s}+\varphi_{s}\right)} \propto h_{1}^{\perp q} \otimes h_{1 T}^{\perp q}.
\end{eqnarray}
Combined with the kinematic information and the target polarization, we can access the TMDs given the experimental asymmetries.  Specifically, for a vector polarized deuteron target at SpinQuest, we can get access to the sea-quark transversity by focusing on the single spin asymmetry  \cite{siss},
\begin{equation}
\left.A_{U T}^{\sin \left(\varphi_{cs}+\varphi_{s}\right) \frac{q_{T}}{M_{N}}}\right|_{p D^{\uparrow} \rightarrow l^{+} l^{-} X} \simeq-\frac{\left[4 h_{1 u}^{\perp(1)}\left(x_{p}\right)+h_{1 d}^{\perp(1)}\left(x_{p}\right)\right]\left[\bar{h}_{1 u}\left(x_{D^{\uparrow}}\right)+\bar{h}_{1 d}\left(x_{D^{\uparrow}}\right)\right]}{\left[4 f_{1 u}\left(x_{p}\right)+f_{1 d}\left(x_{p}\right)\right]\left[\bar{f}_{1 u}\left(x_{D^{\uparrow}}\right)+\bar{f}_{1 d}\left(x_{D^{\uparrow}}\right)\right]}.
\end{equation}
Here, the Boer-Mulders function $(h_{1}^{\perp q})$ portion can also be measured in the $\text{cos}(2\varphi_{cs})$ term of the unpolarized Drell-Yan measurement \cite{sis}.  Using
the strictly vector polarized deuteron target provides a clean probe to the $\bar{d}$-quark transversity $\bar{h}_{1d}$.  This is a primary motivation of this proposal and physically represents the $\bar{d}$-quark polarization in the transversely polarized
deuteron.  To optimize such an experiment, the target should be only vector polarized in the transverse vertical direction, unlike the standard Boltzmann equilibrium spin configured deuteron target required for SpinQuest E1039, which contains a mix of vector and tensor polarized deuterons.  It is necessary to mitigate contamination from the tensor polarized observables to isolate quark polarization contribution to the TSA.  Such a target requires special treatment and is discussed later in Section \ref{rf2}.

Unlike the Sivers function \cite{E1039}, the quark transversity and pretzelosity are predicted to exhibit true, or \textit{genuine}, universality and do not have a sign-change between SIDIS and Drell-Yan exhibited by \textit{conditional} universality.  These universality relations provide a set of fundamental QCD predictions that must be checked experimentally.  These are summarized as,
\begin{eqnarray}
\left.h_{1}^{q}\right|_{SIDIS}&=&~\left.h_{1}^{q}\right|_{DY} \\
\left.h_{1 T}^{\perp q}\right|_{SIDIS}&=&~\left.h_{1 T}^{\perp q}\right|_{DY}\\
\left.h_{1}^{\perp q}\right|_{SIDIS} &=&-\left.h_{1}^{\perp q}\right|_{D Y} \\
\left.f_{1 T}^{\perp q}\right|_{SIDIS} &=&-\left.f_{1 T}^{\perp q}\right|_{D Y}.
\end{eqnarray}
With the combined experimental data from E1039 and these proposed measurements, all of the right-hand side of each of these relations can be measured specifically for the sea-quarks.  There are no other experiments that can directly measure the sea-quark contribution, so this data will be essential for separating the sea and valance contribution for global fits and deepening our general understanding.

In Drell–Yan processes, the transverse motion of quarks in nucleons is particularly important. We measure the transverse momentum of a particle that inherits part of the quark intrinsic transverse momentum.  The quark momentum is given as,
$$k^{\mu}\approx x P^{\mu}+k^{\mu}_T.$$
The most general form of the correlation matrix can be expressed as \cite{tran4},
\begin{equation}
\Phi(k, P, S)=\frac{1}{2}\left[A_{1} \not P+A_{2} \lambda_{N} \gamma_{5} \not P+A_{3} \not P \gamma_{5} \not \phi_{\mathrm{T}}\right].
\end{equation}
When the correlation matrix is integrated over $k$, the result gives,
\begin{equation}
\Phi=\int \mathrm{d}^{4} k \Phi(k, P, S)=\frac{1}{2}\left[g_{\mathrm{V}} \not p+g_{\mathrm{A}} \lambda_{N} \gamma_{5} \not P+g_{\mathrm{T}} \not P \gamma_{5} \not \phi_{\mathrm{T}}\right],
\end{equation}
where the constants $g_v$, $g_A$, and $g_T$ are the vector, axial, and tensor charge.  They can be calculated from the quark and anitquark distribution functions as \cite{charge},
\begin{equation}
\begin{aligned}
g_{\mathrm{V}}^{q} &=\int_{0}^{1} \mathrm{d} x\left[f_{1}^{q}(x)-f_{1}^{\bar{q}}(x)\right] \\
g_{\mathrm{A}}^{q} &=\int_{0}^{1} \mathrm{d} x\left[g_{1}^{q}(x)+g_{1}^{\bar{q}}(x)\right] \\
g_{\mathrm{T}}^{q} &=\int_{0}^{1} \mathrm{d} x\left[h_{1}^{q}(x)-h_{1}^{\bar{q}}(x)\right].
\end{aligned}
\end{equation}
Note that the vector charge is just the valence number. 
As a consequence of the charge conjugation properties of the field bilinears, the vector and tensor charges are the first moments of flavor non-singlet combinations (quarks minus antiquarks), whereas the axial charge is the first moment of a flavor singlet combination (quarks plus antiquarks).  We will come back to the tensor charge later.

The description of the quark-quark correlation matrix at leading twist is,
\begin{equation}
\begin{aligned}
\Phi=\frac{1}{2} & {\left[\not A_{1}+\varepsilon_{\mu \nu \rho \sigma} \gamma^{\mu} \frac{P^{\nu} k_{\mathrm{T}}^{\rho} S_{ \mathrm{T} }^{\sigma}}{M} \tilde{A}_{1}\right.} \\
&+\lambda_{N} \gamma_{5} \not P A_{2}+\frac{k_{\mathrm{T}} \cdot S_{\mathrm{T}}}{M} \gamma_{5} \not p \tilde{A}_{2} \\
&+P \gamma_{5}+A_{3}+\frac{k_{\mathrm{T}} \cdot S_{\mathrm{T}}}{M^{2}} \not p_{\gamma_{5}} k_{\mathrm{T}} \tilde{A}_{3}+\frac{\lambda_{N}}{M} \not P \gamma_{5} k_{\mathrm{T}} \tilde{A}_{4} \\
&\left.+\varepsilon_{\mu \nu \rho \sigma} \gamma^{\mu} \gamma^{\nu} \gamma_{5} \frac{P^{\rho} k_{\mathrm{T}}^{\sigma} \tilde{A}_{5}}{M}\right]
\end{aligned}
\end{equation}

In total, the matrix is
described by 8 functions, where $A_n$ and $\tilde{A}_n$ are real parameters used to simplify the characterization \cite{matousek}. Powers of the nucleon mass $M$ are present to keep the functions dimensionless while considering the quark transverse polarization $S_T$.  

\begin{figure}
    \centering
    \includegraphics[width=\textwidth]{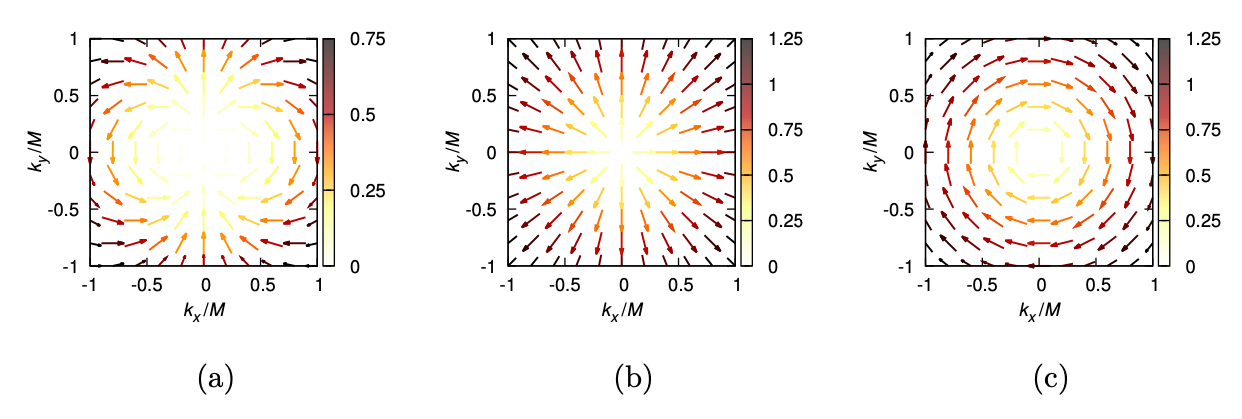}
    \caption{The kinematic factors accompanying $h^\perp_{1T}$(a),~$h^\perp_{1L}$(b), and $h^\perp_1$(c) for ($\boldsymbol{S}$) = (0,1,0).  The position of the center of each arrow corresponds to the quark transverse momentum, its direction denotes the preferred quark polarization, and the color shows the modulus of the factor \cite{matousek}.}
    \label{fig:Kt}
\end{figure}

The pretzelosity ($h_{1T}^\perp$) with the transversity ($h_1$) determine the transverse polarization distribution of quarks in a transversely polarized nucleon.  There is modeling \cite{r1} that shows that the pretzelosity is a direct measurement of the parton orbital angular momentum and can be connected to spin densities \cite{r2}.  Also, model calculations \cite{r3} indicate that different angular momentum components contribute to the deformation of the nucleon shape.
With pretzelosity, there are direct correlations between quark transverse momentum and spin.  As shown in Fig. \ref{fig:Kt}(a), if pretzelosity is nonzero, quarks in a nucleon polarized along the $y$-axis may be polarized in all transverse directions, depending on their momentum.  The position of the center of each arrow corresponds to the quark transverse momentum, its direction denotes the preferred quark polarization, and the color shows the modulus of the factor \cite{matousek}.  

In Fig. \ref{fig:Kt}(b), the transverse polarized worm-gear ($h^\perp_{1L}$) is shown which describes transverse polarization of quarks in a longitudinally
polarized nucleon.  If nucleon helicity ($g_1$) and $h^\perp_{1L}$ are both positive, the quarks are polarized along the direction of their transverse motion.  

The correlation between quark transverse polarization and momentum may exist in an unpolarized nucleon (or even in a spin-less hadron), as shown in Fig. \ref{fig:Kt}(c).  This is, of course, the Boer–Mulders function ($h_1^\perp$).  The quark polarization in this case would be perpendicular to its transverse momentum.  

Much of this can be explored for the $\bar{d}$ case in the deuteron, though we focus on transversity of the mentioned sea and gluons in this proposal.  The trick, of course, is to be able to disentangle the quark observables from the gluons by manifestly isolating either the spin-1/2 or the spin-1 case for the same polarized target.

The decomposition of the correlators in terms of relevant structures allowed by symmetry and scaling by the non-perturbative TMD functions is now a common and advantageous practice. This enables a singling out of the relevant quantities that contribute to the cross-section of a selected process. The complete parametrization of the TMD correlator for quarks, including the $T$-odd structure, is given in \cite{83} for spin-1/2 hadrons, and complemented in \cite{bacch,85} with the addition of spin-1 hadrons with the tensor polarization parts for quarks. For gluons, the first parametrization was performed in \cite{86}, followed by \cite{87}, with extended parameterization in \cite{boer}.  The work on gluons indicate that some distributions are accessible in polarized nuclei.  Exploring nuclei in pursuit of gluonic content of hadrons of spin greater than $1/2$ is highly attractive, especially because they are expected to be accessible at high-$x$.  Looking at novel gluon distributions, not related to the ones from the nucleons, is very interesting in the study of exotic effects in the binding of nuclei, as well as their dynamic contribution to spin and mass.

To consider the application to the full spin-1 target including the tensor polarization components, we have to start with the deuteron polarization density matrix.
In being consistent with the popular work on the subject, the subscript $U$ is used to denote unpolarized hadrons, the subscripts $L$ and $T$ are used to denote respectively longitudinal and transverse vector polarization, and the subscripts $LL$, $LT$, and $TT$ are used to denote longitudinal-longitudinal, longitudinal-transverse, and transverse-transverse tensor polarization.  The tensor polarizations have double index, indicating a specific orientation of the tensor polarized state ($M_J=0$) of the spin-1 target.  It is also necessary to use superscripts to indicate which axis is the axis of quantization.  For example, $S_{LL}$ is the longitudinal component of the spin tensor, and it is oriented longitudinally along the z-axis, or the beam-line.  However, the $S_{TL}^x$ term indicates a tensor polarization pointed $\pi/4$ with respect to the beam line in the xz-plane, where the x-axis is pointing directly vertical transverse to the beam-line, and the y-axis is pointing sideways transverse to the beam-line. 

The density matrix has the form:
 \begin{equation}
\rho(S, T)=\frac{1}{3}\left(I+\frac{3}{2} S^{i} \Sigma^{i}+3 T^{i j} \Sigma^{i j}\right), 
\end{equation} 
where the components $S^i$ of the vector $S$ represent the vector part of the spin. The tensor part of the spin state is represented by the $T^{ij}$ by demanding $P_\mu T^{\mu\nu}$. 
With this notation in mind, the density matrix is parameterized in terms of a spacelike spin vector
 \textbf{\textit{$S$}} and a symmetric traceless spin tensor \textbf{\textit{$T$}} \cite{boer}:
 \begin{equation}
 S^{\mu}=S_{L} \frac{P^{\mu}}{M}+S_{T}^{\mu}-M S_{L} n^{\mu}
\end{equation} 
 and,
\begin{equation}
T^{\mu \nu}=\frac{1}{2}\left[\frac{2}{3} S_{L L} g_{T}^{\mu \nu}+\frac{4}{3} S_{L L} \frac{P^{\mu} P^{\nu}}{M^{2}}+\frac{S_{L T}^{\{\mu} P^{\nu\}}}{M}+S_{T T}^{\mu \nu}\left.-\frac{4}{3} S_{L L} P^{\{\mu} n^{\nu\}}-M S_{L T}^{\{\mu} n^{\nu\}}+\frac{4}{3} M^{2} S_{L L} n^{\mu} n^{\nu}\right.\right]
\end{equation}
The density matrix would take the form,
\begin{eqnarray}
\rho(S, T)&=\left(\begin{array}{ccc}
\frac{1}{3}+\frac{S_{L}}{2}+\frac{S_{L L}}{3} & \frac{S_{T}^{x}-i S_{T}^{y}}{2 \sqrt{2}}+\frac{S_{L T}^{x}-i S_{L T}^{y}}{2 \sqrt{2}} & \frac{S_{T T}^{x x}-i S_{T T}^{x y}}{2} \\
\frac{S_{T}^{x+i S_{T}^{y}}}{2 \sqrt{2}}+\frac{S_{L T}^{x}+i S_{L T}^{y}}{2 \sqrt{2}} & \frac{1}{3}-\frac{2 S_{L L}}{3} & \frac{S_{T}^{x}-i S_{T}^{y}}{2 \sqrt{2}}-\frac{S_{L T}^{x}-i S_{L T}^{y}}{2 \sqrt{2}} \\
\frac{S_{T T}^{x x}+i S_{T T}^{x y}}{2} & \frac{S_{T}^{x}+i S_{T}^{y}}{2 \sqrt{2}}-\frac{S_{L T}^{x}+i S_{L T}^{y}}{2 \sqrt{2}} & \frac{1}{3}-\frac{S_{L}}{2}+\frac{S_{L L}}{3}
\end{array}\right).
\end{eqnarray}

To explore both transversity of quarks and gluons with the same spin-1 target, we must take a closer look at the leading-twist correlators for both.  For parametrization of the quarks, the leading-twist TMD correlator is,
\begin{equation}
\begin{aligned}
\Phi\left(x, \boldsymbol{k}_{T}\right) & \equiv \Phi^{[U]}\left(x, \boldsymbol{k}_{T} ; n, P, S, T\right) \\
&\left.\equiv \int \frac{d(\xi \cdot P) d^{2} \boldsymbol{k}_{T}}{(2 \pi)^{3}} e^{i k \cdot \xi}\langle P, S, T|\bar{\psi}(0) U(0, \xi) \psi(\xi)| P, S, T\rangle\right|_{\xi^{+}=0}.
\end{aligned}
\end{equation}
Using the indicated notation, the quark correlator is organized in terms of target polarization such that,
$$\Phi=\Phi_U+\Phi_L+\Phi_T+\Phi_{LL}+\Phi_{LT}+\Phi_{TT},$$
and the decomposition is expressed as:
\begin{equation}
\begin{aligned}
&\Phi_{U}\left(x, \boldsymbol{p}_{T}\right)=\frac{1}{4}\left\{f_{1}\left(x, p_{T}^{2}\right) \not n_{+}+\left(h_{1}^{\perp}\left(x, p_{T}^{2}\right) \sigma_{\mu \nu} \frac{p_{T}^{\mu}}{M} n_{+}^{\nu}\right)\right\}\\
&\Phi_{L}\left(x, \boldsymbol{p}_{T}\right)=\frac{1}{4}\left\{g_{1 L}\left(x, p_{T}^{2}\right) S_{L} \gamma_{5} \not n_{+}+h_{1 L}^{\perp}\left(x, p_{T}^{2}\right) S_{L} \text { i } \sigma_{\mu \nu} \gamma_{5} n_{+}^{\mu} \frac{p_{T}^{\nu}}{M}\right\}\\
&\Phi_{T}\left(x, \boldsymbol{p}_{T}\right)=\frac{1}{4}\left\{g_{1 T}\left(x, p_{T}^{2}\right) \frac{\boldsymbol{S}_{T} \cdot \boldsymbol{p}_{T}}{M} \gamma_{5} \not n_{+}+h_{1 T}\left(x, p_{T}^{2}\right) \mathrm{i} \sigma_{\mu \nu} \gamma_{5} n_{+}^{\mu} S_{T}^{\nu}\right.\\
&+h_{1 T}^{\perp}\left(x, p_{T}^{2}\right) \frac{\boldsymbol{S}_{T} \cdot \boldsymbol{p}_{T}}{M} \mathrm{i} \sigma_{\mu \nu} \gamma_{5} n_{+}^{\mu} \frac{p_{T}^{\nu}}{M}\\
&\left.+\left(f_{1 T}^{\perp}\left(x, p_{T}^{2}\right) \epsilon_{\mu \nu \rho \sigma} \gamma^{\mu} n_{+}^{\nu} \frac{p_{T}^{\rho}}{M} S_{T}^{\sigma}\right)\right\}\\
&\Phi_{L L}\left(x, \boldsymbol{p}_{T}\right)=\frac{1}{4}\left\{f_{1 L L}\left(x, p_{T}^{2}\right) S_{L L} h_{+}+\left(h_{1 L L}^{\perp}\left(x, p_{T}^{2}\right) S_{L L} \sigma_{\mu \nu} \frac{p_{T}^{\mu}}{M} n_{+}^{\nu}\right)\right\}\\
&\Phi_{L T}\left(x, \boldsymbol{p}_{T}\right)=\frac{1}{4}\left\{f_{1 L T}\left(x, p_{T}^{2}\right) \frac{\boldsymbol{S}_{L T} \cdot \boldsymbol{p}_{T}}{M} \not n_{+}+\left(g_{1 L T}\left(x, p_{T}^{2}\right) \epsilon_{T}^{\mu \nu} S_{L T \mu} \frac{p_{T} \nu}{M} \gamma_{5} h_{+}\right)\right.\\
&+\left(h_{1 L T}^{\prime}\left(x, p_{T}^{2}\right) \mathrm{i} \sigma_{\mu \nu} \gamma_{5} n_{+}^{\mu} \epsilon_{T}^{\nu \rho} S_{L T \rho}\right)\\
&\left.+\left(h_{1 L T}^{\perp}\left(x, p_{T}^{2}\right) \frac{\boldsymbol{S}_{L T} \cdot \boldsymbol{p}_{T}}{M} \sigma_{\mu \nu} \frac{p_{T}^{\mu}}{M} n_{+}^{\nu}\right)\right\}\\
&\Phi_{T T}\left(x, \boldsymbol{p}_{T}\right)=\frac{1}{4}\left\{f_{1 T T}\left(x, p_{T}^{2}\right) \frac{\boldsymbol{p}_{T} \cdot \boldsymbol{S}_{T T} \cdot \boldsymbol{p}_{T}}{M^{2}} h_{+}\right.\\
&-\left(g_{1 T T}\left(x, p_{T}^{2}\right) \epsilon_{T}^{\mu \nu} S_{T T \nu \rho} \frac{p_{T}^{2} p_{T \mu}}{M^{2}} \gamma_{5} \not h_{+}\right)\\
&-\left(h_{1 T T}^{\prime}\left(x, p_{T}^{2}\right) \mathrm{i} \sigma_{\mu \nu} \gamma_{5} n_{+}^{\mu} \epsilon_{T}^{\nu \rho} S_{T T} \rho_{P \sigma} \frac{p_{T}^{\sigma}}{M}\right)\\
&\left.+\left(h_{1 T T}^{\perp}\left(x, p_{T}^{2}\right) \frac{\boldsymbol{p}_{T} \cdot \boldsymbol{S}_{T T} \cdot \boldsymbol{p}_{T}}{M^{2}} \sigma_{\mu \nu} \frac{p_{T}^{\mu}}{M} n_{+}^{\nu}\right)\right\} .
\end{aligned}
\end{equation}

In regards to the phenomenology, the intrinsic motion of partons inside the nucleons is responsible for the specific dependence of the cross-section in the azimuthal angle.  The various correlations encoded in the TMDs translate into the aforementioned azimuthal or spin asymmetries of the measured cross-section, which are calculable and provide the basis for measurements that give access to a physical interpretation of structure and dynamics.

\begin{figure}[ht]
\centering
\includegraphics[width=11cm]{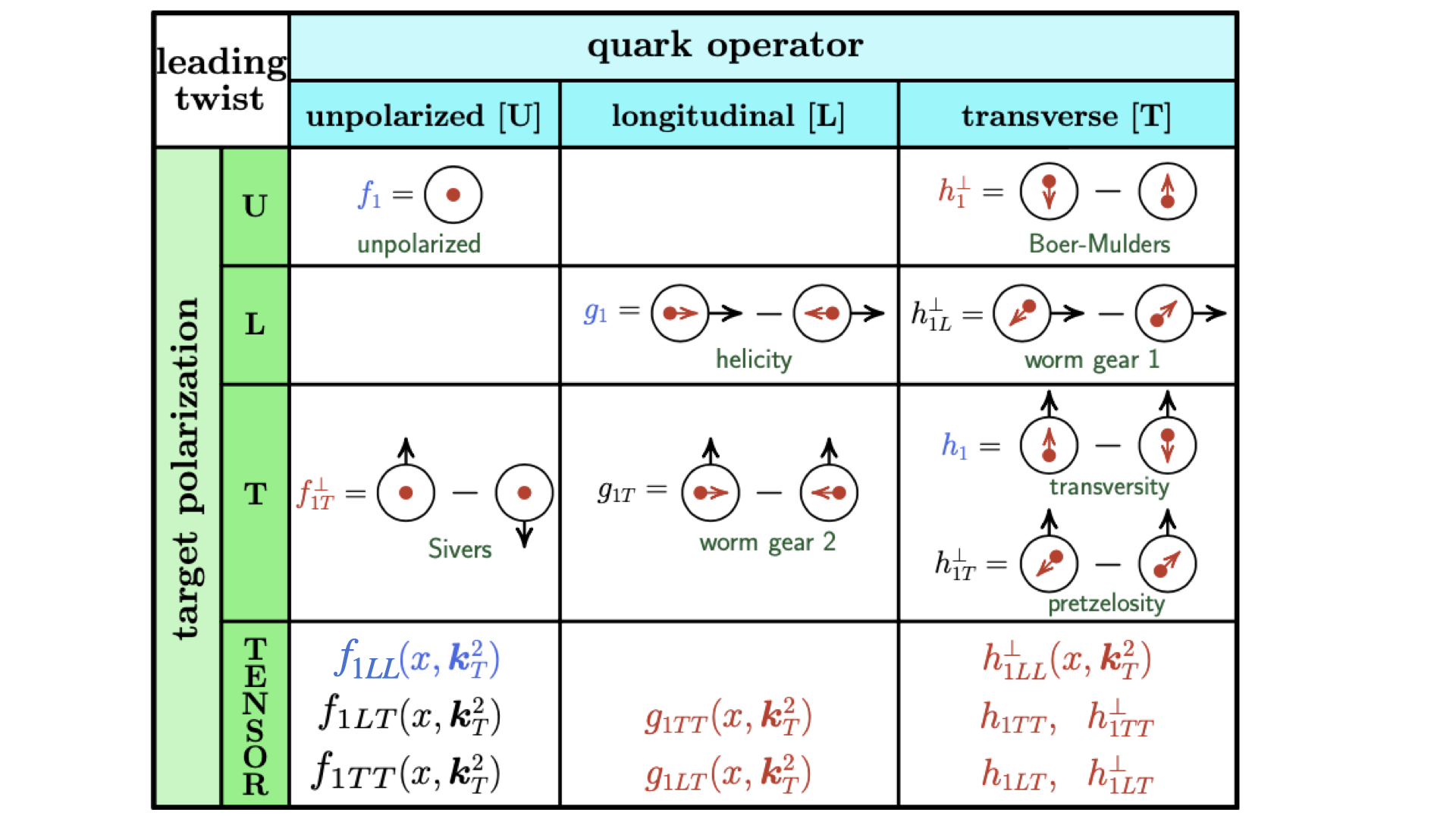}
\caption{The list of leading twist quark TMDs for the spin-1 target, which contain 3 additional $T$-even and 7 additional $T$-odd TMDs (all $T$-odd are shown in red) compared to spin-1/2 nucleons.  The blue indicates collinear PDFs.  Here, the rows indicate target polarization, and the columns indicate quark polarization.  The bold-face functions survive integration over transverse momenta.}
\label{spin1TMDs}
\end{figure}

The unpolarized TMD $f_1$ is quite well known now,
which is helpful because it is critical to the interpretation of other measurements.  The data on the unpolarized function is extracted from facilities worldwide.  For the valance quarks Sivers function $f_{1T}^\perp$, there is also increasingly more information.  SpinQuest E1039 will measure the essential sea-quarks Sivers function.  Some information exists on valance quark transversity $h_1$ and the Boer-Mulders $h_1^\perp$ function, as well as the pretzelosity $h^\perp_{1T}$ \cite{bog}, but considerably less exists for the sea.  Beyond this, there is essentially no experimental information on any of the other functions.  In Fig. \ref{spin1TMDs}, the list is shown of leading twist quark TMDs for the spin-1 target, which contain 3 additional $T$-even and 7 additional $T$-odd TMDs compared to spin-1/2 nucleons.  The rows indicate target polarization, and the columns indicate quark polarization.  The bold-face functions survive integration over transverse momenta.
\vspace{10mm}

To zero in on some observables of interest, we can integrate over transverse momenta and force many functions to vanish. The collinear correlator can then be parametrized as,
\begin{equation}
\begin{aligned}
\Phi(x ; P, S, T)=\frac{1}{2} &\left[\not P f_{1}(x)+S_{L} \gamma_{5} P g_{1}(x)+\frac{\left[\not B_{T}, \not P\right] \gamma_{5}}{2} h_{1}(x)\right.\\
&\left.+S_{L L} P f_{1 L L}(x)+\frac{\left[\not B_{L T}, P\right]}{2} i h_{1 L T}\left(x, k_{T}^{2}\right)\right].
\end{aligned}
\end{equation}
So, even for the spin-1 target (deuteron), the quark PDFs for the spin-1/2 constituents are easiest to access.  They, of course, represent the distribution of quarks in the longitudinal momentum space of unpolarized $f_1$, longitudinally polarized $g_1$, and transversely polarized $h_1$ quarks in the proton and neutron.  For the polarized cases, the neutron dominates in its contribution to the observables, as it caries more than 90\% of the deuteron polarization. 

Then, there are the extra collinear functions from the tensor polarization contributions $f_{1LL}$ and $h_{1LT}$.  For quarks, there is a measurement of $f_{1LL}$ \cite{b1} indirectly as the $b_1$ structure function in DIS.  This observable deserves its own Drell-Yan experimental effort (mentioned later).  Information on sea-quark $f_{1LL}$ specifically is needed \cite{close,b1k}.  This is also an attractive function because it contains non-nucleonic degrees of freedom that are detectable in nuclei.  There is also the tensor polarized observable $h_{1LT}$, which is $T$-odd and simultaneously survives integration over transverse momenta. At first order, the function $h_{1LT}$ vanishes due to the gauge link structure and the behavior under naive time reversal transformations.  In any case, these tensor polarized observables are mitigated when the spin-1 target has zero tensor polarization but some finite vector polarization.

Naturally, valence quarks have been the focus for the last few decades.  There has also been considerable theoretical effort in the last several years to understand the gluonic content of hadrons. Gluon observables can be easily overwhelmed by the valence quarks depending on the target and the kinematics available at the facility.  However, the structure and dynamics produced by the gluons and the quark sea are turning out to be critical to answer many pressing questions, and they must be studied in detail.

There is a clear need for sea-quark specific experiments; however, the information on gluon distributions is far more scarce and essentially restricted to the collinear gluon PDFs for spin-1/2 targets. Gluon TMDs are mostly unknown because it is generally very challenging to access the relevant kinematic regions for a spin-1/2 target.  What little information that is available on gluons comes from the LHC at CERN.

Little GPD or TMD information is available on spin-1 targets, and absolutely no experimental information is available on the tensor polarization contributions in TMDs. However, the interest in the gluon content of nuclei is growing, even if restricted to the collinear quantities. The collinear structure function for gluons in spin-1 targets was first defined by Jaffe and Manohar \cite{jaffe3} and referred to as nuclear gluonometry. This observable is related to a transfer of two units of helicity to the polarized target and vanishes for any target of spin smaller than 1.  A finite value of this observable requires the existence of a tower of gluon operators contributing to the scattering amplitude, where such a double-helicity flip cannot be linked to single nucleons.  This observable is exclusive to hadrons and nuclei of spin $\ge1$, and measures a gluon distribution, providing a clear signature for exotic gluonic components in the target.  In the parton model language, this observable comes from the linearly polarized gluons in targets with transverse tensor polarization and is related to the TMD $h_{1TT}$.  This interesting function is the focal point of our motivation and is one of the least investigated aspects in the gluonic structure linked to the target polarization where non-nucleonic dynamics becomes accessible. TMD $h^g_{1TT}$ is expected to yield new insights into the internal dynamics of hadrons and nuclei.

\begin{figure}[ht]
\centering
\includegraphics[width=10cm]{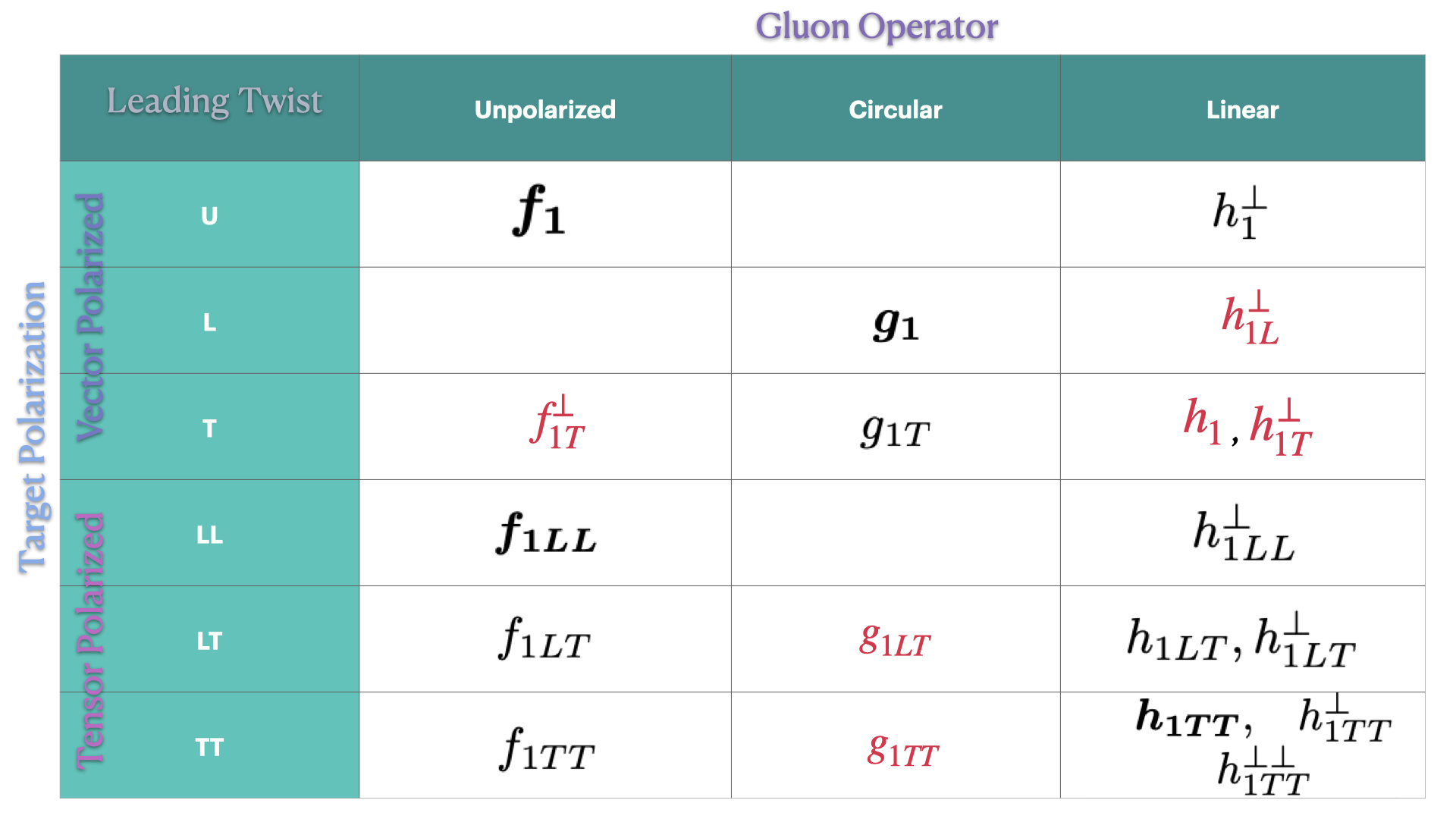}
\caption{The list of leading twist TMDs gluons.  The rows indicate the target polarization, while the columns indicate the gluon polarization.  The bold-face functions survive integration over transverse momenta, and the red indicates $T$-odd.  For
gluons, all the new $h$ TMDs from the tensor polarized target are $T$-even, unlike the quark case.}
\label{GluonTMDs}
\end{figure}

Going  beyond  the  collinear case, one can define new TMDs, see Fig \ref{GluonTMDs}.  These TMDs appear in the parametrization of a TMD correlator, which is a bilocal matrix element containing nonlocal field strength operators and Wilson lines.  The Wilson lines, or gauge links, guarantee color gauge invariance by connecting the nonlocality and give rise to a process dependence of the TMDs.  The description of spin-1 TMDs is presented by Bacchetta and Mulders \cite{bacch} for quarks and Boer{\it et al.} \cite{boer} for gluons.  Additionally, a study of the properties of and the relations between the gluon TMDs for spin-1 hadrons has recently been published \cite{cotogno}.  Positivity bounds were derived that provide model-independent inequalities that help in relating and estimating the magnitude of the gluon TMDs.

In \cite{boer}, the gluon-gluon TMD correlator was parametrized in terms of TMDs for unpolarized, vector, and tensor polarized targets.
We use a decomposition for the gluon momentum $k$ in terms of the hadron momentum $P$ and the lightlike four-vector $n$, such that,
$$k^{\mu}=x P^{\mu}+k_{T}^{\mu}+\left(k \cdot P-x M^{2}\right) n^{\mu},$$
satisfying $P\cdot n=1$ and $P^2=M^2$, where $M$ is the mass of the hadron.  The gluon-gluon TMD correlator for spin-1 hadrons is defined as:
\begin{equation}
\Gamma^{\left[U, U^{\prime}\right] \mu \nu ; \rho \sigma}\left(x, \boldsymbol{k}_{T}, P, n ; S, T\right)
\left.\quad \equiv \int \frac{d \xi \cdot P d^{2} \boldsymbol{\xi}_{T}}{(2 \pi)^{3}} e^{i k \cdot \xi}\left\langle P ; S, T\left|\operatorname{Tr}_{c}\left(F^{\mu \nu}(0) U_{[0, \xi]} F^{\rho \sigma}(\xi) U_{[\xi, 0]}^{\prime}\right)\right| P ; S, T\right\rangle\right|_{\xi \cdot n=0}
\end{equation}
where the process-dependent Wilson lines $U_{[0,\xi]}$ and $U'_{[0,\xi]}$ are required for color gauge invariance. The leading-twist terms are identified as the ones containing the contraction of the field strength tensor with $n$ and one transverse index $(i, j = 1, 2)$, explicitly indicating the dependence of the vector and tensor part of the spin.  The correlator is then expressed as
\begin{equation}
\Gamma^{i j}\left(x, \boldsymbol{k}_{T}\right) \equiv \int \frac{d(\xi \cdot P) d^{2} \boldsymbol{k}_{T}}{(2 \pi)^{3}} e^{i k \cdot \xi}\left\langle P, S, T\left|F^{\mu \nu}(0) U(0, \xi) F^{\rho \sigma}(\xi) U^{\prime}(\xi, 0)\right| P, S, T\right\rangle_{\xi+=0}
\end{equation}
where there is a trace over color, and the dependence on the gauge links is omitted.  After the separation in terms of the possible hadronic polarization states, the correlator can be expressed using the indicated notation as the following,
\begin{equation}
\Gamma^{i j}=\Gamma_{U}^{i j}+\Gamma_{L}^{i j}+\Gamma_{T}^{i j}+\Gamma_{L L}^{i j}+\Gamma_{L T}^{i j}+\Gamma_{T T}^{i j}.
\end{equation}
The parametrization in terms of TMDs with specific polarizations and orientations can then be expressed as,
\begin{equation}
\begin{array}{l}
\Gamma_{U}^{i j}\left(x, \boldsymbol{k}_{T}\right)=\frac{x}{2}\left[-g_{T}^{i j} f_{1}\left(x, \boldsymbol{k}_{T}^{2}\right)+\frac{k_{T}^{i j}}{M^{2}} h_{1}^{\perp}\left(x, \boldsymbol{k}_{T}^{2}\right)\right] \\
\Gamma_{L}^{i j}\left(x, \boldsymbol{k}_{T}\right)=\frac{x}{2}\left[i \epsilon_{T}^{i j} S_{L} g_{1}\left(x, \boldsymbol{k}_{T}^{2}\right)+\frac{\epsilon_{T}^{\{i} \alpha k_{T}^{j\} \alpha} S_{L}}{2 M^{2}} h_{1 L}^{1}\left(x, \boldsymbol{k}_{T}^{2}\right)\right] \\
\Gamma_{T}^{i j}\left(x, \boldsymbol{k}_{T}\right)=\frac{x}{2}\left[-\frac{g_{T}^{i j} \epsilon_{T}^{S_{T} k_{T}}}{M} f_{1 T}^{1}\left(x, \boldsymbol{k}_{T}^{2}\right)+\frac{i \epsilon_{T}^{i j} \boldsymbol{k}_{T} \cdot \boldsymbol{S}_{T}}{M} g_{1 T}\left(x, \boldsymbol{k}_{T}^{2}\right)\right. \\
\left.\quad-\frac{\epsilon_{T}^{k_{T}}\left\{i_{S}^{j\}}+\epsilon_{T}^{S_{T}\{i} k_{T}^{j\}}\right.}{4 M} h_{1}\left(x, \boldsymbol{k}_{T}^{2}\right)-\frac{\epsilon_{T}^{\{i} \alpha_{T}^{j\} \alpha S_{T}}}{2 M^{3}} h_{1 T}^{\perp}\left(x, \boldsymbol{k}_{T}^{2}\right)\right]\\
\Gamma_{L L}^{i j}\left(x, \boldsymbol{k}_{T}\right)=\frac{x}{2}\left[-g_{T}^{i j} S_{L L} f_{1 L L}\left(x, \boldsymbol{k}_{T}^{2}\right)+\frac{k_{T}^{i j} S_{L L}}{M^{2}} h_{1 L L}^{\perp}\left(x, \boldsymbol{k}_{T}^{2}\right)\right] \\
\Gamma_{L T}^{i j}\left(x, \boldsymbol{k}_{T}\right)=\frac{x}{2}\left[-\frac{g_{T}^{i j} \boldsymbol{k}_{T} \cdot \boldsymbol{S}_{L T}}{M} f_{1 L T}\left(x, \boldsymbol{k}_{T}^{2}\right)+\frac{i \epsilon_{T}^{i j} \epsilon_{T}^{S_{L T} k_{T}}}{M} g_{1 L T}\left(x, \boldsymbol{k}_{T}^{2}\right)\right. \\
\left.\quad+\frac{S_{L T}^{\{i} k_{T}^{j\}}}{M} h_{1 L T}\left(x, \boldsymbol{k}_{T}^{2}\right)+\frac{k_{T}^{i j \alpha} S_{L T \alpha}}{M^{3}} h_{1 L T}^{\perp}\left(x, \boldsymbol{k}_{T}^{2}\right)\right] \\
\Gamma_{T T}^{i j}\left(x, \boldsymbol{k}_{T}\right)=\frac{x}{2}\left[-\frac{g_{T}^{i j} k_{T}^{\alpha \beta} S_{T T \alpha \beta}}{M^{2}} f_{1 T T}\left(x, \boldsymbol{k}_{T}^{2}\right)+\frac{i \epsilon_{T}^{i j} \epsilon_{T \gamma}^{\beta} k_{T}^{\gamma \alpha} S_{T T \alpha \beta}}{M^{2}} g_{1 T T}\left(x, \boldsymbol{k}_{T}^{2}\right)\right. \\
\quad+S_{T T}^{i j} h_{1 T T}\left(x, \boldsymbol{k}_{T}^{2}\right)+\frac{S_{T T \alpha}^{(i)} k_{T}^{j\} \alpha}}{M^{2}} h_{1 T T}^{L}\left(x, \boldsymbol{k}_{T}^{2}\right) \\\nonumber
\left.\quad+\frac{k_{T}^{i j \alpha \beta} S_{T T \alpha \beta}}{M^{4}} h_{1 T T}^{\perp \perp}\left(x, \boldsymbol{k}_{T}^{2}\right)\right].
\end{array}
\end{equation}
The resulting list of leading twist TMDs for gluons is shown in Fig. \ref{GluonTMDs}.  The Gluon TMD functions are divided in terms of target polarization and gluon polarization, as shown in the figure. The bold-face functions survive integration over transverse momenta.

The parametrization of this correlator in terms of collinear PDFs is given by,
\begin{equation}
\Gamma^{i j}(x)=\frac{x}{2}\left[-g_{T}^{i j} f_{1}(x)+i \epsilon_{T}^{i j} S_{L} g_{1}(x)-g_{T}^{i j} S_{L L} f_{1 L L}(x)+S_{T T}^{i j} h_{1 T T}(x)\right].
\end{equation}
The surviving collinear PDF for vector polarization is $g^g_1$ or $g^g_{1L}$, while the tensor polarized observables are $f^g_{1LL}$, and the transversity term $h^g_{1TT}$.  The function $f^g_{1LL}$ is expected to be very small \cite{close,b1k} for a transversely polarized target, but this, too, is an interesting observable and deserves its own experimental effort in the future.

The phenomenological studies of gluons generally focus on characterizing the appropriate angular dependencies to access gluon distributions.  The extraction of these functions should rely on all-order TMD factorization, even though, for processes initiated by gluons, factorization breaking effects are often present \cite{rog1,rog2,rog3,rog4,rog5}. Here, complexity can arise in factorization, breaking from color entanglement and color-singlet configurations.
It is worth pointing out that the extraction of the gluon TMDs from different high energy processes requires taking into account the appropriate gauge link structures.  In situations where a higher number of hadrons are involved, the gauge links can be combinations of future and past pointing Wilson lines, with the possibility of additional loops \cite{bom}.

The gluon Sivers function ($f^{g\perp}_{1T}$) can be studied at RHIC and COMPASS and now Fermilab. The Sivers function can be accessed through the measurement of the Sivers asymmetry in $pp^\uparrow \to \pi X$ and in $J$/$\psi$ production \cite{qiu1,qiu2,qiu3}.  As far as the universality of the gluon Sivers function is concerned, we should expect a sign-change analogous to the quark case.

As mentioned, the longitudinal tensor polarized TMD $f^g_{LL}$ can also be measured at Fermilab \cite{LOI}.  This would require a different target magnet, which the University of Virginia presently owns.  This would require the disassembly and reassembly of the target, so it is better to measure everything possible within the transverse case first.  In both the longitudinal and transverse case, these gluon TMDs are related to a transfer of two units of helicity to the nuclear target, and vanish for any target with spin less than 1.  For the transverse case, this becomes strictly a probe of linearly polarized gluons in targets using transverse tensor polarization to access the gluon transversity $h^g_{1TT}$.

The observable $h^g_{1TT}$ provides unique information about gluon distributions and sheds light on the spin correlations between the gluon polarization content and the role it plays in the deuteron structure and wave-function.  This, in essence, yields a novel path for studying gluon based Spin Physics that can be accessible at higher-$x$ and lower $Q^2$ kinematics and is sensitive to momentum and polarization degrees of freedom that arise in nucleons bound inside nuclei.  

We also point out that, unlike the Sivers function, these gluon $T$-even observables can provide an unusually clean test of universality, as the contribution from quarks and sea-quarks in SIDIS is also disentangled from the gluon distributions when using a purely tensor polarized spin-1 target.  This is promising in providing additional fundamental tests of TMD theory and their role in QCD.  Specifically, for gluon transversity a \textit{genuine} universality relation holds,
\begin{equation*}
\left.h^g_{1TT}\right|_{SIDIS} =\left.h^g_{1TT}\right|_{D Y}.
\end{equation*}
All these new tensor polarized TMDs for spin-1 that are $T$-even will have analogous relations for the interpretation to be consistent.  The same corresponding quark distributions are $T$-odd \cite{daal}.  This is an important distinction and will help to impose constraints with data from multiple future experiments. 

\subsection{Transversity}
The transverse-polarization physics of the deuteron can be investigated by the transversity distributions in the twist-2-level collinear framework.  Of specific interest is the sea-quark transversity distribution, as well as the gluon transversity.  To understand how to access the gluon transversity, more explanation is required.  The approach to quark transversity, on the other hand, is relatively well known, and some measurements have already been performed on the valance contributions.  The proposed experiment would provide essential information for such a test specific to the sea-quarks.  Fermilab is unique in its kinematic range, providing some overlap with other high-$x$ facilities, allowing for the critical tests of universality and much more.

\subsubsection{Quark Transversity}

As mentioned, an important channel to investigate the quark transversity distribution is the space-like process to Drell-Yan or SIDIS, where it is necessary to measure the Collins azimuthal spin asymmetries in order to extract the TMDs \cite{col,coll}.  Measurements have been made by the HERMES Collaboration \cite{her1,her2}, the COMPASS Collaboration \cite{comp}, and JLab Hall A \cite{jlab1}. The quark transversity distributions require the Collins fragmentation functions, which are different from the usual unpolarized fragmentation functions.  BELLE and BABAR have attempted some extractions of the observables \cite{belle1,belle2,babar1}.  Due to the universality of the Collins fragmentation function \cite{metz}, it is possible to constrain the fragmentation function and the valance quark transversity, given the multiple data sets and analyses using transversity coupled to the dihadron interference fragmentation functions \cite{radi}.  There has also been effort to apply the appropriate QCD evolution to the phenomenological studies \cite{kang} and improve the global picture of transversity.

\begin{figure}[ht]
\centering  
\includegraphics[width=10cm]{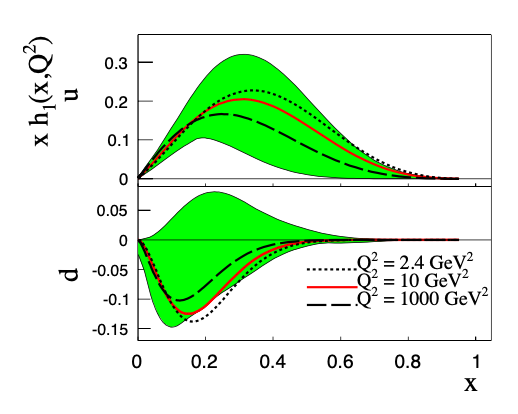}
\caption{The extracted transversity distribution at three different scales in $Q^2$.  The shaded region corresponds to the 90\% confidence level error band at $Q^2=10$ GeV$^2$ \cite{kang}.}
\label{valtran}
\end{figure}

In the global fit with TMD evolution, there are two unknown functions to be extracted using the experimental data.  The collinear transversity distributions $h_1^q$ and the collinear twist-3 fragmentation function $\hat{H}^{(3)}_{h/q}$.  The fit parameterizes \cite{kang} the quark transversity distributions as,
\begin{equation}
h_{1}^{q}\left(x, Q_{0}\right)=N_{q}^{h} x^{a_{q}}(1-x)^{b_{q}} \frac{\left(a_{q}+b_{q}\right)^{a_{q}+b_{q}}}{a_{q}^{a_{q}} b_{q}^{b_{q}}} \frac{1}{2}\left(f_{1}^{q}\left(x, Q_{0}\right)+g_{1}^{q}\left(x, Q_{0}\right)\right),
\end{equation}
where $Q_0$ is the initial scale for up and down quarks $q=u,~d$, respectively, and $f_1^q$ are the unpolarized CT10NLO quark distributions \cite{flor} and the NLO DSSV quark helicity distributions.  In this parametrization, the so-called Soffer positivity bound \cite{soffer} of transversity distribution at the initial scale was applied. This bound is known to be valid \cite{evol3} up to NLO order in perturbative QCD.  This study assumes that all the sea-quark transversity distributions are negligible.  It is well understood that this is a less than ideal place to start for such an extraction; however, there is no data for the sea-quark contribution.  With more data available from SpinQuest and this proposed experiment, it would be possible to constrain the sea-quarks as well.  The resulting extracted transversity distributions for the valance $u$ and $d$ quarks are shown in Fig. \ref{valtran}.  Other extractions have been made using the two-hadron production in electron-positron annihilation $e^+e^-\to h_1 h_2 X$ where the Collins effect is observed in the combination of the fragmenting processes of a quark and an antiquark, resulting in the product of two Collins functions with an overall modulation of the azimuthal angles of the final hadrons around the quark-antiquark axis \cite{anna}.
 Similarly, extraction of the transversity distribution in the framework of collinear factorization was made based on the global analysis of pion-pair production in deep-inelastic scattering and in proton-proton collisions with a transversely polarized proton \cite{radi2}. 
For the transversely polarized nucleon, transversity distributions are expressed as $\Delta Tq(x) = q_{\uparrow}(x)-q_{\downarrow}(x)$, where $\uparrow$ and $\downarrow$ indicate parallel and anti-parallel quark polarizations to the transversely-polarized nucleon spin.
\begin{figure}[ht]
\centering  
\includegraphics[width=9cm]{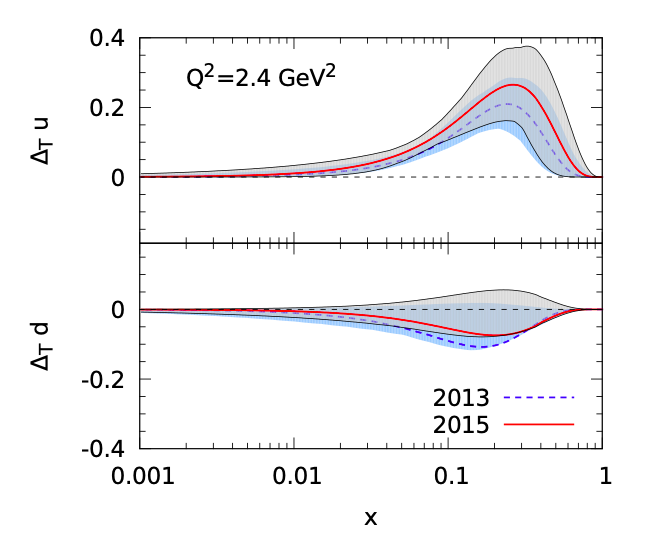}
\caption{The extracted $u$ and $d$ transversity distributions in a comparison of the best fit results  (red, solid lines). \cite{radi2}.}
\label{valtran2}
\end{figure}

There has been one attempt made to extract something about the sea-quark contribution \cite{anna}, but this extraction largely lacks constraints for any realistic determination.  The results in this case imply that the sea-quark transversity is compatible with zero, but, especially in the case of the $\bar{d}$, the error is simply too large to say anything definite.  We will use these results to demonstrate the possible constraints that this proposal could add in Section \ref{results}.

To take a closer look at the quark transversity and the physics specific to the deuteron, consider that the unpolarized, longitudinally-polarized, and transversity distribution functions are defined for quarks by the following matrix elements \cite{tran5},
\begin{eqnarray}
&q(x)&=\int \frac{d \xi^{-}}{4 \pi} e^{i x p^{+} \xi^{-}}\left\langle p\left|\bar{\psi}(0) \gamma^{+} \psi(\xi)\right| p\right\rangle_{\xi^{+}=\vec{\xi}_{\perp}=0} \\
&\Delta q(x)&=\int \frac{d \xi^{-}}{4 \pi} e^{i x p^{+} \xi^{-}}\left\langle p s_{L}\left|\bar{\psi}(0) \gamma^{+} \gamma_{5} \psi(\xi)\right| p s_{L}\right\rangle_{\xi^{+}=\bar{\xi}_{\perp}=0} \\
&\Delta_{T} q(x)&=\int \frac{d \xi^{-}}{4 \pi} e^{i x p^{+} \xi^{-}}
\left\langle p s_{T j}\left|\bar{\psi}(0) i \gamma_{5} \sigma^{j+} \psi(\xi)\right| p s_{T j}\right\rangle_{\xi^{+}=\bar{\xi}_{\perp}=0}
\end{eqnarray}
where $s_L$ and $s_{Tj}$ ($j = 1$ or $2$) indicate longitudinal and transverse polarizations of the nucleon, and $\psi$ is the quark field. These distribution functions are leading twist.  The structure function $g_T$, associated with the transverse spin, can be written in an operator matrix element in a similar way as,
\begin{eqnarray}
g_{T,q}(x)= \frac{p^{+}}{M_{N}} \int \frac{d \xi^{-}}{4 \pi} e^{i x p^{+} \xi^{-}}\left\langle p s_{T}\left|\bar{\psi}(0) \gamma_{\perp} \gamma_{5} \psi(\xi)\right| p s_{T}\right\rangle_{\xi^{+}=\bar{\xi}_{\perp}=0},
\end{eqnarray}
which is a twist-3 structure function.

\begin{figure}[ht]
\centering
\includegraphics[width=8cm]{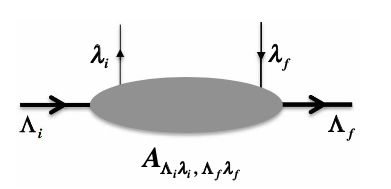}
\caption{Parton-hadron forward scattering amplitude $A_{\Lambda_i\lambda_i,\Lambda_f\lambda_f}$ with the
hadron helicities $\Lambda_i$ and $\Lambda_f$ and parton helicities $\lambda_i$ and $\lambda_f$.}
\label{amp_1}
\end{figure}

The structure functions of the nucleon are given by the imaginary part of forward scattering amplitudes by the optical theorem.  Figure \ref{amp_1} shows the parton-hadron forward scattering amplitudes. 

The amplitude is written as $A_{\Lambda_i\lambda_i,\Lambda_f\lambda_f}$ with the initial and final hadron helicities $\Lambda_{i}$ and $\Lambda_{f}$, and with parton helicities $\lambda_i$ and $\lambda_f$, such that the PDFs can be related to
the helicity amplitudes by \cite{tran5,jaffe1},
\begin{eqnarray*}
f_1(x)=q(x) &=&q_{+}(x)+q_{-}(x) \sim \operatorname{Im}\left(A_{++,++}+A_{+-,+-}\right) \\
g_1(x)=\Delta q(x) &=&q_{+}(x)-q_{-}(x) \sim \operatorname{Im}\left(A_{++,++}-A_{+-,+-}\right) \\
-h_1(x)=\Delta_{T} q(x) &=&q_{\uparrow}(x)-q_{\downarrow}(x) \sim \operatorname{Im} A_{++,--}.
\label{amps}
\end{eqnarray*}
where the direction of the polarization is perpendicular to the beam, and the amplitudes are defined by the transversely-polarized states, so the transversity distribution is
\begin{equation}
\Delta_{T} q(x)=q_{\uparrow}(x)-q_{\downarrow}(x) \sim \operatorname{Im}\left(A_{\uparrow \uparrow, \uparrow \uparrow}-A_{\uparrow \downarrow, \uparrow \downarrow}\right).
\label{qt}
\end{equation}
The SpinQuest polarized target configuration can be used to probe the sea-quark transversity distributions and help determine the tensor charge in the nucleon.  The already proposed experiment E1039 will take data on both transversely polarized protons and neutrons; however, without additional data to separate the vector and tensor polarization contributions, the neutron transversity will be very difficult to decipher.  This proposal is specific to the control of the deuteron polarization states where a large part of the vector polarization actually comes from the neutron.  Transversity is an important physical quantity for clarifying the nature of the nucleon spin and also for exploring possible signatures beyond the standard model \cite{edm8,edm9,edm10} by observing electric dipole moments of the neutron.  There is also considerable theoretical work in lattice QCD \cite{edm9, edm11, edm12, edm13, edm14, edm15, edm16, edm17, edm18, edm19} as well as the Dyson-Schwinger Equation (DSE) \cite{edm20,edm21}. 

The neutron electromagnetic current can be written as\cite{czar, hecht1, hecht2, liu, posp, chupp, qin},
\begin{equation}
\begin{aligned}
\left\langle n\left|J_{\mu}^{\mathrm{em}}\right| n\right\rangle=& \bar{u}\left(p^{\prime}\right)\left[\gamma_{\mu} F_{1}\left(q^{2}\right)+\frac{\kappa}{2 M_{N}} i \sigma_{\mu \nu} q^{\nu} F_{2}\left(q^{2}\right)\right.\\
&\left.+\frac{d_{n}}{2 M_{N}} \gamma_{5} \sigma_{\mu \nu} q^{\nu} F_{3}\left(q^{2}\right)\right] u(p).
\end{aligned}
\end{equation}
Here, the time-reversal odd term is included with the form factor $F_3$, in addition to the ordinary parity and time-reversal even terms with $F_1$ and $F_2$, the Dirac and Pauli form factors, respectively, and $\kappa$ as the anomalous magnetic moment.  The initial and final neutron momenta are denoted as $p$ and $p_0$, where $q$ is the momentum transfer given by $q=p-p'$, and $u(p)$ is the Dirac spinor for the neutron. Finally, $F_3$ is the time-reversal odd form factor, in combination with the electromagnetic field $A^{\mu}$ in the Hamiltonian, with the factor of the neutron electric dipole moment (EDM) $d_n$ in units of $e/(2M_n)$.  

The nucleon tensor charge is a fundamental nuclear property, and its determination is among the main goals of several experiments \cite{edm1,edm2,edm3,edm4,edm5,edm6,edm7}.  In terms of the partonic structure of the neutron, the tensor charge, for a particular quark type $q$, is constructed from the quark transversity distribution, $h_1(x, Q^2)$ \cite{edm1,edm2,edm3,edm4,edm5}, where the neutron EDM is expressed by integrals of the transversity distributions to obtain the tensor charge,
\begin{equation}
d_{n} =\sum_{q} d_{q} \delta q(Q^2)   
\end{equation}
\begin{equation}
\delta q\left(Q^{2}\right) \equiv \int_{0}^{1} d x\left(h_{1}^{q}\left(x, Q^{2}\right)-h_{1}^{\bar{q}}\left(x, Q^{2}\right)\right)
\end{equation}
where $d_q$ is the quark EDM.  The neutron EDM is investigated theoretically by calculating the quark EDMs in the standard model, or theories that deviate from the standard model.  The EDM is multiplied by the tensor charge in order to compare with experimental measurements. The contributions from the sea-quarks to the transversity distributions of the neutron are critical to a detailed understanding and physics interpretation.

\subsubsection{Gluon Transversity}
Equation \ref{amps} shows that the transversity distribution $h_1(x)$ is associated with the quark spin flip ($\lambda_i=+$, $\lambda_f=-$), a chiral-odd distribution.  Clearly, the gluon transversity cannot exist in the nucleon where the spin flip $\Delta s=2$ is not possible.  The
quark transversity distributions evolve without the corresponding gluon distribution in the nucleon \cite{evol3} which differs from the longitudinally-polarized PDFs, where the quarks and gluon distributions couple with each other in the $Q^2$ evolution.  This is a subtle yet critical point because it provide a crucial test of the perturbative QCD in Spin Physics.

\begin{figure}[ht]
\centering
\includegraphics[width=8cm]{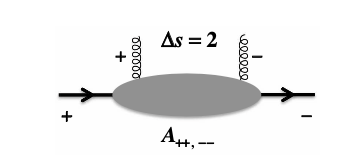}
\caption{Gluon-deuteron forward scattering amplitude $A_{++,--}$ with the spin flip of 2 ($\Delta s=2$) for gluon transversity.}
\label{amp_2}
\end{figure}

Similarly to the quark transversity, Eq. \ref{qt}, the gluon transversity is written as,
\begin{equation}
h^g_{1TT}(x)\sim \operatorname{Im} A_{++,--},
\end{equation}
where the spin flip of $\Delta s=2(|\lambda_f-\lambda_i|=|\Lambda_f-\Lambda_i|=2)$ is necessary for
gluon transversity, see Fig \ref{amp_2}.  The most simple and stable spin-1 hadron or nucleus is the deuteron, which is our choice for the future experiment to study gluon transversity.
By angular momentum conservation, the linear polarization of a gluon is zero for the spin-1/2 hadron. Naturally, linear polarization is measured by an operator that flips helicity by two units. Since no helicity is absorbed by the space-time part of the definition of the parton densities (the integrals are azimuthally symmetric), the helicity flip in the operator must correspond to a helicity flip term in the density.
The gluon correlation function is defined as,
\begin{equation}
\begin{aligned}
\Phi_{g / H}^{\alpha \beta}\left(p_{h}, p_{H}, s_{H}\right)=& N_{g / H} \int \frac{d^{4} \xi}{(2 \pi)^{4}} e^{i p_{h} \cdot \xi_{h}} \\
& \times\left\langle p_{H} s_{H}\left|A^{\alpha}(0) A^{\beta}(\xi)\right| p_{H} s_{H}\right\rangle
\end{aligned}
\end{equation}where $A^{\alpha}$ is given by $A^{\alpha}=A^{\alpha}_a t^{\alpha}$, and $N_{h/H}$ is the normalization constant.
The gluon correlation function in the deuteron at twist-2 is,
\begin{equation}
\begin{array}{l}
\Phi_{g / B}^{\alpha \beta}\left(x_{b}\right) \equiv \int d^{2} p_{b T} \Phi_{g / B}^{\alpha \beta}\left(x, \vec{p}_{b T}\right) \\
=\frac{1}{2}\left[-g_{T}^{\alpha \beta} f^g_{1, B}\left(x_{b}\right)+i \epsilon_{T}^{\alpha \beta} S_{L} g^g_{1,B}\left(x_{b}\right)\right.
\left.-g^{g_{T}\alpha \beta} S_{L L} f^g_{1 L L, B}\left(x_{b}\right)+S_{T T}^{\alpha \beta} h^g_{1 T T,B}\left(x_{b}\right)\right]
\end{array}
\end{equation}
where $f^g_{1,B}$ is the unpolarized gluon distribution function, $g^g_{1,B}$ is the longitudinally-polarized distribution function, $f^g_{1LL,B}$ is the longitudinally tensor polarized distribution function, and $h^g_{1TT,B}$ is the transversely tensor polarized distribution function, or the gluon transversity.  It is clear that the matrix elements $S_{TT}^{\alpha\beta}$ must be finite in order to measure this observable.

The matrix element form of the gluon transversity is,
\begin{equation}
\begin{aligned}
-h^g_{1TT,B}=\Delta_{T} g(x)=& \varepsilon_{T T, \alpha \beta} \int \frac{d \xi^{-}}{2 \pi} x p^{+} e^{i x p^{+} \xi^{-}} \\
& \times\left\langle p E_{x}\left|A^{\alpha}(0) A^{\beta}(\xi)\right| p E_{x}\right\rangle_{\xi^{+}=\vec{\xi}_{\perp}=0}
\end{aligned}
\end{equation}
where $\epsilon^{\alpha\beta}_{TT}=+1$ for $\alpha=\beta=1$, $\epsilon^{\alpha\beta}_{TT}=-1$ for $\alpha=\beta=2$, and all else is zero.  The linear polarization of the gluons requires a tensor polarized target oriented along the x-axis or the vertical direction transverse to the beam direction.  This is indicated by the $E_x$ in the above equation.

The cross-section can be written in terms of parton correlation functions by considering the subprocess
assuming a quark from the proton beam and an antiquark from the deuteron target ($q(p)+\bar{q}(d)\to\gamma+g$),
\begin{equation}
    \begin{aligned}
\left.d \sigma_{p d \rightarrow \gamma X}\right|_{q \bar{q} \rightarrow \gamma^{*} g}=& \frac{1}{4 p_{A} \cdot p_{B}} \int \frac{d^{4} p_{a}}{(2 \pi)^{4}} \int \frac{d^{4} \gamma_{b}}{(2 \pi)^{4}} \sum_{\mathrm{spjin}, \mathrm{favor}} \sum_{X_{A}, X_{B}}(2 \pi)^{4} \delta^{4}\left(p_{A}-p_{a}-p_{A X}\right)(2 \pi)^{4} \delta^{4}\left(p_{B}-p_{b}-p_{B X}\right) \\
& \times\left|\left\langle X_{B}\left|\bar{\psi}_{b, l}(0)\right| p_{B} s_{B}\right\rangle\left(\Gamma_{q \bar{q} \rightarrow \gamma^{*} g, \mu}\right)_{l k}\left\langle X_{A}\left|\psi_{a, k}(0)\right| p_{A} s_{A}\right\rangle M_{\gamma^{*} \rightarrow \mu^{+} \mu^{-}}^{\mu}\right|^{2} \\
& \times\left(\frac{-e}{Q^{2}}\right)^{2}(2 \pi)^{4} \delta^{4}\left(p_{a}+p_{b}-k_{1}-k_{2}-p_{d}\right) \frac{d^{3} k_{1}}{2 E_{1}(2 \pi)^{3}} \frac{d^{3} k_{2}}{2 E_{2}(2 \pi)^{3}} \frac{d^{3} p_{d}}{2 E_{d}(2 \pi)^{3}},
\end{aligned}
\end{equation}
where the spin summations are over the muons, quark, antiquark, and gluon.  The parton-interaction part is $\Gamma_{q\bar{q}}\to \gamma^* g,\mu=e_q \epsilon^{*\alpha}(p_d,\lambda_b)\Gamma_{\mu\alpha}$ by extraction of the quark charge $e_q$, and the gluon-polarization vector $\epsilon^{*\alpha}(p_d,\lambda_d)$ from $\Gamma_{q\bar{q}\to\gamma^* g, \mu}$.
\begin{figure}[ht]
\centering
\includegraphics[width=8cm]{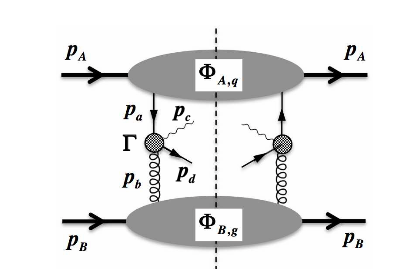}
\caption{The Quark-gluon process contribution to the cross-section.}
\label{g-p}
\end{figure}
By changing from three to two-body phase space and recalculating the cross-section using the lepton and hadron tensor, we get,
\begin{equation}
    \frac{d \sigma_{p d \rightarrow \mu^{+} \mu^{-} X}}{d \tau d q_{T}^{2} d \phi d y}=\frac{\alpha^{2}}{3(2 \pi)^{2} Q^{4}}\left(\int d \Phi_{2}\left(q ; k_{1}, k_{2}\right) 2 L^{\mu \nu}\right) W_{\mu \nu},
\end{equation}
where the hadron tensor for the $q\bar{q}\to\gamma^* g$ is,
\small{
\begin{equation}
    \begin{aligned}
&W_{\mu \nu}(q \bar{q})=\int \frac{d^{4} p_{a}}{(2 \pi)^{4}} \int \frac{d^{4} p_{b}}{(2 \pi)^{4}} \sum_{\operatorname{spin}_{c 0 l b r}} \sum_{q} \sum_{X_{A}, X_{B}} e_{q}^{2}(2 \pi)^{4} \delta^{4}\left(p_{A}-p_{a}-p_{A X}\right)(2 \pi)^{4} \delta^{4}\left(p_{B}-p_{b}-p_{B X}\right)\\
&\times\left[\left\langle X_{B}\left|\bar{\psi}_{b, j}(0)\right| p_{B} s_{B}\right\rangle\left(\Gamma_{\left.q \bar{q} \rightarrow \gamma^{\prime} g_{, \mu}\right) j i}\left\langle X_{A}\left|\psi_{a, i}(0)\right| p_{A} s_{A}\right\rangle\right]^{\dagger}\left[\left\langle X_{B}\left|\bar{\psi}_{b, l}(0)\right| p_{B} s_{B}\right\rangle\left(\Gamma_{q \bar{q} \rightarrow \gamma^{*} g, \nu}\right)_{l k}\left\langle X_{A}\left|\psi_{a, k}(0)\right| p_{A} s_{A}\right\rangle\right]\right.\\
&\times(2 \pi)^{4} \delta^{4}\left(p_{a}+p_{b}-q-p_{d}\right) \frac{d^{3} p_{d}}{2 E_{d}(2 \pi)^{3}}.
\end{aligned}
\end{equation}
}
Here, the hadron tensor can be expressed in terms of the correlation functions.  The quark-gluon process contribution to the cross-section diagram is shown in Fig. \ref{g-p}, indicating the quark in the proton beam $A$ and the gluon in the deuteron target $B$.  The $\delta$ function $\delta^4(p_H-p_h-p_{H_x})$ ($H=A$ or $B$, $h=a$ or $b$) is expressed by the integrals of exponential function: $(2\pi)^4\delta^4(p_H-p_h-p_{H_x})=\int d^4\xi_h e^{-i(p_H-p_h-p_{H_x})\dot\xi_h}$.  The quark field is given at $\xi_h$ in the matrix elements, with the exponential factor $e^{-i(p_H)\dot\xi_h}\psi(0)e^{-i(p_H)\dot\xi_h}=\psi(\xi_h)$.

\section{The Measurement}

To measure transversity of both the sea-quarks and gluons in a polarized deuteron, a set of unique target spin asymmetries must be measured.  For the sea-quark transversity, ideally what is needed is a transversely vector polarized target system which mitigates any tensor polarized contributions.

The gluon transversity is ideally measured with a vector and tensor polarized target as to isolate linearly polarized gluons in the deuteron.  To understand this configuration, we start again with the spin vector ($\boldsymbol{S}$) and tensor ($\boldsymbol{T}$) which are parameterized in the rest frame of the deuteron,
\begin{equation}
\boldsymbol{S}=\left(S_{T}^{x}, S_{T}^{y}, S_{L}\right)
\end{equation}

\begin{equation}
\begin{array}{l}
\boldsymbol{T}=\frac{1}{2}\left(\begin{array}{ccc}
-\frac{2}{3} S_{L L}+S_{\mathrm{TT}}^{x x} & S_{\mathrm{TT}}^{x y} & S_{L T}^{x} \\
S_{\mathrm{TT}}^{x y} & -\frac{2}{3} S_{L L}-S_{\mathrm{TT}}^{x x} & S_{L T}^{y} \\
S_{L T}^{x} & S_{L T}^{y} & \frac{4}{3} S_{L L}
\end{array}\right)
\end{array}
\end{equation}
\vspace{10mm}
where $S^x_T$, $S^y_T$, $S^{xx}_{TT}$, $S^{xy}_{LT}$, and $S^y_{LT}$ are the parameters to indicate the deuteron's vector and tensor polarizations.  The deuteron polarization vector $\vec{E}$ is,
\begin{equation}
\begin{array}{l}
\vec{E}_{0}=(0,0,1)\\
\vec{E}_{\pm}=\frac{1}{\sqrt{2}}(\mp 1,-i, 0) \\
\vec{E}_{x}=\frac{1}{\sqrt{2}}\left(\vec{E}_{-}-\vec{E}_{+}\right)=(1,0,0) \\
\vec{E}_{y}=\frac{i}{\sqrt{2}}\left(\vec{E}_{-}+\vec{E}_{+}\right)=(0,1,0)
\end{array}
\end{equation}
where $\vec{E}_+$, $\vec{E}_0$, and $\vec{E}_-$ indicate the three possible spin states of the deuteron.
Here, the polarizations $\vec{E}_x$ and $\vec{E}_y$ are spin-1 alignment dependent states and can be used to orient the gluons in a linearly polarized configuration in the target based on the gluon transversity distributions defined by the matrix elements between linearly-polarized states.
The spin vector and tensor are written in terms of the polarization vector $\vec{E}$ of the deuteron as,
\begin{equation}
\vec{S}=\operatorname{Im}\left(\vec{E}^{*} \times \vec{E}\right), \quad T_{i j}=\frac{1}{3} \delta_{i j}-\operatorname{Re}\left(E_{i}^{*} E_{j}\right).
\label{ex}
\end{equation}
For optimal gluon transversity extraction, the key is in the target configuration utilized to selectively reduce all unneeded terms in the spin tensor to zero, preserving only the terms that relate to the observable of interest.  In this case, having a finite $S_{TT}^{xx}$ gives the desired access to the gluon transversity.  Making the other terms zero or negligible is advantageous to a clean measurement.  In this case, the polarization vectors $E_x$ and $E_y$ can be used to provide linear polarization, and both consist of a deuteron tensor polarized in the transverse plane to the beam-line.  The difference in the cross-section from these polarization states can be used in an asymmetry to build an observable to extract gluon transversity.

The polarization vectors $\vec{E}_x$, $\vec{E}_0$, and $\vec{E}_y$ are all indicative of a purely tensor polarized target with spin quantization axis along the $x$, $z$, and $y$ axis respectively.  From Eq. \ref{ex}, we get for $\vec{E}_x$ a vector polarization of $S^x_T=S^y_T=S_L=0$, with $S_{LL}=1/2$, $S^{xx}_{T}=-1$, and $S^{xy}_{TT}=S^{x}_{LT}=S^{y}_{LT}=0$.  For $\vec{E}_y$, a vector polarization of $S^x_T=S^y_T=S_L=0$, with $S_{LL}=1/2$, $S^{xx}_{TT}=+1$ and $S^{xy}_{TT}=S^{x}_{LT}=S^{y}_{LT}=0$ is obtained.  For $\vec{E}_0$, a vector polarization of $S^x_T=S^y_T=S_L=0$, with $S_{LL}=-1$, $S^{xx}_{TT}=0$ and $S^{xy}_{TT}=S^{x}_{LT}=S^{y}_{LT}=0$ is obtained.  We can then use combinations to optimize such that $\vec{E}_x-\vec{E}_y$ yields $S^x_T=S^y_T=S_L=0$, with $S_{LL}=0$, $S^{xx}_{TT}=-2$ and $S^{xy}_{TT}=S^{x}_{LT}=S^{y}_{LT}=0$.  Also, $2\vec{E}_x-\vec{E}_0$ yields $S^x_T=S^x_T=S^x_T=S^x_T=0$, with $S_{LL}=0$, $S^{xx}_{T}=-2$, and $S^{xy}_{TT}=S^{x}_{LT}=S^{y}_{LT}=0$.
With either of these configurations, the longitudinal tensor polarization is zero as well as any vector polarization contributions, and the critical term $S^{xx}_{TT}$ is also maximized.
\begin{figure}
    \centering
    \includegraphics[width=10cm]{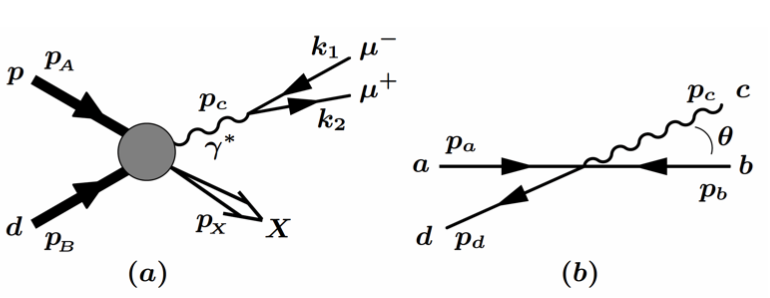}
    \caption{(a) The proton-deuteron Drell-Yan process $p+d\to \mu^+ \mu^- + X$ showing the notation for each momentum index. (b)  The parton reaction with corresponding index for process $a+b\to c+d$ in the center-of-momentum frame.}
    \label{fig:my_label}
\end{figure}

To exploit the observables, we rely on the correlation functions in the collinear formalism.  For the difference in the $\vec{E}_x$ and $\vec{E}_y$ polarized cross-section, the hadron tensor is given by,
\begin{equation}
\begin{array}{l}
W_{\mu \nu}\left(E_{x}-E_{y}\right)
=\sum_{\lambda_{d}} \sum_{\text {color }} \sum_{q} e_{q}^{2} \int_{\min \left(x_{a}\right)}^{1} d x_{a}\frac{\pi}{p_{g}^{-}\left(x_{a}-x_{1}\right)}
\operatorname{Tr}\left[\Gamma_{\nu \beta}\left\{\Phi_{q / A}\left(x_{a}\right)+\Phi_{\bar{q} / A}\left(x_{a}\right)\right\} \hat{\Gamma}_{\mu \alpha} \Phi_{g / B}^{\alpha \beta}\left(x_{b}\right)\right].
\end{array}
\end{equation}
Here, the summation is taken over the quark spin $\lambda_d$, and all spin tensor matrix elements are zero, except for the gluon transversity in the target.  There is no equivalent polarization term in the quark and antiquark distributions of the spin-1 target, so the transversity of the sea-quarks and the gluons can be separated through the strategic use of vector and tensor polarizations.  This is because the virtual photon in the intermediate stage interacts with a charge parton, so only quark and antiquark correlation functions contribute as the leading process from the spin-1/2 nucleons inside the spin-1 deuteron.  This implies that the geometric shape the deuteron in the different $M_J$ spin states are highly correlated to the transverse gluon and sea-quark observables.

To build an asymmetry, the cross-section difference is written as,
\begin{equation}
\begin{aligned}
\frac{d \sigma_{p d \rightarrow \mu^{+} \mu^{-} X}}{d \tau d q_{T}^{2} d \phi d y}\left(E_{x}-E_{y}\right)=&\frac{\alpha^{2} \alpha_{s} C_{F} q_{T}^{2}}{6 \pi s^{3}} \cos (2 \phi) \int_{\min \left(x_{a}\right)}^{1} d x_{a} \frac{1}{\left(x_{a} x_{b}\right)^{2}\left(x_{a}-x_{1}\right)\left(\tau-x_{a} x_{2}\right)^{2}} \\
& \times \sum_{q} e_{q}^{2} x_{a}\left[q_{A}\left(x_{a}\right)+\bar{q}_{A}\left(x_{a}\right)\right] x_{b} h^g_{1TT,B}\left(x_{b}\right).
\end{aligned}
\label{2phi}
\end{equation}
The cross-section sum of these polarization states can also be calculated where $\bar{q}q \to \gamma^* g$ and $q/\bar{q}\to\gamma^{*}q/\bar{q}$.  This leads to the cross-section,
\begin{equation}
\begin{aligned}
\frac{d \sigma_{p d \rightarrow \mu^{+} \mu^{-} X}}{d \tau d q_{T}^{2} d \phi d y}\left(E_{x}+E_{y}\right)=& \frac{\alpha^{2} \alpha_{s} C_{F}}{2 \pi \tau s^{2}} \int_{\min \left(x_{a}\right)}^{1} d x_{a} \frac{1}{\left(x_{a}-x_{1}\right) x_{a}^{2} x_{b}^{2}} \\
& \times \sum_{q} e_{q}^{2}\left[\frac{4}{9}\left\{q_{A}\left(x_{a}\right) \bar{q}_{B}\left(x_{b}\right)+\bar{q}_{A}\left(x_{a}\right) q_{B}\left(x_{b}\right)\right\}\right.\\
& \times \frac{2 \tau\left\{\tau-\left(-2 x_{a} x_{b}+x_{1} x_{b}+x_{2} x_{a}\right)\right\}+x_{b}^{2}\left(x_{a}-x_{1}\right)^{2}+x_{a}^{2}\left(x_{b}-x_{2}\right)^{2}}{\left(x_{a}-x_{1}\right)\left(x_{b}-x_{2}\right)} \\
&+\frac{1}{6}\left\{q_{A}\left(x_{a}\right)+\bar{q}_{A}\left(x_{a}\right)\right\} g_{B}\left(x_{b}\right) \frac{2 \tau\left(\tau-x_{1} x_{b}\right)+x_{b}^{2}\left\{\left(x_{a}-x_{1}\right)^{2}+x_{a}^{2}\right\}}{x_{b}\left(x_{a}-x_{1}\right)} \\
&\left.+\frac{1}{6} g_{A}\left(x_{a}\right)\left\{q_{B}\left(x_{b}\right)+\bar{q}_{B}\left(x_{b}\right)\right\} \frac{2 \tau\left(\tau-x_{2} x_{a}\right)+x_{a}^{2}\left\{\left(x_{b}-x_{2}\right)^{2}+x_{b}^{2}\right\}}{x_{a}\left(x_{b}-x_{2}\right)}\right]
\end{aligned}
\end{equation}
This provides the necessary numerator to construct a gluon transversity asymmetry, which can be written as,
\begin{equation}
A_{E_{x y}}=\frac{d \sigma_{p d \rightarrow \mu^{+} \mu^{-} X}\left(E_{x}-E_{y}\right) /\left(d \tau d q_{T}^{2} d \phi d y\right)}{d \sigma_{p d \rightarrow \mu^{+} \mu^{-} X}\left(E_{x}+E_{y}\right) /\left(d \tau d q_{T}^{2} d \phi d y\right)}.
\end{equation}
Based on the polarization vector difference, an equivalency can be derived using the unpolarized combination vector $\vec{E}_x+\vec{E}_y+\vec{E}_z:=U$, resulting is zeros for all terms in the spin polarization vector and tensor.  We can then write $\vec{E}_x-\vec{E}_y\equiv 2\vec{E}_x+\vec{E}_0-U$ and $\vec{E}_x+\vec{E}_y\equiv U-\vec{E}_0$. If we use $f^g_{1LL} \approx0$ for gluons \cite{b1k} such that the differential cross-section from the longitudinal tensor polarized part is small compared to the transverse tensor polarized part, we can write,
\begin{equation}
A_{E_{x y}}=\frac{d \sigma_{p d \rightarrow \mu^{+} \mu^{-} X}\left(E_{x}-E_{y}\right) /\left(d \tau d q_{T}^{2} d \phi d y\right)}{d \sigma_{p d \rightarrow \mu^{+} \mu^{-} X}\left(E_{x}+E_{y}\right) /\left(d \tau d q_{T}^{2} d \phi d y\right)}=\frac{d \sigma_{p d \rightarrow \mu^{+} \mu^{-} X}\left(2E_{x}-U\right) /\left(d \tau d q_{T}^{2} d \phi d y\right)}{d \sigma_{p d \rightarrow \mu^{+} \mu^{-} X} U  /\left(d \tau d q_{T}^{2} d \phi d y\right)} 
\end{equation}
The generalized experimental gluon transversity asymmetry can then be written as,
\begin{equation}
A_{E_{x y}}=\frac{2\sigma^{E_x}_{p d \rightarrow \mu^{+} \mu^{-} X}-\sigma^{U}_{p d \rightarrow \mu^{+} \mu^{-} X}}{\sigma^{U}_{p d \rightarrow \mu^{+} \mu^{-} X}} =\frac{1}{fP_{zz}}\frac{2N^{E_x}_{p d \rightarrow \mu^{+} \mu^{-} X}-N^{U}_{p d \rightarrow \mu^{+} \mu^{-} X}}{N^{U}_{p d \rightarrow \mu^{+} \mu^{-} X}},
\end{equation}
where $P_{zz}$ is the target ensemble tensor polarization pertaining to the tensor polarized cross-section events $N^{E_x}$, $f$ is the correction for the presence of unpolarized nuclei the beam interacts with, and $N$ is the number of counts in that spin state.
There are several ways to build a gluon transversity asymmetry using different quantization axes and polarized target configurations, but this equivalence provides a way to compare directly with predictions and requires the same polarized target magnet and orientation already in place in the SpinQuest experimental hall.  We point out here that $\sigma^{E_x}$ can be measured with either a purely tensor polarized target or as the difference between a enhanced tensor polarized target with high tensor polarization and some vector polarization subtracted from a purely vector polarized target.  A purely vector polarized target is significantly easier to make compared to a purely tensor polarized target, so this is our preferred method.  This term in the asymmetry then becomes,
\begin{equation}
\frac{1}{P_{zz}}N^{E_x} =\frac{1}{P'}N^{E_{\pm},E_x}-\frac{1}{P}N^{E_{\pm}}.
\end{equation}
Here, $P'$ represents the vector polarization when the target is tensor enhanced using the ss-RF method (see Section \ref{target}).  This is a different vector polarization value than when the tensor polarization is mitigated in the subtracted term.  In that case, we label the vector polarization $P$.  Both $P'$ and $P$ should be as high as possible to optimize statistical significance.

Also, due to the $\text{cos}2\phi$ term in Eq. \ref{2phi}, it is possible to extract a tensor polarization contribution in the azimuthal angle produced by gluon transversity.  This would show up even from the $\vec{E}_x$ polarized state alone, and the difference between a target with some tensor polarization and with no-tensor polarization can be use to measure the whole coefficient while exploring any azimuthal dependence.

As mentioned previously, the quark transversity is easiest to measure in the neutron/deuteron by mitigating any contribution from the tensor polarization.  The best possible target system would then
alternate between vector polarized, tensor polarized, and unpolarized.  With the UVA RF technology, it is possible to start with a target that is in Boltzmann equilibrium, which has both tensor and vector polarization, and then, on the scale of milliseconds, use the selective RF in the NMR frequency domain to remove tensor polarization in the target ensemble, as well as to create an unpolarized target and then flip back to the original spin state.  These alterations to the target spin configurations can be done between beam spills, allowing data collection in the different spin states while minimizing time dependant false asymmetries.

As pointed out earlier, the SpinQuest polarized target system can already accommodate most of the needs of this proposal.  Only slight modification must be made to the target cell to add a selective RF manipulation coil and adapt the polarization measuring NMR system to be optimized to function with the two competing RF sources.  For the purpose of the proposed measurements, one needs to separately measure different target spin configurations but with the field always pointing transverse vertical as it is now for SpinQuest. The experimental setup and data taking approach we will follow is similar to that used previously by experiments E866, E906, and E1039.

The material ND$_3$ is our preferred target for polarized deuterons.  Here, the dilution factor is higher (0.3) than that of NH$_3$, with a maximum vector polarization of up to 50\% and with a tensor polarization of 20\% under Boltzmann equilibrium.  This target can be RF manipulated to have a tensor polarization of over 35\% or 0\%.  The ND$_3$ target material is highly radiation resistant and has been used for decades, yet there are still new target systems being developed to leverage its full potential.  The ND$_3$ is our source for tensor observables as the spin-1 system but also our source for neutron vector polarized observables.  The neutron polarization is always over 90\% of the vector polarization of the deuteron.  This means the deuteron target is a very good source of neutron polarized TMDs when the tensor polarization is mitigated.

\section{Experimental Setup}
\subsection{The Spectrometer}
The experimental apparatus leans heavily on the E605, E772, E789, E866, and E906 experience for the best technique to handle high luminosities in fixed target Drell-Yan experiments. The key features of the apparatus are two independent magnetic field volumes: one to focus the high transverse momentum muons and defocus low transverse momentum muons and one to measure the muon momenta.  There is also a hadron absorber to remove high transverse momentum hadrons, a beam dump at the entrance of the first magnet, and Zinc/concrete walls for muon identification at the rear of the apparatus.  We intend to employ maximum use of existing equipment consistent with the physics goals.

\begin{figure}[ht!]
  \centering
  \includegraphics[width=150mm]{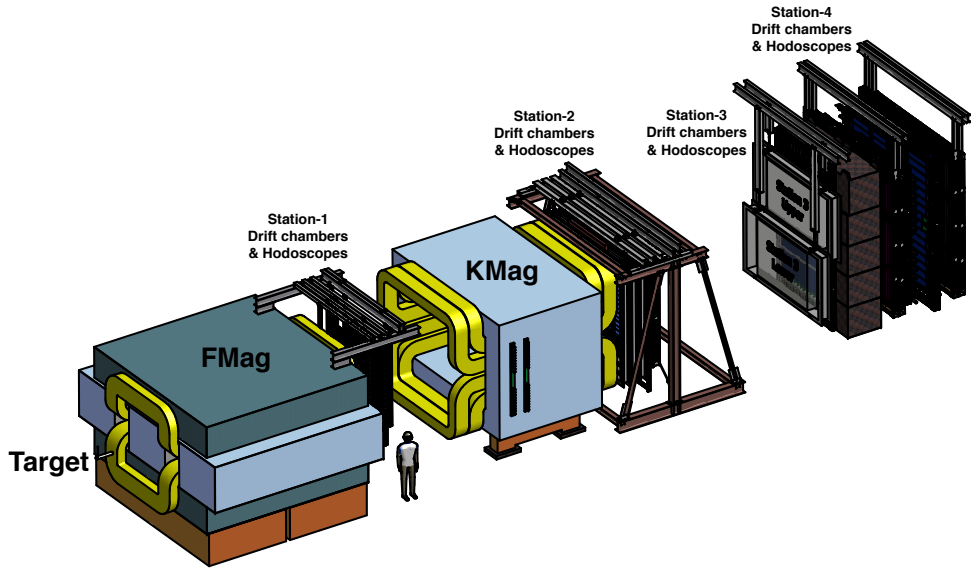}
  \caption{The SeaQuest Spectrometer}
  \label{E906Spec}
\end{figure}

In preparation for the SpinQuest/E1039 experiment, the existing spectrometer \cite{E906spec} shown in Fig.\ref{E906Spec} has been significantly updated with multiple repairs to a number of dead channels, optimization of bias voltage, studies of threshold/width, and repairs to a number of cables/cliplines, amplifiers, power-supplies, and discriminators.  As mentioned, the spectrometer consists of two magnets, FMAG and KMAG, and four tracking stations, where the last one serves as a muon identifier. The first magnet (FMAG) is now almost entirely surrounded in shielding blocks for use in SpinQuest and future experiments.  This magnet is a closed-aperture, solid iron
magnet. The beam protons that do not interact
in the targets are absorbed in the iron of the first
magnet, which allows only muons to traverse the
remaining spectrometer. The downstream magnet (KMag) is a large, open-aperture magnet that was previously used in the Fermilab KTeV experiment.  Each of the tracking/triggering stations consists of a set of scintillator hodoscopes to provide fast signals for the
FPGA-based trigger system and drift chambers.

Muon identification is accomplished with station 4, which is located downstream of a 1 m thick iron wall. Like the other stations, this station contains both triggering hodoscopes and tracking detectors. The station 4 tracking detectors consist of 4 layers of proportional tube
planes. Each plane is made of 9 proportional tube modules, with each module assembled from 16 proportional tubes, each 3.66 m long
with a 5.08 cm diameter, staggered to form two sub-layers. 

\begin{figure}[ht!]
  \centering
  \includegraphics[width=100mm]{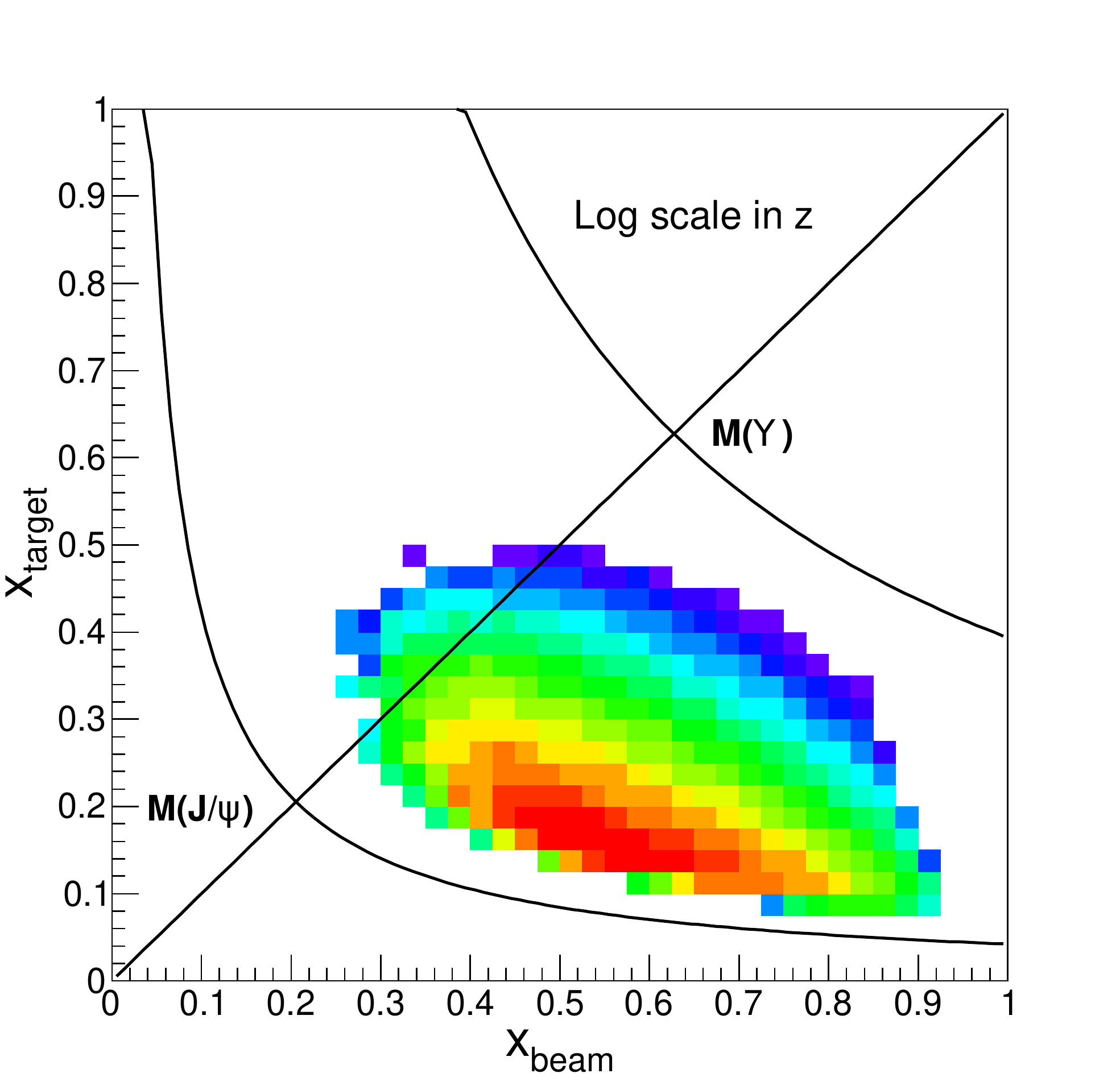}
  \caption{The acceptance of SeaQuest Spectrometer with polarized target.}
  \label{E906Acc}
\end{figure}

This spectrometer was designed to perform Drell-Yan measurements at large \xB. This is illustrated in Fig. \ref{E906Acc}, where the acceptance of the SpinQuest detector is plotted as function of \xB (x-axis) and \xT (y-axis).

This is an excellent kinematic range for the proposed sea-quark and gluon transversity measurements, covering the region of large anti-down quark excess observed by E866, where large pion-cloud effects may be expected. The contributions from target valence quarks at large \xT\ are then negligible.

The experiment will use the Fermilab main injector beam with an energy of 120 GeV and a 4.4 second spill every minute. The maximum beam intensity will be $3 \times 10^{12}$ protons per spill, which is defined by the polarized target and spectrometer.  We will use  $1 \times 10^{12}$ protons per spill for conservative beam-time estimates.

\subsection{SpinQuest Construction Status}
The entire SpinQuest shielding assessment is complete and approved, and the construction is nearly complete as well.  All that remains for the shielding construction is to place the top of the target alcove in place.  This can be completed after the piping in utilities in the west penetration are installed.  This effort is underway.  The polarized target system is in place, as is the helium liquefaction system.  Much work has also been put into the ODH analysis and safety assessment.  Considerable effort has been put into understanding the safety protocol and cryogenic safety regulations required from the Fermilab Environment, Safety, and Health Manual (FESHM).  The final steps of fabrication of the safety relief and plumbing system are also underway.

\begin{figure}[ht!]
  \centering
  \includegraphics[width=120mm]{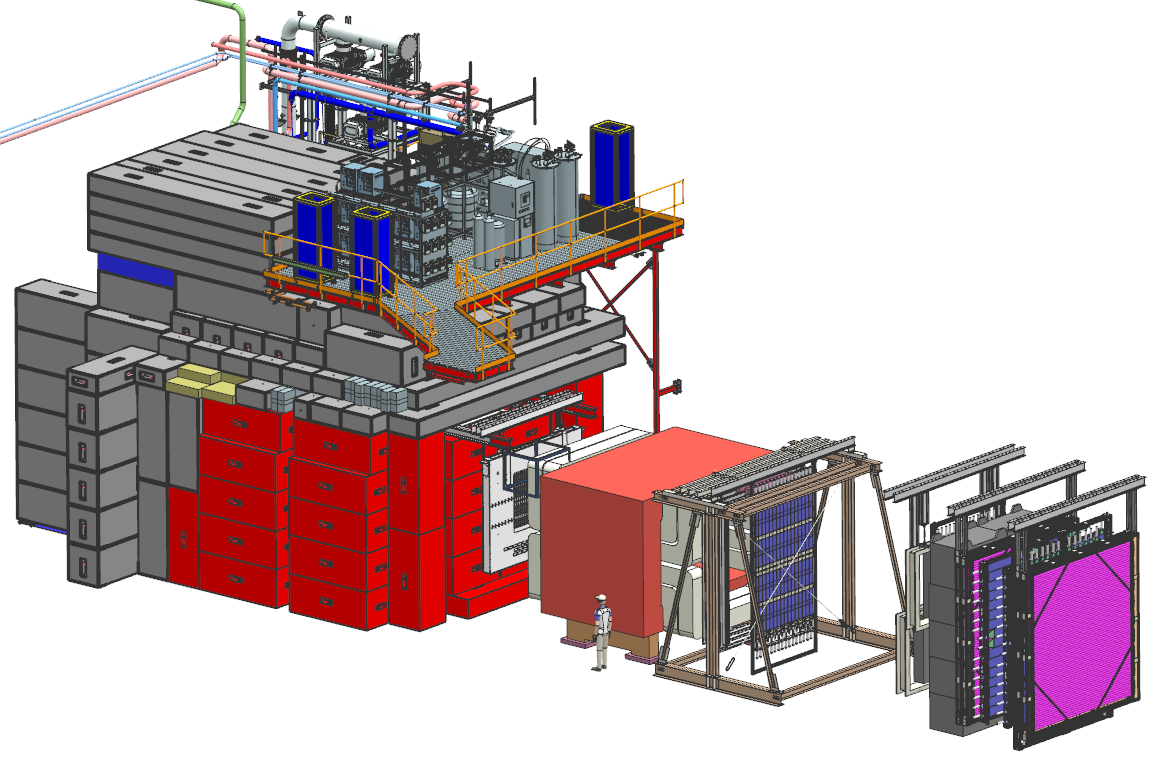}
  \caption{The target Cave and spectrometer down stream.  The red shielding blocks are surrounding FMag.  The new UVA helium liquifier is setup on top of the new cryo-platform.  Image courtesy of Don Mitchell of Fermilab.}
  \label{cave1}
\end{figure}

\begin{figure}[ht!]
  \centering
  \includegraphics[width=120mm]{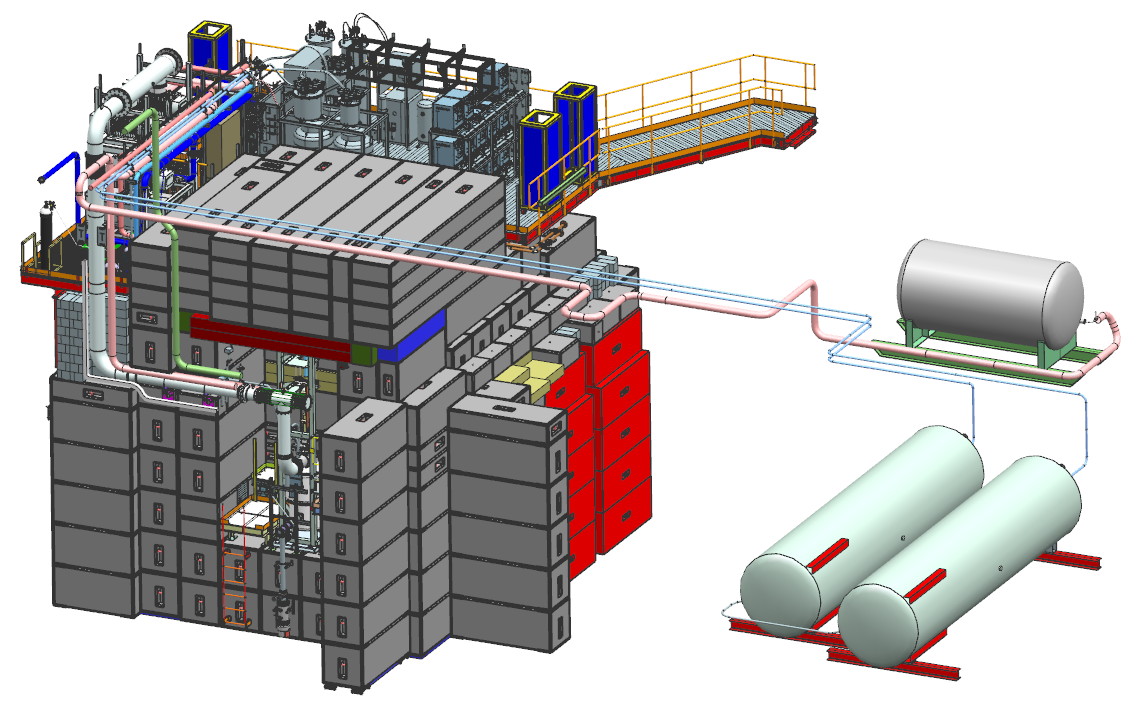}
  \caption{The target cave shown from the upstream end with the SpinQuest target system shown in place in the alcove.  The connecting storage tanks for Helium gas and liquid nitrogen are outside of the NM4 building.}
  \label{cave2}
\end{figure}

The helium liquefaction system and roots pumps for the high power evaporation refrigerator are in place and leak-checked as is the vacuum space for the magnet, the superconducting magnet vessel, and the refrigerator space.  The installation of these systems was a major achievement and makes NM4 a specialized, world-class spin physics facility.  The superconducting magnet and the evaporation refrigerator require a local liquid helium supply. The liquid He consumption and the collection of the exhaust gas is absolutely essential for this type of polarized target.  Keeping the target at 1 K while irradiating it with the microwave requires evaporation of over 100 liters of liquid Helium per day through pumping.  Combined with the heat-load of the magnet, we require about 130 liquid liters per day for sustainable running.  Not recovering the Helium gas after the roots pumps would lead to wasting an unacceptable amount of a nonrenewable resource, adding considerable running costs.  New DOE guidelines required the installation of the closed loop Helium liquefier system now part of the SpinQuest infrastructure.  Furthermore, such a system needs a special plumbing and recovery system consisting of Helium and Nitrogen transfer lines, pumping lines from the target to the roots pumps as well as a special quench line, which would handle the Helium exhaust gas during magnet quenches.  With considerable effort from Fermilab, the University of Virginia, and Quantum Technologies, a closed loop system with all appropriate safety requirements was designed and installed.  Fig. \ref{cave1} shows the target system and shielding alcove with the spectrometer down stream to run SpinQuest.  The liquefaction system is sitting on the cryo-platform.  The entire cryo-platform, all shielding, and the target system inside are all new for SpinQuest.  The present proposal will use this new infrastructure as is.  The upstream perspective is shown in Fig. \ref{cave2}.  

The beam line and collimator have also been installed and approved.  The collimator is required for operation of the target polarizing magnet.  The superconducting coils will become resistive if any part of the coil temperature exceeds the critical value.  The collimator was designed to minimize beam tails going into the magnet coils.  Because we are pushing the proton intensity frontier on a polarized target system with this type of magnet, special care has been taken to try to minimize needless excess heat-load to the coils.

There have also been considerable updates to the NM4 hall infrastructure, including replacing hoses on the cooling water lines for the kTeV magnet and adding in low conductivity water utilities to cool the liquifier, roots pumps, and microwave generator.  And, of course, all of the power utilities to support the target infrastructure in the target cave and cryo-platform have also been installed.

\clearpage

\subsection{The Polarized Target}
\label{target}
\begin{figure}[!tbph]
  \centering
  \begin{minipage}[b]{0.4\textwidth}
    \includegraphics[width=\textwidth]{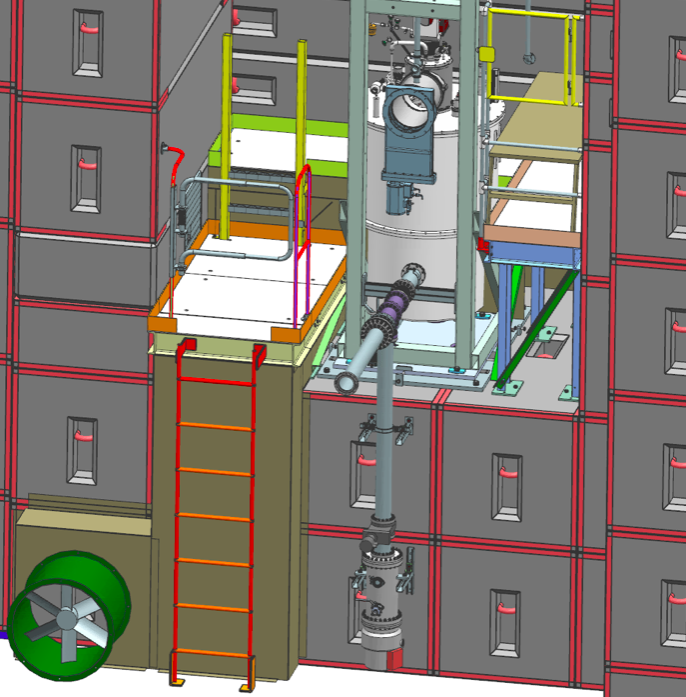}
  \end{minipage}
  \hspace{20mm}
  \begin{minipage}[b]{0.45\textwidth}
    \includegraphics[width=\textwidth]{Capture2.PNG}
  \end{minipage}
    \caption{\it{(Left) The target cave entrance and the polarized target with beam line connection from the upstream perspective.  (Right) Cross-sectional drawing of the polarized target showing the target insert, the evaporation refrigerator, and the superconducting magnet.} }
    \label{TGTschem}
\end{figure}

This proposal requires the same SpinQuest polarized target which has been rebuilt and tested at UVA and recently installed in the NM4 experimental hall at Fermilab. The target system consists of a 5T superconducting split pair magnet, a $^{4}$He evaporation refrigerator, a 140 GHz microwave source, and a large 17,000 m$^{3}$/hr pumping system. The target is polarized using Dynamic Nuclear Polarization (DNP) \cite{crabb1} and is shown schematically in Fig. \ref{TGTschem}.  In the left hand picture, the target cave entrance is shown, and the polarized target with beam line connection from the upstream perspective can be seen (Figure provided by Don Mitchell of FNAL).  In the right hand picture, the cross-sectional drawing of the polarized target shows the target insert, the evaporation refrigerator, and the superconducting magnet. The beam direction is from right to left, and the field direction is vertical along the symmetry axis so that the target polarization is transverse to the beam direction. The UVA refrigerator is also shown with the target insert holding the polarized target material (ND$_3$) with the top cell in the center of the split coils.

While the magnetic moment of the deuteron is too small to lead to a sizable polarization in a 5 T field through the Zeeman effect, electrons in that field at 1 K are better than 99\% polarized. By doping a suitable solid target material with paramagnetic radicals to provide unpaired electron spins, one can make use of the highly polarized state of the electrons.  The dipole-dipole interaction between the nucleon and the electron leads to hyperfine splitting, providing the coupling between the two spin species. By applying a suitable microwave signal, one can populate the desired spin states.
As mentioned, we will use frozen deuterated ammonia beads \cite{crabb2} (ND$_{3}$) as the target material and create the paramagnetic radicals (roughly $10^{19}$ spins/ml) through irradiation with a high intensity electron beam at NIST.  The cryogenic refrigerator, which works on the principle of liquid $^{4}$He evaporation, can cool the bath  to ~1 K by lowering the $^{4}$He vapor pressure down to less than 0.118 Torr. The polarization will be measured with NMR techniques with three NMR coils per cell, placed inside each target cell. The maximum polarization achieved with the deuteron target is around 50\% vector polarization with a packing fraction of about 60\%.  In our estimate for the statistical precision, we have assumed an average of 32\% vector polarization.  The polarization dilution factor, which is the ratio of free polarized deuterons to the total number of nucleons, is 3/10 for ND$_{3}$, due to the presence of nitrogen.  The target material will need to be replaced approximately every 8-10 days in all three target cells due to the beam induced radiation damage. This work will involve replacing the target material in the target insert, cooling down the target and performing multiple thermal equilibrium measurements. From previous experience, we estimate that this will take about a shift to accomplish. Careful planning of these changes will reduce their impact on the beam time. Furthermore, we will be running with three active targets on one target insert, thus reducing any additional loss of beam time. The target cells are about 80 mm long and hold about 12 grams of ND$_3$. Each cell contains 3 NMR coils spaced evenly over the whole length.

In our spin-1 target, the deuterons have nonzero quadrupole moments, and the structural arrangement of the nuclei in the solid generate electric field gradients (EFG) which couple to the quadrupole moment.  This results in an additional degree of freedom in polarization that the spin-1/2 nucleons do not possess. The target spins in the ensemble can be aligned in both a vector ($P$) and tensor ($P_{zz}$) polarization.  Defined
in terms of the relative occupation of the three magnetic substates of the spin-1 system ($M_J=0$, $\pm1$) which correspond to the three energy levels of a solid polarized spin-1 target ($m=0$, $\pm1$),
$$P=\frac{n_+ - n_-}{N}$$ and
$$P_{zz}=\frac{n_+ -2n_0 + n_-}{N},$$
with $n_i$ being the relative occupation of the energy levels with $m=i$, and $N=n_+ + n_0 + n_-$.

Recent advancement in tensor polarized target technology and spin-1 NMR measurements of these targets make the proposed experiment possible \cite{kel5}.  In the next few sections these details are outlined.

\subsubsection{NMR Measurements}

The deuteron spin polarization is measured with a continuous-wave NMR system based
on the Liverpool Q-meter design \cite{court}
and recently upgraded by LANL/UVA. The Q-meter
works as part of a circuit with phase sensitivity designed to respond to the
change of the impedance in the NMR coil. The radio-frequent (RF) susceptibility
of the material is inductively coupled to the NMR coil which is part of a
series LCR circuit, tuned to the Larmor frequency of the nuclei being probed.
The output, consisting of a DC level digitized and recorded as a target event \cite{crabb1}
in the target data acquisition system.

The polarized target NMR and data acquisition include the software control
system, the Rohde \& Schwarz RF generator (R\&S), the Q-meter enclosure, and
the target cavity insert. The Q-meter enclosure contains a series of Q-meter
circuits with separate connection cables which are used for different target cup cells during the
experiment.  The target material and NMR coil are held in
polychlorotrifluorethylene (Kel-F) cells with the whole target insert
cryogenically cooled to 1 K. Kel-F is used because it contains no free protons.

The R\&S generator produces a RF signal which is frequency modulated to sweep
over the frequency range of interest. Typically, the R\&S responds to an
external modulation, sweeping linearly from 400 kHz below to 400 kHz above the
Larmor frequency. The signal from the R\&S is connected to the NMR coils
within the target material. To avoid degrading reflections in the long
connection from the NMR coil to the electronics, a standing wave can be
created in the transmission cable by selecting a length of cable that is an
integer multiple of the half-wavelength of the resonant frequency. This
specialized connection cable is known as the $\lambda/2$ cable and is a
semi-rigid cable with a teflon based dielectric. The NMR coil is a set of loops
made of 70/30 copper-nickel tube, which minimizes interaction with the
proton beam.  The coil opens up into an oval shape spanning approximately
2 cm inside the cup.  It is possible to enhance signal-to-noise information
through the software control system by making multiple frequency sweeps and
averaging the signals. A completion of the set number of sweeps results in a
single target event with a time stamp. The averaged signal is integrated to
obtain a NMR polarization area for that event. Each target event written
contains all NMR system parameters and the target environment variables needed
to calculate the final polarization. The on-line target data and conditions
are analyzed over the experiment's set of target events to return a final
polarization and associated uncertainty for each run.

A target NMR calibration measurement or Thermal Equilibrium measurement (TE)
is used to find a proportionality relation to determine the enhanced
polarization under a range of thermal conditions given the area of the
\textquotedblleft Q-curve\textquotedblright\ NMR signal at the same magnetic
field. The magnetic moment in the external field results in a set of 2$J$+1
energy sublevels through Zeeman interaction, where $J$ is the particle spin.
The thermal equilibrium (TE) natural polarization can be calculated
from Curie's Law \cite{kahn} with knowledge of
the external field strength and the
temperature at thermal equilibrium. Measuring $P_{TE}$ at low temperature increases stability and the
polarization signal. This is favorable because the uncertainty in the NMR
signal increases as the area of the signal decreases. In fact, much of the
target uncertainty comes from error in the calibration. 

The dynamic polarization value is derived by comparing the enhanced signal
$S_{E}$ integrated over the driving frequency $\omega$, with that of the (TE)
signal,
\begin{equation}
P_{E} = G\frac{\int S_{E}(\omega)d\omega}{\int S_{TE}(\omega)d\omega}%
P_{TE}=GC_{TE}A_{E}, \label{pe}%
\end{equation}
and calibration constant defined as,
\begin{equation}
C_{TE}=\frac{P_{TE}}{A_{TE}} \label{cc}%
\end{equation}
where $P_{E}$ ($A_{E}$) is the polarization (area) of the enhanced signal, and
$P_{TE}$ ($A_{TE}$) is the polarization (area) from the thermal equilibrium
measurement. The uncertainty in the calibration constant, $\delta
C_{TE}/C_{TE}$, can easily be calculated using the fractional error from
$P_{TE}$ and $A_{TE}$. The ratio of gains from the Yale card used during the
thermal equilibrium measurement to the enhanced signal is represented as $G$.
For more detail see, \cite{kel2}.

\subsubsection{Neutron Polarization Measurements}

The deuteron polarization will be monitored by our continuous wave NMR
system as used for the proton with one small change. There are two means
whereby the polarization can be extracted from the NMR signal: the area method
and the peak-height method. We intend to use both.

First, the total area of the NMR absorption signal is proportional to the
vector polarization of the sample, and the constant of proportionality can be
calibrated against the polarization of the sample measured under thermal
equilibrium (TE) conditions. This is the standard method used for polarized
proton targets but can be more problematic for deuteron targets. Typical
conditions for the TE measurements are 5 T and 1.4 K, where the deuteron
polarization is only 0.075\%, compared to 0.36\% for protons. This smaller
polarization, along with quadrupolar broadening, makes the deuteron TE signal
more difficult to measure with high accuracy. A cold NMR system can be used
to improve the signal-to-noise ratio of the NMR signal
\cite{court1}.

The deuteron polarization can also be extracted from the shape of the NMR
signal. The deuteron is a spin-1 nucleus with the three energy levels,
$m=-1,0,+1$, and the NMR absorption signal lineshape is the sum of the two
overlapping absorption lines consisting of the $-1\rightarrow0$ and
$0\rightarrow+1$ transitions. In the case of $^{14}$ND$_{3}$, the deuteron's
electric quadrupole moment interacts with electric field gradients within the
molecule and splits the degeneracy of the two transitions. The degree of
splitting depends on the angle between the magnetic field and the direction of
the electric field gradient. The resultant lineshape, integrated over a sample
of many polycrystalline beads, has the form of a Pake doublet \cite{pake}. It has been
experimentally demonstrated that, at or near steady-state conditions, the
magnetic substates of deuterons in dynamically polarized $^{14}$ND$_{3}$ are
populated according to the Boltzmann distribution with a characteristic spin
temperature $T$ that can be either positive or negative, depending on the sign
of the polarization.

When the system is at thermal equilibrium with the solid lattice, the deuteron
polarization is known from:
\begin{equation}
\label{VECT}P_{z} = \frac{4+\tanh\frac{\mu B}{2 k T}} {3+\tanh^{2}\frac{\mu
B}{2 k T} }%
\end{equation}
where $\mu$ is the magnetic moment, and $k$ is Boltzmann's constant. The
vector polarization can be determined by comparing the enhanced signal with
that of the TE signal (which has known polarization). This polarimetry method
is typically reliable to about 5\% relative uncertainty.

Similarly, the tensor polarization is given by:
\begin{equation}
\label{TENS}P_{zz} = \frac{4+\tanh^{2}\frac{\mu B}{2 k T}} {3+\tanh^{2}%
\frac{\mu B}{2 k T} }%
\end{equation}.

From Eqs.~\ref{VECT} and~\ref{TENS}, we find:
\begin{equation}
\label{PZZEQN}P_{zz}= 2 - \sqrt{4-3 P_{z}^{2}}
\end{equation}

In addition to the TE method, polarizations can be determined by analyzing NMR
lineshapes as described in~\cite{dulya} with a typical 5-7\% relative
uncertainty. At high polarizations, the intensities of the two transitions
differ, and the NMR signal shows an asymmetry $R$ in the value of the two
peaks. The vector polarization is then given by,
\begin{equation}
\label{RVECT}P_{z} = \frac{R^{2}-1}{R^{2}+R+1}%
\end{equation}
and the tensor polarization is given by,
\begin{equation}
\label{TVECT}P_{zz} = \frac{R^{2}-2 R +1}{R^{2}+R+1}.%
\end{equation}
This measuring technique can be used as a compliment to the TE method,
resulting in reduced uncertainty in polarization for vector polarizations over 28\%.

The measurement of the neutron polarization ($P_{n}$) is achieved by a
calculation using the NMR measured polarization of the deuteron ($P_{d}$). The
quantum mechanical calculation using Clebsch-Gordan coefficients show 75\% of
the neutron spins in the $D$-state are antiparallel to the deuteron spins. The
resulting neutron polarization is,
\[
P_{n}=(1-1.5\alpha_{D})P_{d}\approx0.91P_{d},
\]
where $\alpha_{D}$ is the probability of the deuteron to be in a $D$-state.

\subsubsection{The Deuteron NMR Lineshape}
\label{signal}
The quadrupole moment of the spin-1 nuclei results from the nonspherically symmetric charge distribution in the quadrupolar nucleus.  For materials without cubic symmetry (e.g. C$_4$D$_9$OH or ND$_3$), the interaction of the quadrupole moment with the EFG breaks the degeneracy of the energy transitions, leading to two overlapping absorption lines in the NMR spectra.
\begin{figure}
\begin{center}
\includegraphics[height=67mm, angle=0]{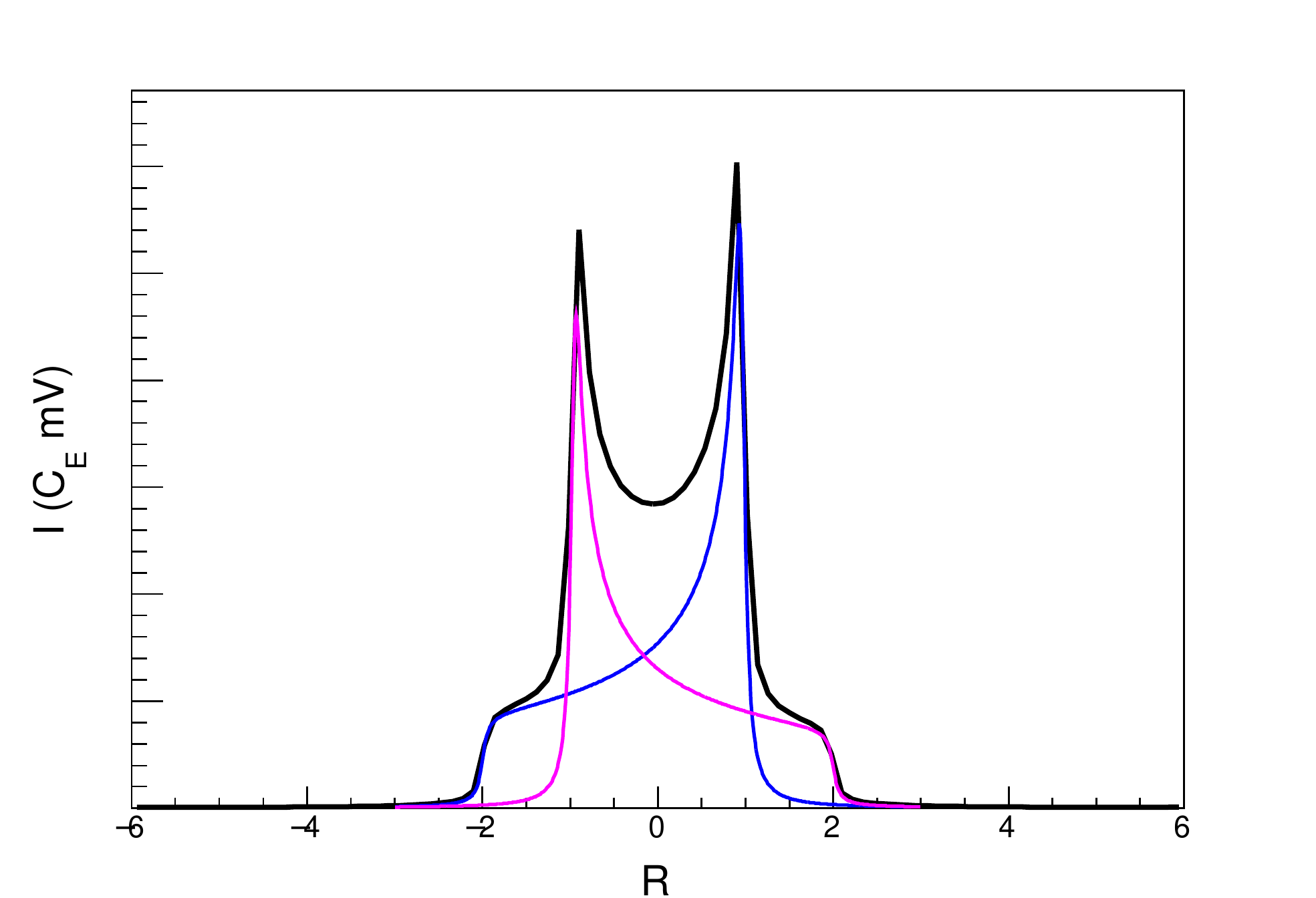}
\end{center}
\caption{An example of the NMR lineshape of a spin-1 target with a non-cubic symmetry demonstrating the two overlapping absorption lines.  The two intensities of the signal $I_{+}$ and $I_{-}$ are shown in blue and pink respectively.  Figure from \cite{kel5}. }
\label{pake}
\end{figure}
The spin-1 NMR lineshape is shown in Fig. \ref{pake}, demonstrating the two intensities $I_+$ (in blue) and $I_-$ (in pink).  In terms of population,$$I_+=C(\rho_+-\rho_0)$$ and $$I_-=C(\rho_0-\rho_-),$$ where $\rho_x$ is the population density in the $m=x$ energy level, and
$C$ is the calibration constant.
The term intensity is used here to indicate both the height and area of these two individual regions.  The frequency is indicated by a dimensionless position in the NMR line $R=(\omega-\omega_D)/3\omega_Q$, which spans the domain of the NMR signal where $\omega_Q$ is the quadrupolar coupling constant.  In these units, $R=0$ corresponds to the Larmor frequency of the deuteron at 5 T ($\omega_D=32.679$ MHz).
The local electric field gradients that couple to the quadrupole moments of the spin-1 system cause an asymmetric splitting of the energy levels into two overlapping absorption lines.  The energy levels of
the non-cubic symmetry spin-1 system can be expressed
as,
$$E_m=-\hbar \omega_D m + \hbar \omega_Q (3\text{cos}^2\theta-1+\eta \text{sin}^2\theta \text{cos}2\phi)(3m^2-2),$$
where $\theta$ is the polar angle between the axis of the deuteron bond and the magnetic field, see Fig. \ref{levels}.  The azimuthal angle $\phi$ and parameter $\eta$ are fixed parameters used to characterize the electric field gradient with respect to the deuteron bond axis.
The degree of axial symmetry and dependence on the polar angle can be understood from the basis lineshape for an isotropic rigid solid, which is known as a Pake doublet.  The polarization information can be extracted from a fit of the NMR data providing the areas of the two intensities \cite{dulya,kel1}.
\begin{figure}
\begin{center}
\includegraphics[height=67mm, angle=0]{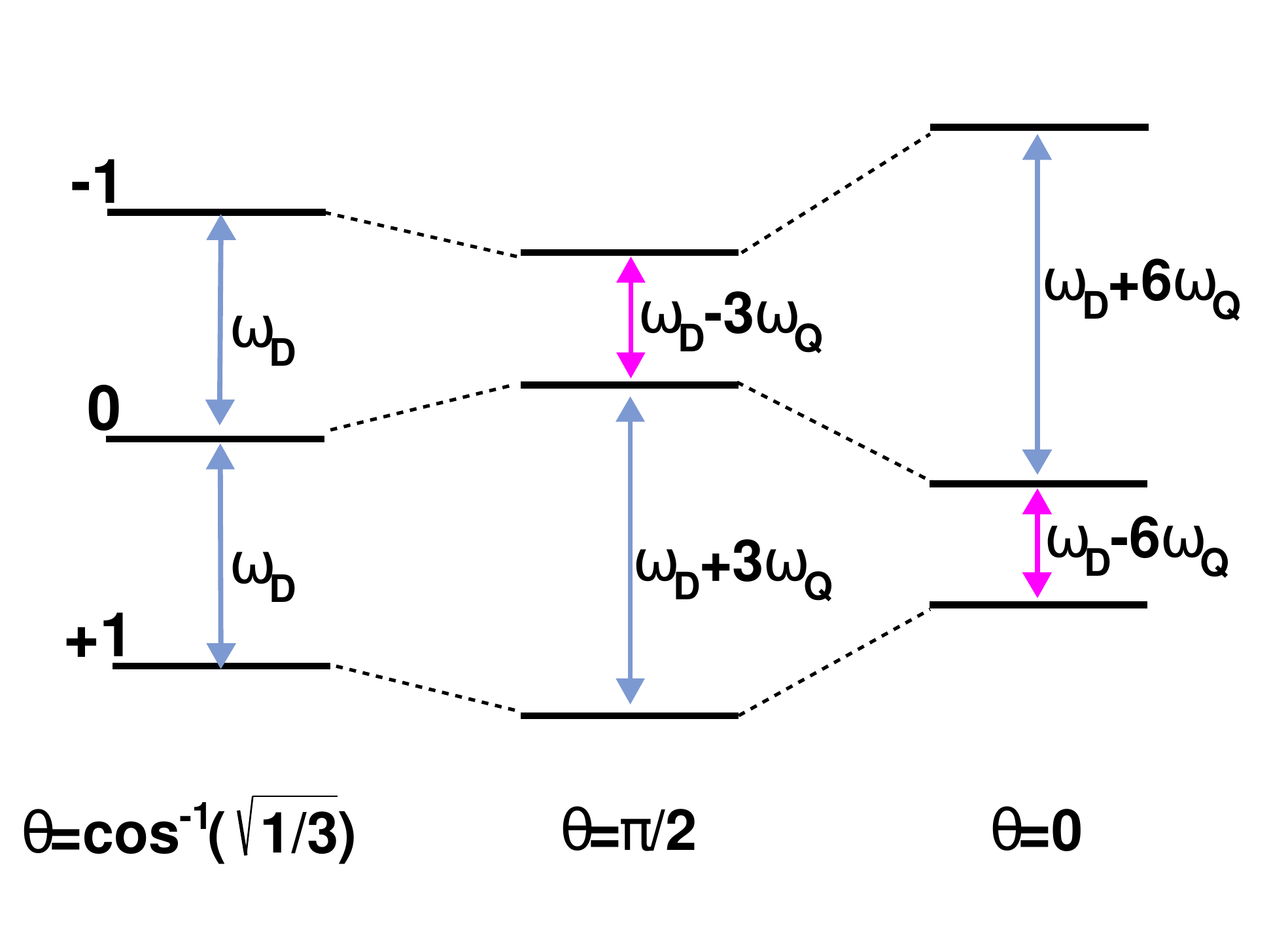}
\end{center}
\caption{The energy level diagram for deuterons in a magnetic field for three values of $\theta$ where $\hbar \omega_D$ is the deuteron Zeeman energy, and $\hbar \omega_Q$ is the quadrupole energy.  The color indicates which transition corresponds to which peak shown in Fig. \ref{pake}.  Figure from \cite{kel5}.}
\label{levels}
\end{figure}
The peaks of the Pake doublet ($R\sim\pm1$) correspond to the principal axis of the coupling interaction, being perpendicular ($\theta=\pi/2$) to the magnetic field. This is the most probable configuration within each transition, as indicated by the height in the intensity of each peak. The opposing end in each absorption line, called the pedestal, corresponds to the configuration when the principal axis of the coupling interaction is parallel ($\theta=0$) to the magnetic field, which has much less statistical significance, as indicated by the small relative height in the intensities in each transition around ($R\sim\mp2$).

If the ensemble of the spin system is in thermodynamic equilibrium, the ratio of the intensities ($r=I_{+}/I_{-}$) can be used to extract the polarizations directly \cite{dulya}.
\begin{equation}
P=\frac{r^2-1}{r^2+r+1}~~~~~~P_{zz}=\frac{r^2-2r+1}{r^2+r+1}
\label{one}
\end{equation}
or simply,
\begin{equation}
\frac{P_{zz}}{P}=\frac{r-1}{r+1}.
\label{two}
\end{equation}
The extracted information from the fit also gives the sum of the two intensities, which provides the vector polarization $P=C(I_{+}+I_{-})$, while the difference provides the tensor polarization $P_{zz}=C(I_+-I_-)$. It is important to note that these two expressions remain true even if the system is not in thermodynamic equilibrium, unlike Eq. \ref{one} and \ref{two}.  Once the calibration constant $C$ is measured, these expressions can be used to extract the averaged polarizations of the ensemble over the course of the experiment \cite{kel2}.

\subsubsection{Tensor Polarization Enhancement}
\label{rf1}

To manipulate the magnitude of tensor polarization during DNP pumping, a separate source of coil generated RF irradiation is used to selectively saturate some
portion of the deuteron NMR line. By applying RF irradiation at a frequency or over a frequency range (hole burning \cite{jeffries, del}), transitions are induced between the
magnetic sublevels within the frequency domain of the applied RF.  A spin-diffusion rate that is
small compared to the effective nuclear relaxation rate allows for significant changes to the NMR line via the RF, which can be strategically applied
to manipulate the spin-1 tensor polarization.  In the presented set of measurements, DNP microwaves were used, as well as an additional RF source that used semi-saturating RF (ss-RF) irradiation to maximize the tensor polarization for the 1 K and 5 T system \cite{kel1}. A semi-saturated steady-state condition is used which
manipulates and holds the magnetic sublevels responsible for polarization enhancement. The continuous wave NMR (CW-NMR) lineshape is measured and manipulated to maximize tensor polarization.  The technique of ss-RF requires using a power profile that is sensitive to the intensity distributions with the correct modulation time signature over the frequency domain to optimally enhance. To be useful in a scattering experiment setting, the target
ensemble averaged tensor polarization must be increased and held during the beam spill.  Temporarily enhanced states during beam-target interactions are much more plausible at facilities that have a short beam spill per cycle such as Fermilab.  This allows significant beam intensity on a target that is RF-manipulated for a short period, and then gives significant recovery time to build up polarization again for the next spills. 

The source of ss-RF comes from a dedicated coil
with a field $\mathbf{B}_{\nu}$, resulting in an induced transition rate proportional to \cite{kel1},
\begin{equation}
\xi=2\pi\frac{\mathbf{B}_{\nu}a_{\nu}}{\mathbf{B}_0}\delta(\omega_D-\omega_{\nu}),
\end{equation}
where $\omega_D$ is the Larmor frequency, $\omega_{\nu}$ is the ss-RF frequency, $a_{\nu}$ is the coupling constant, and $\mathbf{B}_0$ is the strength of the holding field.  The ss-RF can only play a role between nuclear spin energy levels that differ by the spin of the mediating photon.  The ss-RF drives transitions that lead to equalization of the populations in the energy levels at the applied frequencies.  This implies that the change in intensities at the location $R$ in the NMR line due to the ss-RF can be expressed as,
\begin{equation}
\frac{I_{\pm}(R)}{dt}=-2\xi\omega_1 I_{\pm}(R),
\label{dos1}
\end{equation}
\begin{equation}
\frac{I_{\mp}(-R)}{dt}=\xi\omega_1 I_{\pm}(R).
\label{dos2}
\end{equation}
In other words, the rate at which any one of the intensities changes due to ss-RF is only dependent on
the intensity level and the strength of the $\mathbf{B}_{\nu}$ field, or RF power.  Here, $\omega_1$ is the reciprocal of the electron longitudinal relaxation rate, used to be consistent with previous work \cite{kel1}.  The total polarization can only be decreased at the ss-RF location in $R$, so strategic implementation is required to enhance the difference in the integrated $I_+$ and $I_-$ regions.  Equation \ref{dos2} indicates that for any region at $R$ in the intensity reduced by the ss-RF also results in the increase in the opposite signed intensity growing at $-R$ at half the rate as the decrease seen at $R$.  This also implies that the region at $-R$ increases in area by half of the lost area at $R$.

Equations \ref{dos1} and \ref{dos2} affirm that materials with the same lineshape can be treated exactly the same under ss-RF tensor enhancement.  This expression does not change for different materials' relaxation rates.
\begin{figure}
\begin{center}
    \includegraphics[width=0.50\textwidth]{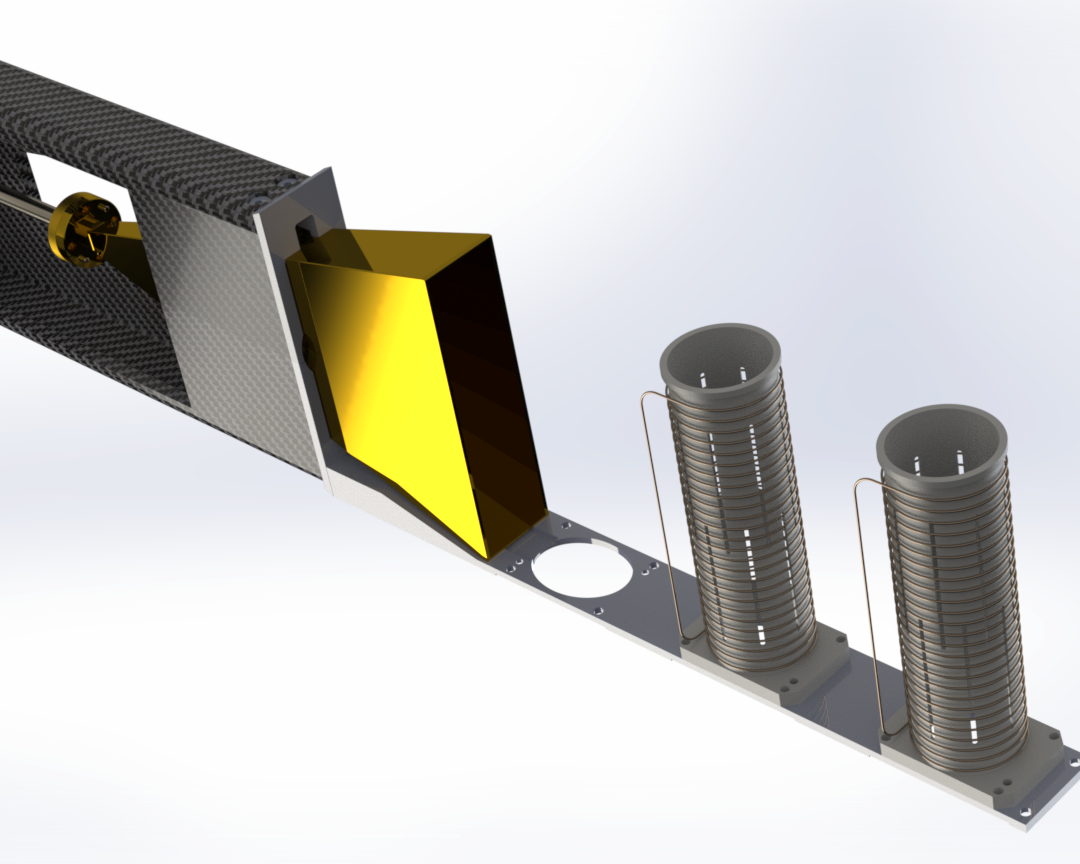}
\end{center}
\caption{Drawing of the ss-RF cup and coil used for the set of experiments discussed.  The microwaves come from the gold plated horn shown on the target insert.  The coils are designed to produce a homogeneous RF-field pointing orthogonal to the holding field.  These specialized coils can perform AFP and ss-RF and still allow the DNP microwave to penetrate the target cell.  Figure from Carlos Ramirez of UVA polarized target group.}
\label{cup}
\end{figure}
The nuclear spin polarization is measured with an CW-NMR system \cite{court}. With this system, the RF
susceptibility of the material is inductively coupled to the NMR coil, which is part of a series LCR Q-meter circuit that is tuned to the Larmor frequency of the nuclei of interest.
The Q-meter based NMR provides a non-destructive polarization probe of the nuclear spin ensemble in the solid-state target.  

For the selective excitation using ss-RF, an additional coil around the target cup is necessary.  This additional RF coil is connected to an RF-generator and amplifier. The ss-RF coil consists of multiple turns of silver-covered copper-clad non-magnetic steel with a diameter of $\sim$0.2 mm.  The coil is constructed to approximate a homogeneous RF field around the target material that is perpendicular to the holding field.  For optimal performance, the coil is both impedance matched and tuned as an LCR circuit to maximize power and reduce reflection at the coil.  The ss-RF is modulated over the frequency domain of interest at the appropriate RF power to semi-saturate the NMR line to the intensity level of interest.  This same RF circuit can be used to perform an Adiabatic fast passage (AFP) on the target.  A half AFP can be used to unpolarize the target, putting about 50\% of the target polarization in the opposing direction.  The ss-RF can then be used to tune the tensor and vector polarization to exactly zero during beam target interaction time.  The RF sweep rate to achieve an AFP is on the order of a few milliseconds.  One can then apply another AFP to return to the maximized polarization state.

The distribution of the power profile delivered is also dependant on the coil.  A high quality factor and optimized tune in the ss-RF coil delivers a more precise and localized RF load in the signal domain.  The power profile of the ss-RF in the CW-NMR is characterized by the Voigt function \cite{kel1}.

The material of choice is ND$_3$, primarily because it has the desired lineshape, is highly radiation resistant, and offers the best polarization and dilution factor out of all possible polarized target materials.  However, because of its long relaxation rate, ND$_3$ does take a while to fully polarize (several hours), and it will maximize at only about 50\%.  This experiment would require tensor polarization as well, which is calculated from the equilibrium relation between the energy level.  Under Boltzmann equilibrium (no ss-RF) for a vector polarzation of 50\%, the resulting tensor polarization is,
\begin{equation}
P_{zz}=2-\sqrt{4-3P^2}=19.7\%.
\label{boltz}
\end{equation}
Enhancement beyond this level requires the application of selective excitation \cite{kel1} using the ss-RF to maximize the difference in the two intensities $I_+$ and $I_-$ such
that $P_{zz}=C(I_{+}-I_{-})$ is maximized.

\begin{figure}[!tbp]
  \centering
    \includegraphics[width=100mm]{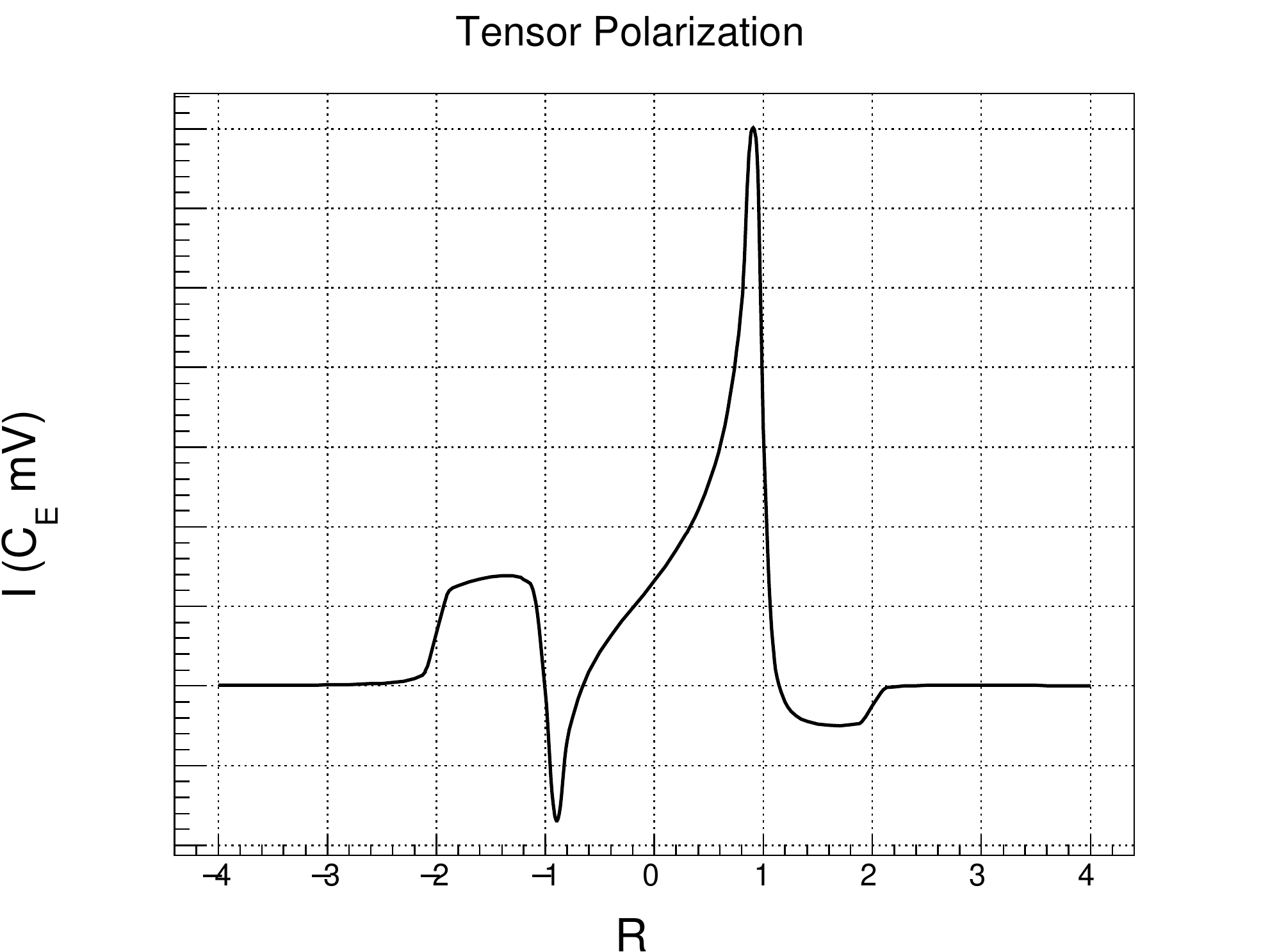}
    \caption{The tensor polarization shown from the
    difference of the intensities $I_+(R)$ and $I_-(R)$.  Figure from \cite{kel5}.}
    \label{tens}
\end{figure}

Vector polarization is the sum of $I_{+}(R)$ and $I_{-}$(R) over the frequency domain in $R$.  Similarly, a tensor polarization plot is shown in 
Figure \ref{tens} represents the difference of $I_{+}(R)$ and $I_{-}(R)$ over the frequency domain in $R$.  By selectively applying the ss-RF, it is possible to reduce the regions in the $P_{zz}$ line that drop below the x-axis.  When this is done simultaneously over all negative regions in the domain, the tensor polarization is enhanced.

\subsubsection{Semi-Saturating RF Enhancement}
\label{rf2}
To optimize the enhancement, the ss-RF excitation must minimize the negative tensor polarization for all $R$ while minimizing the reduction to the overall area of the NMR signal from the process.  The two critical regions lie around $R\sim \mp1$ ($\theta\approx\pi/2$) and $\pm1<R<\pm2$  ($\theta \approx 0$). For positive vector polarization, the greatest integrated tensor polarization enhancement is achieved through selective excitation to reduce the size of the smaller transition area with intensity $I_{-}$.  This can be thought of as minimizing the negative parts of the tensor polarization, shown in Fig. \ref{tens}.  In both figures, the y-axis would normally be millivolts scaled by a multiplicative factor $C_E$, which is sensitive to the characteristics of the NMR coil, such as inductance, geometry, and orientation.  We therefore leave these units generalized and provide a scale for relative change when necessary.  For negative vector polarization, the greatest enhancement comes from the reduction of the transition area with intensity $I_{+}$; otherwise, the treatment of both cases is identical, so it is convenient to focus on positive vector and tensor polarization.

The target must first be polarized with DNP to achieve the highest vector polarization possible for that material.  This maximizes the signal area to be used in the ss-RF manipulation. 

The ss-RF is then applied as described.  Because of the power amplification in the ss-RF, damage to the Q-meter may result if these systems are run simultaneously.  Cycling between RF manipulation and NMR measurement can result in additional uncertainty in the NMR measurement due to the delayed sampling and the evolution of the spin state in the ensemble over time.

The ss-RF is applied by modulating the frequency over the domain of interest. The pedestal region of the smaller intensity ($I_-$) can be brought to near saturation to optimize.  Saturation occurs when the RF drives the population of the magnetic sublevels to equalize.  However, the peak is highly sensitive to power as it is higher in magnitude, and it is necessary to preserve as much of the larger intensity ($I_+$) underneath the $I_-$ peak as possible.  Optimization requires just the right amount of RF power to reduce the area in $I_-$ without depleting $I_+$.

\begin{figure}[!tbp]
  \centering
    \includegraphics[width=100mm]{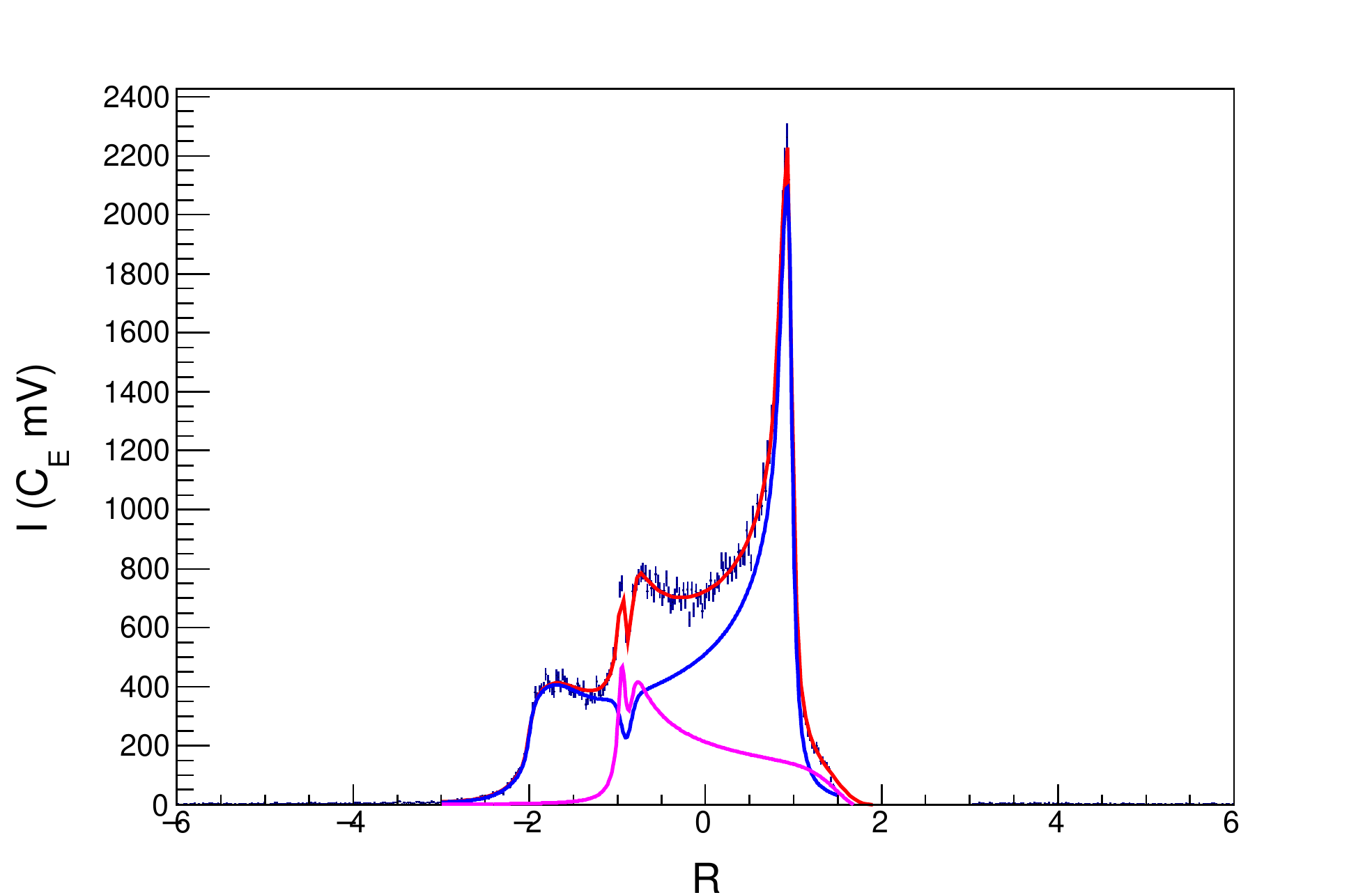}
    \caption{NMR measurement with fit result after the ss-RF has been applied to the two negative tensor regions.  The tensor
    polarization from this example is about 32\%. Figure from \cite{kel5}.}
    \label{dat2}
\end{figure}

To maximize tensor polarization, first the DNP process is used to build up the available polarization as much as possible, and then the spin system can be optimized for either the vector polarization observable $A_{UT}^{\text{sin}(\varphi_{sc}+\varphi_c)}$ or the tensor polarized observable $A_{E_{xy}}$.  Once the DNP process has maximized the NMR signal area (to around 50\%), then the ss-RF is imposed to maximize tensor polarization.  A fit to the data is also shown, indicating the $I_+$ intensity in blue and the $I_-$ intensity in pink.  Figure \ref{dat2} shows the NMR measurement after the ss-RF has been applied to the two negative tensor regions in $R$.  A fit to the data is shown, which uses only the constraints from Eq. \ref{dos1} and Eq. \ref{dos2}, resulting in a tensor polarization measurement of about 30\%.  In optimal circumstances, using a combination of AFP and ss-RF, much more than 30\% tensor polarization can be achieved (over 35\% has been demonstrated using multiple RF techniques).  Fortunately, as the target polarized decays due to radiation damage, the enhancement potential per total NMR signal area increases.  This is simply because the two peaks in the pake doublet become closer in area.

\begin{figure}
    \centering
    \includegraphics[width=100mm]{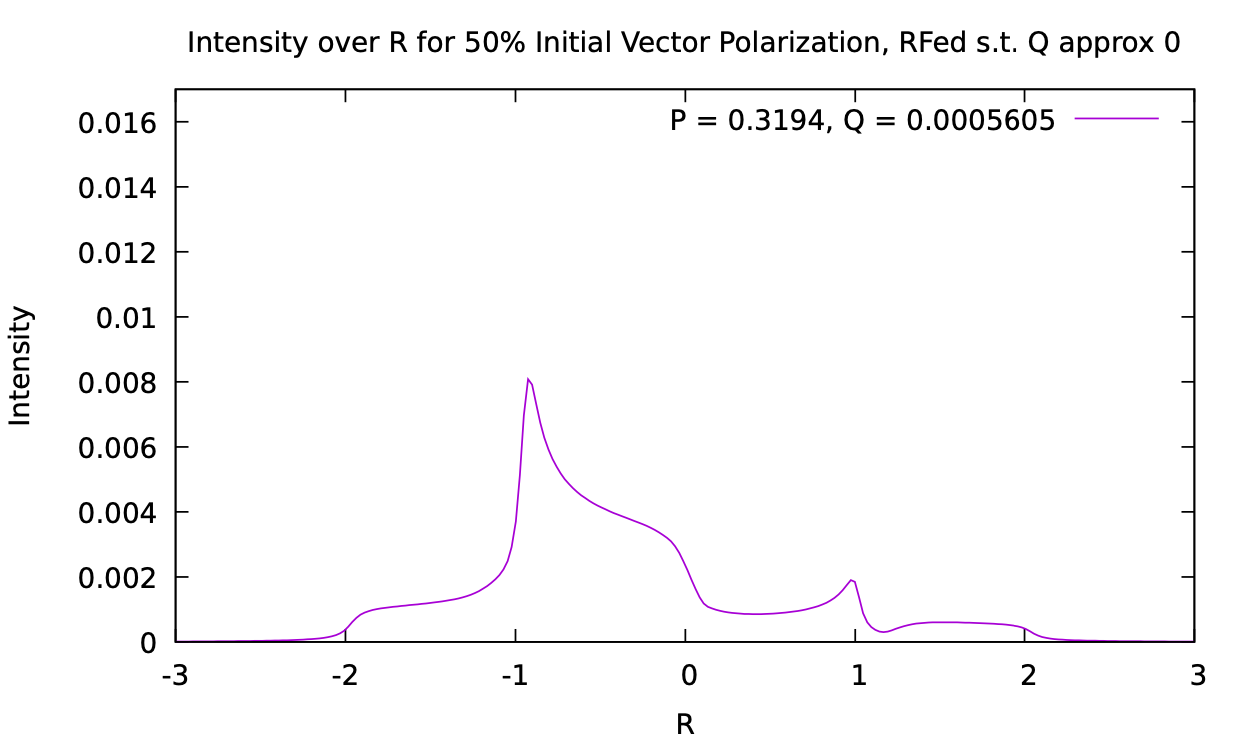}
    \caption{The NMR lineshape for ss-RF manipulated deuteron signal with zero tensor polarization and roughly 35\% vector polarization.}
    \label{zerozz}
\end{figure}

 The same techniques that are used to enhance the tensor polarization can also be used to reduce it. This is done by applying the ss-RF to the larger peak and manipulating the signal so that $I_-$ and $I_+$ are equal.  The ss-RF manipulated signal lineshape is show in Fig. \ref{zerozz} for vector polarization of $P=32\%$ and $P_{zz}=0$. 

The proposed experiment will take advantage of this novel polarized target system, providing a farther physics reach by improving the figure of merit of the polarized observables and providing a method to disentangle the polarized neutron observables associated with quark transversity and the tensor polarized observable associated with gluon transversity.  This is achieved best by polarizing under Boltzmann equilibrium with DNP, then enhancing the tensor polarization, and then alternating for subsequent beam spills to zero tensor polarization (with a pure vector polarization), and then back again.  The ss-RF manipulation can be done on the order of several seconds, so these target spin flips can be done in-between the beam spills and continuously cycled to reduce the over all systematics.

\subsubsection{Target Polarization Uncertainty}

The lower limit for polarization uncertainty is set by the Q-meter style NMR,
which can not be expected to preform better than 1\% relative error. UVA test
lab studies have gone down as far as 1.5\%, but typically in an experiment
2-4\% is achieved for the proton. The Deuteron/neutron has much larger error,
but, with the use of the cold NMR system \cite{kel1,kel2} in combination with the
multiple measurement techniques, it is also possible to get down into the same
uncertainty region as the proton.  There is work on reducing this uncertainty even further at UVA, using the characterization of the Q-meter electronics and modern signal extraction techniques.

\subsubsection{FNAL Auspicious Beam Cycle}
Fermilab's unique beam cycle from the Main Injector, 4.4 seconds of high intensity proton beam with 55.6 seconds before the next spill, allows applications for polarized fixed target experiments not otherwise achievable.  The superconducting polarizing magnet cannot withstand such a continuous heat-load at such high beam intensity.  We intend to run at the highest intensity that the SpinQuest target magnet can withstand without creating local hot spots in the coils that go over the superconductor critical temperature, resulting in magnet quenching.
 The time between spills allows for higher instantaneous intensity than previously achieved by allowing time for the coil temperature to decrease back to baseline before another spill.  That combined with the length (8 cm) of the SpinQuest target means that SpinQuest is operating at the polarized target \textit{intensity frontier}.  
 
 The time between spills also allows for RF manipulation of the target spin configuration using ss-RF and AFP.  The manipulation can be applied between the spill to prime the target for an enhanced state during the spill.
 Since these RF-generated spin states are not in thermal equilibrium, holding these states without a decrease in polarization is not possible.  The time between beam spills allows for recovery before the next spill.  For this experiment, tensor enhancement using a combination of AFP and ss-RF will increase the tensor polarization over the beam spill and then rebuild polarization when waiting for the next spill.  This also makes it possible to cycle between a vector enhanced and a tensor enhanced target.  This drastically reduces the uncertainty associated with time dependent drifts in the systematics.
 
 Temporarily enhanced spin states during beam-target interactions are much more plausible at facilities that have a short beam spill per cycle such as Fermilab.  The unique beam cycle of the high intensity proton beam at Fermilab allows for special characteristics of the thermal properties of the solid-state polarized target system to be employed, allowing significant improvement over any other facility to run intense proton beams on novel RF-manipulated target systems. The combination of high luminosity, large $x$-coverage, and a high-intensity beam with significant time between proton spills is paramount for this new approach to measuring polarized target asymmetries in Drell-Yan scattering with high precision.  This makes Fermilab very unique in this regard and provides a great potential for this proposal and for future projects.

\subsubsection{Kinematic Dilution Factor}

The figure of merit for this type of polarized target experiment is
proportional to the active target contribution squared times polarization
squared. The active target contribution is made of the dilution factor and
the packing fraction over the length of the target. The packing fraction can
be measured using a method of cryogenic volume displacement measurement which
compares an empty target cell to the full target cell used in the experiment.
The target cell is filled with beads of solid ND$_3$ material with a typical
packing factor of about 60\%, with the rest of the space filled with liquid helium.

The dilution factor is the ratio of the number of polarizable nucleons to
the total number of nucleons in the target material and can be defined as,
\begin{equation}
f=\frac{N_{D}\sigma_{D}}{N_{N}\sigma_{N}+N_{D}\sigma_{D}+\Sigma N_{A}%
\sigma_{A}},
\end{equation}
where $N_{D}$ is the number of deuteron nuclei in the target and $\sigma_{D}$
is the corresponding inclusive double differential scattering cross-section,
$N_{N}$ is the nitrogen number of scattered nuclei with cross-section
$\sigma_{N}$, and $N_{A}$ is the number of other scattering nuclei of mass
number $A$ with cross-section $\sigma_{A}$. The denominator of the dilution
factor can be written in terms of the relative volume ratio of ND$_{3}$ to LHe
in the target cell, the packing fraction $p_{f}$. For the case of a
cylindrical target cell oriented along the magnetic field, the packing
fraction is exactly equivalent to the percentage of the cell length filled
with ND$_{3}$. The dilution factor for ND$_{3}$ is 0.3 in terms of target nuclei accounting when considering the three deuterons to the one nitrogen per molecule.  The material density for ND$_3$ is 1.007 g/cm$^3$, and the target radiation length is about 5.7\%. The uncertainty in these factors from irreducible background is typically 2-3\%, considering the ammonia density and packing fraction as well.

There are more materials in the experimental beam path than just the ND$_3$, which means the amounts and cross-sections of those materials must also be accounted
for when calculating the kinematically sensitive dilution factor. Due to this, the dilution factor of the target will actually be given by the equation
\begin{equation}
f=\frac{3 d^{4} \sigma^{DY}_{D}\left(x_2, x_F, \phi, Q^2\right)}{\sum_{A} N_{A} d^{4} \sigma_{A}^{DY} \left(x_2, x_F, \phi, Q^2\right)}.
\end{equation}
Here, $A$ is required for each nuclei in the beam path.  Background that are not from Drell-Yan must also be considered.

\begin{table}[h!]
  \begin{tabular}{|c|c|c|c|c|c|c|} \hline
	\centering
 \multirow{2}{*}{$x_2$-bin} &  \multirow{2}{*}{$<x_2>$} & \multicolumn{2}{c|}{ND$_3$ }& \multicolumn{2}{c|}{ND$_3$+$A$} & Full  \\ \cline{3-7}
	                          &                           & $f$                        & $\delta f$ (\%) & $f$ & $\delta f$ (\%)  &  $\delta f (\%)$ \\ \hline
        0.10 - 0.16           &        0.139              &  0.305 & 0.3\%             & 0.310& 0.6\%    &  2.3\% \\
        0.16 - 0.19           &        0.175              & 0.304  & 0.5\%             & 0.319 & 0.7\%   &  2.4\%\\
        0.19 - 0.24           &        0.213              & 0.303  & 0.5\%             & 0.327 & 0.7\%   &  2.4\%\\
        0.24 - 0.60           &        0.295              &  0.306 & 1.6\%             & 0.341& 1.8\%    &  2.5\%\\ \hline
	\end{tabular}
  \caption{Here we list the dilution factor based on the MCFM simulations with kinematic sensitivity in $x_2$ for our four kinematic bins for pure deuterated ammonia (ND$_3$), as well as for
  the contribution from all materials (ND$_3$+$A$), as well as the total with contribution from packing fraction and target density in (Full).  Errors contain contributions from both statistical and systematic uncertainty estimates.}
  \label{dil}
\end{table}

To estimate the contribution from other materials in the beam path, such as the aluminum windows, the target cell material, the NMR coils, the liquid Helium, and the target ladder, a cross-section generator called Monte Carlo Femtobarn (MCFM) \cite{mcfm} was used.  The MCFM program was designed to calculate cross-sections based off of the parton distribution functions for various femtobarn-level processes in hadron-hadron collisions.  A number of processes can be calculated at next-to-next-to-leading order in QCD.  We use this software to estimate the Drell-Yan cross-section for our kinematics.  Table \ref{dil} shows the resulting dilution factor from pure ND$_3$ as well as from the combination of all material in the beamline, shown in column (ND$_3$+$A$).  The total percentage of dimuon yield is obtained by using the GEANT4 with our target and spectrometer geometries constructed in the simulations.  We also show the error estimate associated with relying purely on MCFM to provide the necessary cross-section to calculate the dilution factor for that kinematic bin.  We also show in column (Full) the combined error estimation from MFCM, the target packing fraction, and the ND$_3$ density.  

\section{Beamline}

The Neutrino-Muon (NM) beamline currently supporting the E1039 Drell-Yan experiment delivers a high-intensity (up to 10$^{13}$ protons/4.4-sec spill over 60 sec), 120-GeV proton beam. The FNAL power supplies and timeline cannot handle a longer spill cycle. The experimental beam has the 53 MHz microbunch characteristics of the Fermilab Main Injector RF structure with 19 nsec substructure and the longer microstructure of consecutively injected Fermilab Booster beam batches, with appropriate intervening kicker gaps separating the injected batches.  There is a 40\% duty cycle with an expectation of about 46\% beam time.  After a lengthy beamline of a couple of kilometers interspersed with vacuum windows and in-beam diagnostics, such as Secondary Emission Monitors (SEMs), the beam is distinctly Gaussian with Lorentzian tails. These tails are problematic for the superconducting cryogenic coils that polarize the E1039 target.  However, this beamline is uniquely suited to tailor and customize beam properties -- upstream beam collimation allows for both matching the beam profile to the dimensions of the polarized target vertically and horizontally and protection against a quench of the magnet, without creating increased backgrounds at the experiment.

\subsection{Current Beamline Configuration}

The  beam is slow-spill extracted from the Fermilab Main Injector on the half integer resonance and travels a couple of kilometers to the SpinQuest target area in NM3.  Losses in the couple of hundred meters upstream of the target are on the percent level and large backgrounds are not created in the experimental area. Although slow spill produces an asymmetric, non-elliptical phase space in the horizontal plane, after traveling through vacuum windows, diagnostics, and other sources of scattering, the beam in both planes becomes Gaussian-like (with Lorentzian tails) and even symmetry. (The vertical split of beam to the MTEST and MCENTER lines is at such low intensity that the beam profile in this high-intensity line is not observably impacted.) 
The NM/E906 beam properties have been extensively studied to determine how to achieve the requested beam profile on the polarized target.  A minimal spot size of $\sigma = 3 - 4 $ mm is the smallest obtainable in both planes simultaneously with the present beamline magnet configuration and distances involved.  The polarized target is 2 meters upstream from the E906 targets, so these measurements apply.  

The new experimental beam line with collimator has been constructed for SpinQuest.  The beam has been tailored to the target cell dimensions and has been modified to be more flat with drift mitigation (minimizing beam role) with FNAL quality assurance.  No magnet reconfiguration or additions are required with beam collimation, greatly reducing the cost and lab resources required.  The present beamline magnet configuration can thus be used for this proposed experiment.

\begin{figure}
    \centering
    \includegraphics[width=150mm]{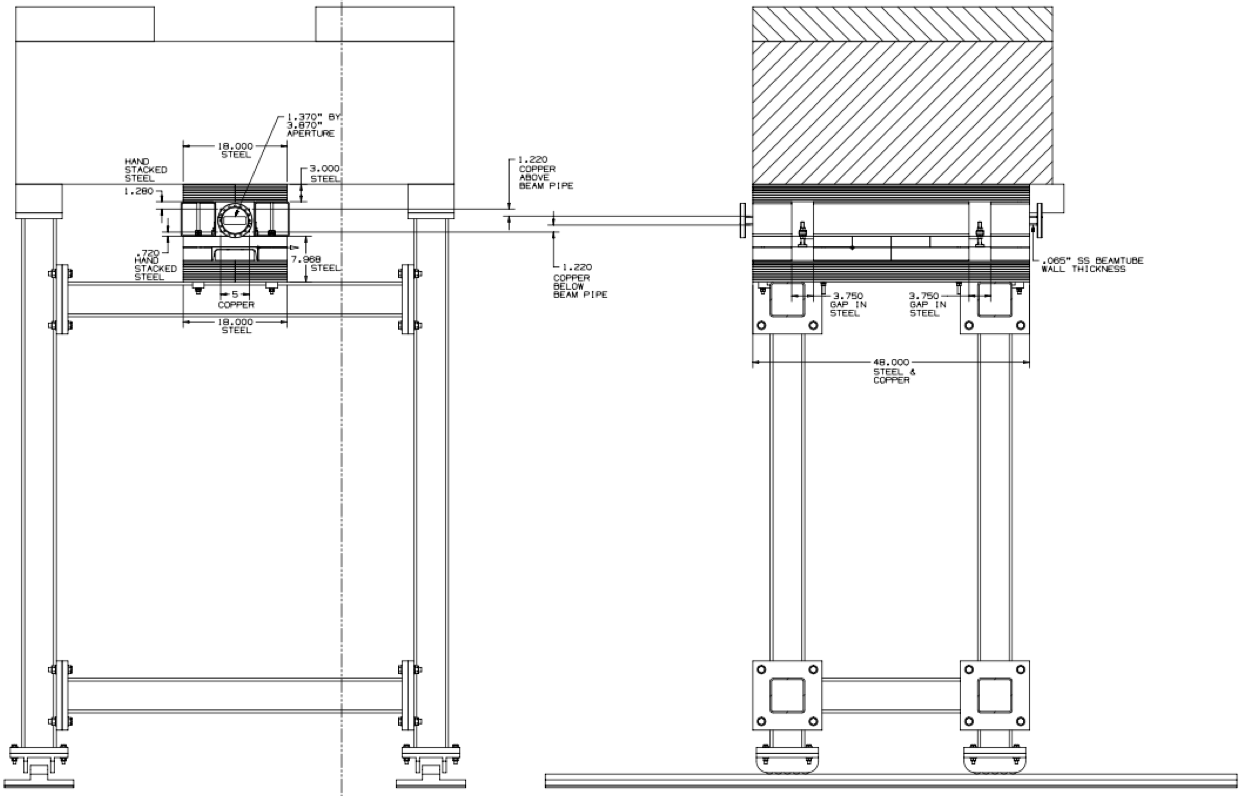}
    \caption{The SpinQuest collimator that is presently installed in NM4.  The collimator provides the needed protection to the superconducting magnet coils of the target.}
    \label{coll_1}
\end{figure}

\begin{figure}
    \centering
    \includegraphics[width=60mm]{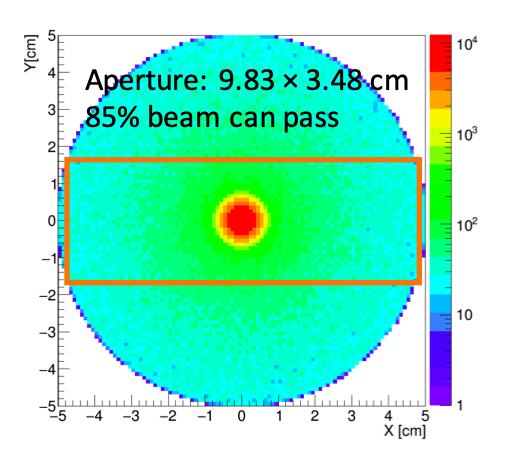}
    \caption{The simulation results after passing through the collimator.  As indicated, approximately 85\% of the beam passes through, cutting the tails in both the x and y direction.}
    \label{coll_2}
\end{figure}

The SpinQuest beam has been collimated by about $\sim$ 10\%, well upstream of the polarized target, to reduce the potential for quenching the superconducting magnet and also to more evenly distribute beam across the target.  To do this, the NM2 target pile from the kTeV experiment was used to absorb beam scattered by collimators (Palmer-style) positioned upstream of this pile.  The collimator and beam line modification have been designed by Carol Johnstone.  The collimator design can is shown in Fig. \ref{coll_1}.  Our simulations showing the surviving beam on the target center are shown in Fig. \ref{coll_2}.  The collimation scheme requires 90$^\circ$ vertical and horizontal collimators embedded in two FODO cells.  Quad triplets are at the exit and entrance of upstream/downstream beam pipes.  This scheme with optics has been approved and constructed.

By installing the collimators upstream, the beam can be collimated and tails clipped, scattered, and completely absorbed by the NM2 target pile with little background reaching the experiment.  A MARS study is planned for this configuration. 
Finally, a fixed collimator in the NM3 enclosure will shadow the SC coils of the polarized target to protect it, not only from any residual halo but also beam steering, allowing target scans.
The present E906 beamline is ideal for the proposed E1039 experiment, and especially for a polarized target.  No modification to the present beamline magnet component configuration or new optics is required outside of replacing two NM2 4Q120s (not in service) with collimators.  These collimators are available and already located within the NM2 enclosure, so no extensive rigging and drop hatch work is required.   This beamline represents the most cost effective approach to the proposed polarized target Drell-Yan Experiment.

\section{Count Rates and Statistical Errors}

The total Drell-Yan count rates on the ND$_3$ target are calculated using our full GEANT4 based Monte Carlo simulation package with Drell-Yan events generated by the NLO calculations with performance of the Fermilab Main Injector imposed, as well as the geometry and detector characteristics from the SpinQuest spectrometer.

The primary physics interest of this proposal is to measure a broad \xT\ range, as in E1039, of transversely tensor polarized Drell-Yan production.  The target position will be the same as it is for E1039, providing the necessary kinematics of the asymmetries of interest with vector polarized asymmetry $A^{\text{Sin}(\varphi_{sc}-\varphi_s)}_{UT}$ (we will use $A_{UT}$ for short) and tensor polarized asymmetry $A_{E_{xy}}$.  
Through $A_{UT}$, we can access the sea-quark contribution to transversity from the vector polarized neutron.  The neutron caries 0.91\% of the polarization of the deuteron, which is why the information from this asymmetry is largely dominated by the neutron.  The statistical error from a $A_{UT}$ measurement is,
\begin{equation}
\delta A_{UT}=\frac{\sqrt{2}}{fP_n}\frac{1}{\sqrt{N_{E_{\pm}}}},
\end{equation}
where $P_n$ is the polarization pertaining to the neutron, and $N_{E_{\pm}}$ is the number of counts from the vector polarized deuteron target.

The $A_{E_{xy}}$ observable represents the linearly polarized gluon asymmetry and comes from a tensor polarized target aligned transversely with respects to the beamline.  The statistical error from a $A_{E_{xy}}$ measurement is,
\begin{equation}
\delta A_{E_{xy}}=\frac{1}{fP_{zz}}\sqrt{\frac{N_{E_{xy}} (N_u+N_{E_{xy}}) }{N^3_u} }.
\end{equation}

The primary bottleneck of the data collection efficiency at E906/SeaQuest was the slow Data Acquisition System (DAQ). A very tight trigger level selection was implemented as to accommodate as many events in the limited DAQ bandwidth as possible. The DAQ system was
upgraded for SpinQuest to reduce this bottleneck as much as possible.  We anticipate using the same SpinQuest DAQ.

Another limiting factor of the data collection efficiency at E906/SeaQuest was the unstable instantaneous beam intensity, which is sometimes more than one order of magnitude larger than average. To prevent the spectrometer from being completely saturated, the total number of protons delivered to the target has to be limited to no more than $3\times10^{12}$ per spill to indirectly constrain the instantaneous beam intensity. This also requires the data taking to be inhibited on all neighboring RF buckets when a high intensity bucket arrives. After careful optimization, E906/SeaQuest has been able to record on average $2.67\times10^{12}$ protons per spill, but the polarized target limit has been studied carefully and is limited to about half of that for the most \textit{helium conservative} type of running.  The theoretical limit has been calculated to be as high as $4\times10^{12}$ protons per spill, but this would require pumping on the superconducting target magnet and would result in significantly more helium consumption.  So, considering the \textit{helium conservative} approach, the requested count would be $4\times10^{17}$ protons per calendar year on the polarized target.  It is also necessary to include beam for unpolarized target measurements.  This can be done while irradiating the fresh ND$_3$ target material to be optimized for polarization.  This can be done under much higher intensity when the superconducting target magnet is not energized.  The total protons requested is $8\times10^{17}$ per calendar year.

For estimating the rates, we focus on vector and tensor polarization cycles.  The associated error is driven by the counts and the polarization as well as the dilution factor.  We use the difference between purely vector polarized and a tensor polarization enhanced state.  This requires a statistical contribution from both spin states.  The associated error estimate is shown for each state with vector for $\delta A_{E_{xy}}$ as well as tensor. We also estimate the quark transversity asymmetry error based on the neutron polarization indicated as $A_{UT}$.

After running for 2 years with beam time shared over the vector and tensor target configuration, the integrated luminosity on \ND target is expected to be $1.82\times10^{42}$($1.05\times10^{42}$) cm$^{-2}$. With the various assumed efficiencies shown in Table \ref{VarEff}, the final event yield and statistical precision of $A_{UT}$ measurement in each \xT bin is summarized in Table \ref{EventYields}. Here, the statistical precision is calculated by $\delta A_{UT} = \frac{1}{f}\frac{1}{P}\frac{1}{\sqrt{N}}$, where $f$ denotes the dilution factor, $P$ denotes the average polarization, and $N$ denotes the event yield in each \xT\ bin.
\begin{table}[h]
  \scalebox{0.8}{
  \begin{tabular}{|c||c|c|c|c|c|c|c|c|} \hline
	\centering

    Sources &  Target/Accelerator & Spectrometer & Acceptance & Trigger & Reconstruction \\ \hline
	  Efficiency (\%) & 50 & 80 & 2.2 & 90 & 60 \\ \hline
	\end{tabular}}
  \caption{Various efficiencies assumed for the count rate estimates based on previous experience with E906 and polarized target operations.}
  \label{VarEff}
\end{table}
\begin{table}[h!]
  \begin{tabular}{|c|c|c|c|c|c|c|} \hline
	\centering
 \multirow{2}{*}{$x_2$-bin} &  \multirow{2}{*}{$<x_2>$} & \multicolumn{2}{c|}{Vector-\ND($d^{\uparrow}$)} & \multicolumn{2}{c|}{Tensor-\ND($d^{\uparrow}$)} & $n^{\uparrow}$  \\ \cline{3-7}
	                            &                           & $N$              & $\delta A_{E_{\pm}}$ & $N$ & $\delta A_{E_{xy}} $&  $\delta A_{UT}$ \\ \hline
        0.10 - 0.16           &        0.139              & $2.1\times10^4$  & 7.1             & $2.2\times10^4$ & 15.8  &  7.4\\
        0.16 - 0.19           &        0.175              & $1.9\times10^4$  & 7.5             & $2.0\times10^4$ & 16.7  &  7.8\\
        0.19 - 0.24           &        0.213              & $2.4\times10^4$  & 6.7             & $2.5\times10^4$ & 14.9  &  7.1\\
        0.24 - 0.60           &        0.295              & $2.4\times10^4$  & 6.7             & $2.5\times10^4$ & 14.9  &  7.1\\ \hline
	\end{tabular}
  \caption{Event yield and statistical precision of the $A_{E_{xy}}$ and $A_{UT}$ measurement in each of the \xT\ bins for the vector polarized \ND($d^{\uparrow}$) and the tensor polarized \ND($d^{\uparrow}$) targets, as well as the deduced $A_{UT}$ measurement precision for polarized $n$.}
  \label{EventYields}
\end{table}
\pagebreak

\section{Luminosity and Beam Intensity}
\subsection{Beam Profile}

The typical profile of the beam delivered to the target is a two dimensional Gaussian with a width of $\sigma_{x}$\,=\,6.8~mm, $\sigma_{y}$\,=\,7.6~mm.  The beam will be clipped with collimators at $\pm$1.25$\sigma$, giving a beam profile of $\Delta x$\,=\,17~mm,  $\Delta y$\,=\,19~mm.  The beam is expected to drift no more than $\pm$2~mm in the $x$-direction before collimation.  The change of the luminosity of the beam due to the beam drifting is $\left(N_{beam} - N_{drift} \right)/N_{beam}$.   The change in the delivered Luminosity is $\Delta \mathcal{L}$\,=\,2.8$\%$.

\subsection{Luminosity measurement \label{sec:lummeas}}
Several detector and measurement techniques can be used in order to control systematic uncertainties from changing beam conditions, such as position, luminosity, and shape. The absolute beam intensity will be determined by Unser Monitors, which are upstream of the target. The accuracy of Unser Monitors has been established to be 0.05$\%$ \cite{unser}. 

The beam cherenkov and the $90^{\circ}$ luminosity monitor will help monitor the instantaneous luminosity. The $90^{\circ}$ monitor (constructed and installed by ACU) consists of four plastic scintillators in coincidence and positioned outside of the shielding wall, pointing through a small hole in the shielding at the target.  Fast MC simulations show that these detectors will detect normal $\pi^{\pm}$s, $\mu^{\pm}$s, $\gamma$s with $E$\,$>$\,100~MeV on the order of $\sim$200~kHz. 

The ratio of the $90^{\circ}$ monitor  detector over the Unser Monitor measurement ($N_{90^{\circ}}/N_{unser}$) will provide a fast relative luminosity measurement.  If part of the beam profile deviates off the target after the Unser measurement, the $90^{\circ}$ detector will be able detect luminosity changes to $>$1$\%$. This error comes from the efficiency of the four fold coincidence scintillators in each $90^{\circ}$ detector.  If each scintillator paddle is $\epsilon_{scint}$\,=\,99.8$\%$ efficient, the total efficiency goes as  $\epsilon_{scint}^{4}$\,$\approx$\,99$\%$.

As an additional check on the relative beam intensity, a four plate RF cavity can be installed, which can also determine relative changes in the beam position, $N_{90^{\circ}}/N_{RF}$\,$\propto$\,($N_{90^{\circ}}/N_{unser}$). 

\subsection{Consistency in Delivered Luminosity}
Since extracting the Sivers asymmetry for the \dbar requires measuring the ratio of  $\frac{\sigma^{pd}}{2\sigma^{pp}}$ , care has to be taken that the running conditions for both targets are as identical as possible. Our target system will have three identical cells: two filled with ND$_{3}$ and one used for empty target studies required for
background subtraction, which can be replaced with a Carbon disk, to study false asymmetries.

\clearpage

\section{Overall Systematic Error}
There are several contributions to the systematic error for this type of polarized target experiment.  There are, of course, several novel approaches to this experiment which makes some of the systematic errors hard to estimate, but we can at least make conservative projections.  From the target, we expect contributions from the polarization calibration measurements, where a large part of this can be from the temperature/pressure measurements. Systematic errors associated with the Q-meter baseline measurements can be induced by changes in the RF ambience.  There can also be target magnet field drifts, errors in the charge averaging, as well as NMR tune drifts.  All of these have been studied thoroughly, and the full combination of these have been brought below 3\% relative uncertainty.  There are also errors in the packing fraction, previously mentioned.  Error from having a long target cell can also occur leading to polarization homogeneity.  This can also happen from microwave induced differences along the length of the cell.  Some polarization in-homogeneity is also expected from beam heating and secondaries scattering off the down stream end of the target cell, as well as producing uneven decay in polarization due to radiation damage.  There are also the contributions from the ss-RF measurements and the NMR delays to produce the ss-RF irradiation.  We estimate that all of these can be carefully managed to result in an additional relative uncertainty of less than 3.5\%.  The total target contributions, including dilutions factor, density, and packing fraction, we expect to be around 5\% relative.

Also considering the beam contributions, such as relative luminosity and alignment, as well as DAQ and background, we estimate at total relative uncertainty of just over 6\% (see Table \ref{SysErr}).  We also expect an absolute systematic error due to the muon spectrometer of $<$1.0\%.
\vspace{1cm}
\begin{table}[h]
	\begin{tabular}{|l||c|} \hline
	\centering
	Quantity & Error \\ \hline
	Target Contributions & 5.0\% \\
	Beam Contributions & 2.5\% \\
	DAQ and Dead Time & 1.5\% \\
	Background  & 2 \% \\
		\hline
	\end{tabular}
\caption{\it {Estimates for the systematic errors}}
\label{SysErr}
\end{table}

\clearpage

\section{Expected Results}
\label{results}
For numerical estimates of transversity extracted from the deuteron, we must be careful about the scaling variable as the momentum fraction for a parton $q$ is in the deuteron which is defined by $p_q=x_2(p_d/2)$, by using the deuteron momentum per nucleon in our kinematical range $0\le x_2 \le 2$.  In the range $1\le x_2 \le 2$, the PDFs are very small, and there is no approach to extract information experimentally, so this part of the range can be neglected.  The definition of $x_2$ corresponds to the Bjorken scaling variable $x=Q^2(2M_N\mu)$ used in lepton DIS experiments with the deuteron.  The cross-sections and the linearly polarized asymmetry is expressed in $x_b$, which is the momentum fraction of the partons in the deuteron and can just as easily be expressed in terms of $x_{Bj}$ by scaling the variable accordingly.

\subsection{Quark Transversity}
As mentioned previously, there has been an effort made to extract some information on the sea-quark contribution \cite{anna}, but there are no constraints from experiment to apply to make any realistic determination of sea-quark transversity.  Especially in the case of the $\bar{d}$ quark, the error is very large.  We use these results in \cite{anna} to demonstrate the possible constraints that this proposal could add by superimposing our potential measured points over the band of uncertainty presently surrounding the sea-quark transversity observable.  The error bars on our points are estimates based on the bound suggested in \cite{anna} for the statistic requested for the vector polarized part of the experiment.  The systematic error is also included in this projection.
\begin{figure}[ht!]
  \centering
  \includegraphics[width=100mm]{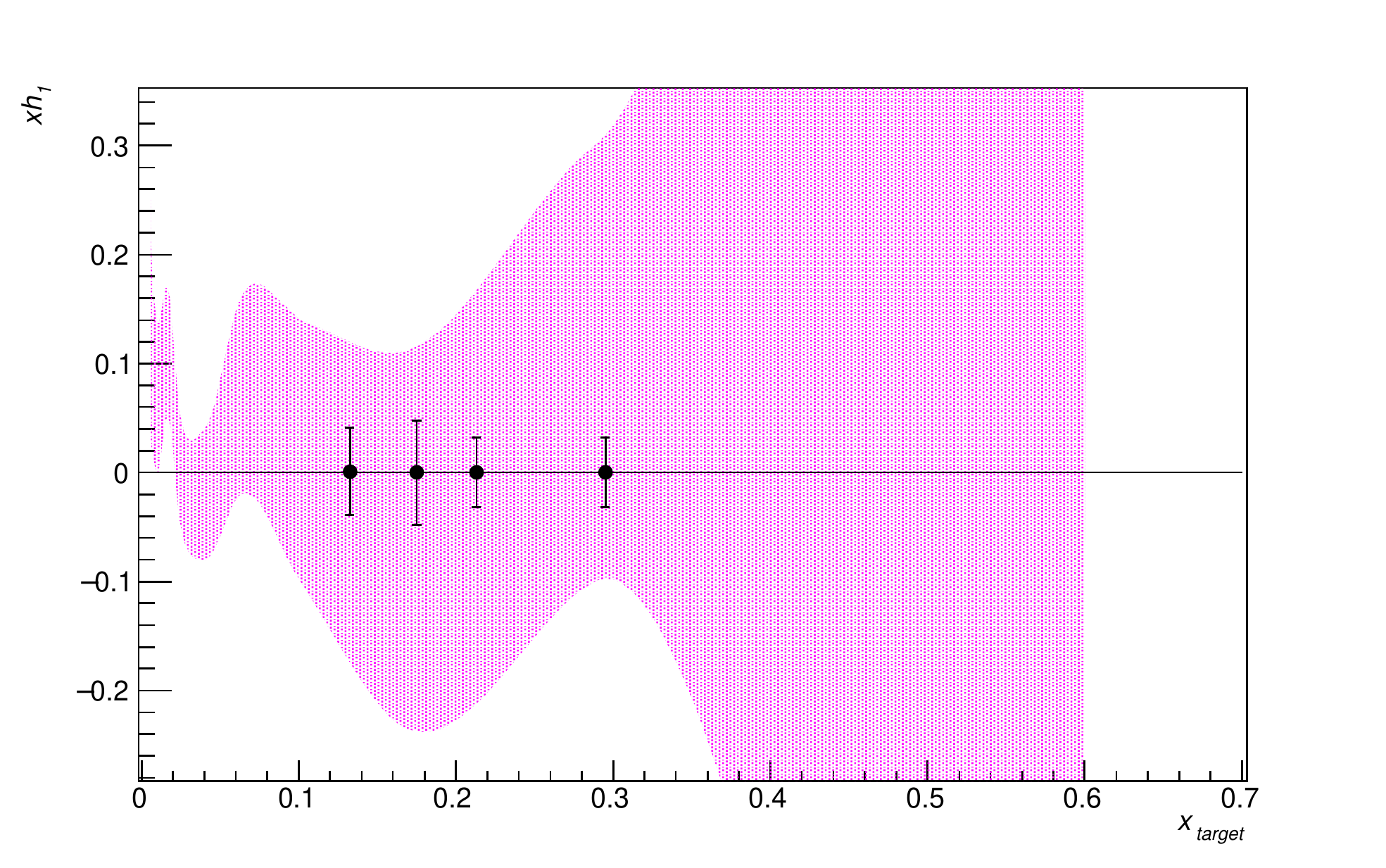}
  \caption{The model error band for sea-quark $\bar{d}$ transversity deduced from \cite{anna} for our range of kinematics.  This projection is based on SIDIS data which is insensitive to sea-quark contributions, hence the large error.}
  \label{staterror}
\end{figure}
The calculations are based on global fits to the available SIDIS data.  It is worth noting the fact that the current SIDIS data is insensitive to the sea-quark contribution, thus leading to large uncertainties in the calculations. This is also reflected in the width of the uncertainty bands. 

\subsection{Gluon Transversity}
For numerical estimates of gluon transversity, it is useful to employ the positivity bounds on gluon TMDs \cite{cotogno}.  The gluon polarization depends on a $6\times6$ matrix in gluon$\otimes$hadron spin space.  This matrix is positive semidefinite, a property which allows for setting constraints on the gluon distributions.  Given the limited amount of information there is currently on gluon functions, we use these Soffer-like bounds, providing a range with any realistic model lying within it.  The resulting bound for gluon transversity is,
\begin{equation}
    |h^g_{1TT}|\le\frac{1}{2}\left(f^g_1+\frac{f^g_{1LL}}{2}-g^g_1 \right).
\end{equation}
We neglect $f^g_{1LL}$, assuming the value is quite small as no gluon contributions have been seen in previous measurements \cite{b1}.

\begin{figure}[ht!]
  \centering
  \includegraphics[width=95mm]{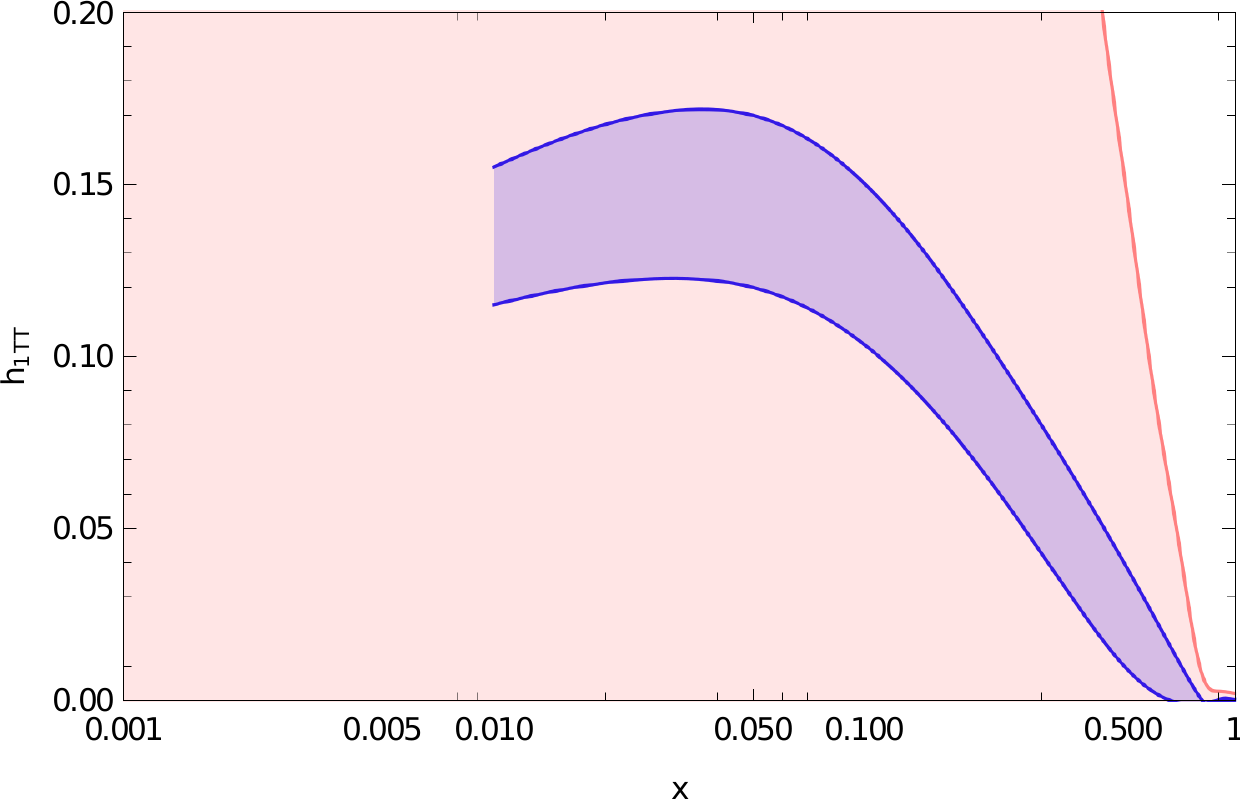}
  \caption{The model for $h^g_{1TT}$, suggested by Kumano, for our range of kinematics is shown in violet.  The Soffer-like positivity bound is also shown in the same $x$ region. (This figure maybe removed or added with projection points) }
  \label{t1}
\end{figure}

We also evaluate the model suggested in \cite{qin} over the kinematic range accessible at SpinQuest.  The model shown takes an average over the SpinQuest $q_T$ range weighted by cross section and with the azimuthal angle at $\phi=0$.  In Fig. \ref{t1}, we show the model prediction of $h^g_{1TT}$ in purple and the Soffer-like bound in pink.

\begin{figure}[ht!]
  \centering
  \includegraphics[width=95mm]{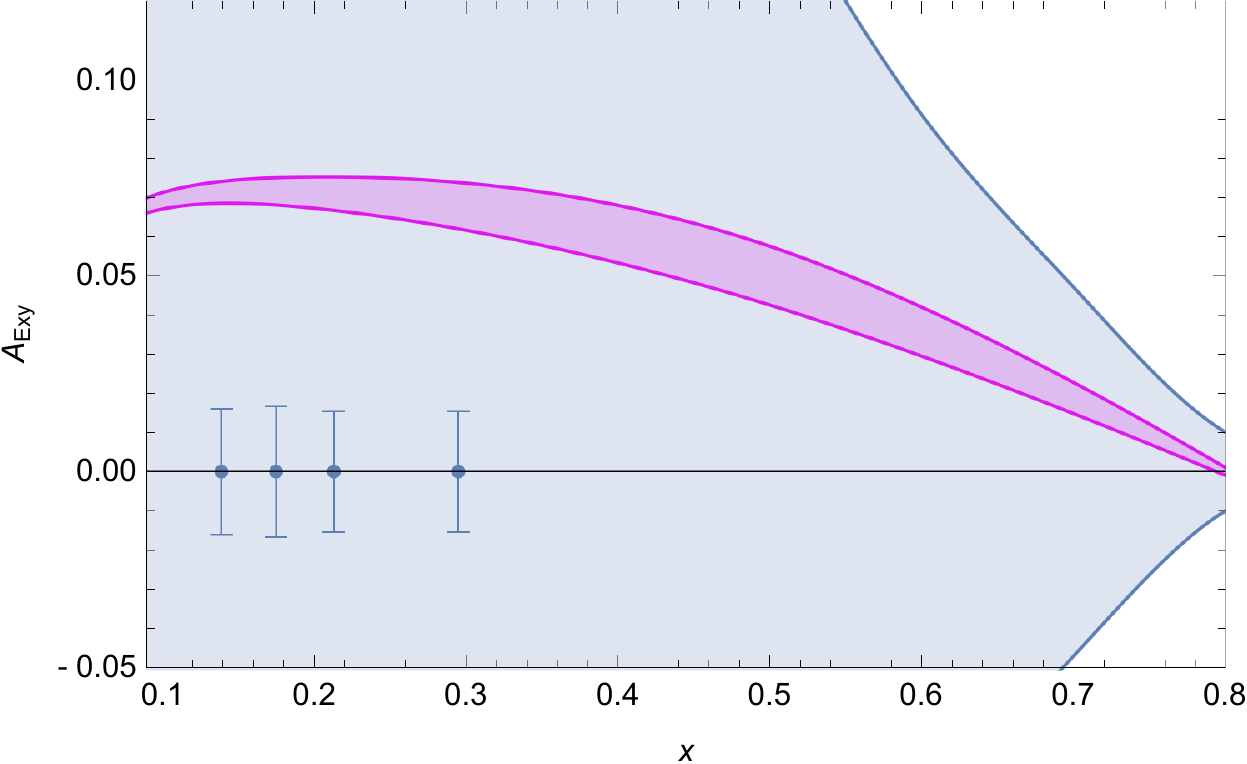}
  \caption{Projections of the linear polarized gluon asymmetry with
  expected errors from our proposed measurements.  We are assuming an additional 0.05\% absolute error as a conservative estimate in addition to the expected statistical and relative systematic contributions.  An average over our range in $q_T$ and $y$ is used, while $\phi$ is set to zero.  The Soffer-like positivity bound is also shown in grey and used for the upper limit, which provides a scale for demonstrating the information gained from the proposed measurement.}
  \label{t2}
\end{figure}

In Fig \ref{t2}, we show the expected results after two years of running with the \ND\ targets for the four selected kinematic bins with the linearly polarized asymmetry $A_{E_{xy}}$. The errors displayed is the statistical precision listed in Table \ref{EventYields}, with the systematic error also taken into account. Here, an additional 2\% absolute error is included to be conservative.

With the proposed experiment, we will be able to analyze the dependence of gluon transversity in $\phi$, $q_T$, $x_2$, and $Q^2$, providing substantial information on the dynamics and spatial gluon structure of the deuteron.

From the analytical scope of QCD, there is a certain ubiquity of gluons to consider in almost any relevant process.  However, probing the gluonic structure of hadrons and nuclei is considerably more difficult than that of quarks.  To some extent, this can be accredited to the significant innate challenges in measurements of gluon observables, which are usually  $\mathcal{O}(\alpha_s)$-suppressed relative to the quark observables.  Here, we suggest a measurement that can provide significant information.  A finite value of the gluon $h_{1TT}$ is likely to trigger a multitude of new experiments to probe the full kinematic range of this observable to help map out and detail the relationship between the nuclear geometry and the gluonic structure.  It has also been suggested \cite{jaffe} that the magnitude of this observable should increase with atomic number ($z$).  There is ongoing polarized target research \cite{kel3} to polarized higher $z$ solid-state targets. 

\clearpage

\section{Cost and Schedule}
The source and operating responsibility of each component of the apparatus is expected to be similar to that of E1039 with the designated responsibilities organized according to the SpinQuest bylaws.  Additional collaborators are also expected as E1039 begins to ramp up.  We have assumed that all the existing equipment from E906/E1039 is available for this experiment. The critical path of the timeline is clearly the successful completion of E1039.  Following approval of the proposal, we will request from DOE-NP the necessary funds for operations.  We expect that the DOE-HEP will match the DOE-NP support, similar to the E1039 budgeting scheme. The expense required for the target system updates and modifications will be covered by UVA.  We anticipate the rest of the collaboration contributions to come from continuing research grants.

The time-line for the running of this experiment is set by the successful completion of E1039.  The FNAL response and beam-time scheduling play a significant role in the expected timeline for this project as well.  Assuming E1039 ends before summer of 2023, then this project can start in Fall of that same year.  We also realize much of this depends on the progression and safety response to COVID-19. 

\subsection{Requests for Fermilab}
No additional requests are made beyond what was required for operations of experiment E1039.  We specifically request Fermilab to provide the following items: 

\begin{itemize}
  \item Provide/maintain beam dump with closed loop water recirculating cooling system. 
  \item Provide magnet power supplies.
  \item Provide all needed utilities (power and cooling water) for magnets, power supplies, Helium liquifier, and target EIO microwave tube. 
  \item Provide beam-line instrumentation.
  \item Provide chamber gas distribution system plumbing.
  \item Provide counting house and electronics areas with appropriate utilities installed.
  \item Provide beam-line staffing.
  \item Provide FNAL liaison necessary for operation.
\end{itemize}

\section{Summary}
We propose using the 120 GeV primary proton beam from the Main Injector to measure Drell-Yan yields for a vector and tensor polarization optimized deuteron (ND$_3$) target.  These measurements will provide precise information on the sea-quark transversity and the never before measured gluon transversity.  Despite the pervasiveness of gluons in QCD, there is truely a very limited understanding of the gluonic structure of hadrons and nuclei.  This can be attributed to the significant experimental challenges inherent in measurements of gluon observables that are typically $\mathcal{O}(\alpha_s)$-suppressed relative to quark observables.  Here, we suggest a novel approach to accessing gluon information in the non-perturbative kinematics.  We request two calendar years of running, or a total of $8\times10^{17}$ protons on target.

\clearpage
\section{Appendix}
\subsection{Statistical Uncertainty Calculation}
\subsubsection{Sea-Quark Transversity}
To calculate the statistical uncertainty for the sea-quark transversity asymmetry $A_{UT}$, we propagator the error from each
set of counts from each state,
\begin{equation}
\delta A_{UT} = \frac{\sqrt{2}}{fP_n\sqrt{N_{tot}}}.
\end{equation}
Now, to calculate the number of events needed to have for a given error in $\delta A$,which turns out to be
\begin{equation}
N_{tot} = \frac{2}{f^2 P^2_n\delta A_{UT}^{2}}.
\end{equation}
The counting time goes with the square of the dilution factor as well as the polarization. We can connect this to luminosity $L$, target thickness $t$, the cross-section $\sigma_{DY}$, and the measuring time $T_{meas}$ for a given statistical accumulation $N_{tot}$, written as,
\begin{equation}
N_{tot} = \frac{2}{f^{2}}\frac{1}{P_n^{2}}\frac{1}{\delta A_{UT}^{2}}= L t T_{meas},
\end{equation}
which we can solve for $T_{meas}$,
\begin{equation}
T_{meas} = \frac{2}{\sigma_{DY}}\frac{1}{\delta A_{UT}^{2}}\frac{1}{Lt f^{2} P^{2}_n}.
\end{equation}

\subsubsection{Gluon Transversity}
For the difference between the tensor enhanced and the pure vector polarized states, the number of counts are,
\begin{equation}
    N_{E_x}=P_{zz}\left( \frac{N_{E_{\pm,x}}}{P'} - \frac{N_{E_\pm}}{P}\right).
\end{equation}
The average vector polarization measured over the course of the run in the tensor enhanced state is $P'$, while the average vector polarization measured over the course of the run in the purely vector polarized state is $P$.  The average value of the tensor polarization in the tensor enhanced state is $P_{zz}$.  
Propagation of error leads to,
\begin{equation}
\delta N_{E_{x}}=\sqrt{\left(\frac{P_{z z}}{P^{\prime}} \delta N_{E_{\pm}}\right)^{2}+\left(\frac{P_{z z}}{P} \delta N_{E_{\pm, x}}\right)^{2}}=\sqrt{\left(\frac{P_{z z}}{P^{\prime}} \sqrt{N_{E_{\pm}}}\right)^{2}+\left(\frac{P_{z z}}{P} \sqrt{N_{E_{\pm, x}}}\right)^{2}}
\end{equation}

So, if we take an average value of the vector polarizations in the two states, then we can assume $P'=P$.  If we also assume these cross-sections are about the same, we can write,
\begin{equation}
    \delta N_{E_x}=\frac{P_{zz}}{P}\sqrt{N_{E_{x}}}.
\end{equation}

Now, turning our attention to the statistical uncertainty in the $A_{E_{xy}}$ asymmetry itself, we start with the experimental asymmetry expression,
\begin{equation}
    A_{E_{xy}}=\frac{2}{f P_{zz}}\frac{2N_{E_x}-N_U}{N_U}.
\end{equation}
The propagation of error leads to,
\begin{equation}
    \delta A_{E_{xy}}=\frac{4}{f P_{zz}} \sqrt{\left(\frac{\delta N_{E_{x}}}{N_U}\right)^2 + \left( \frac{N_{E_x}}{N_U^2}\sqrt{N_U} \right)^2  }.
\end{equation}
We can now plug in $\delta N_{E_x}$ from before, leading to,
\begin{equation}
    \delta A_{E_{xy}}\approx \frac{4}{f P_{zz}} \sqrt{\frac{1}{N^2_U}\left( \frac{P_{zz}}{P} \sqrt{N_{E_x}}\right)^2 + \frac{1}{N_U}}.
\end{equation}
If we assume $P'$ is close to $P$ and all the polarized cross-sections are on a similar order as the unpolarized case, we can simplify to,
\begin{equation}
    \delta A_{E_{xy}}\approx\frac{4}{f P_{zz}} \sqrt{\frac{P^2_{zz}+P^2}{P^2}\frac{1}{N_{E_x}}}.
\end{equation}

\end{document}